\PassOptionsToPackage{table}{xcolor}
\documentclass{article}
\usepackage[T1]{fontenc}
\usepackage{iclr2027_conference,times}
\usepackage{amsmath,amsfonts,bm}

\def\eqref#1{equation~\ref{#1}}
\def\1{\bm{1}}

\DeclareMathAlphabet{\mathsfit}{\encodingdefault}{\sfdefault}{m}{sl}
\SetMathAlphabet{\mathsfit}{bold}{\encodingdefault}{\sfdefault}{bx}{n}

\newcommand{\R}{\mathbb{R}}

\usepackage[hypertexnames=false]{hyperref}
\usepackage{url}
\usepackage{booktabs}
\usepackage{multirow}
\usepackage{makecell}
\usepackage{tabularx}
\usepackage{graphicx}
\usepackage{tikz}
\usepackage{float}
\usepackage{placeins}
\definecolor{paperlink}{RGB}{0,72,120}

\iclrfinalcopy
\hypersetup{
  pdftitle={AneumoBench: A Source-Linked Benchmark for Synthetic-Geometry Transfer in Aneurysm CFD},
  pdfauthor={Xigui Li, Yuanye Zhou, Feiyang Xiao, Xin Guo, Chen Jiang, Tan Pan, Xingmeng Zhang, Cenyu Liu, Zeyun Miao, Xiansheng Wang, Qimeng Wang, Yichi Zhang, Wenbo Zhang, Hongwei Zhang, Ruoxi Jiang, Fengping Zhu, Limei Han, Chensen Lin, Yuan Cheng}
}

\title{\raggedright AneumoBench: A Source-Linked Benchmark for Synthetic-Geometry Transfer in Aneurysm CFD}

\author{%
  \parbox[t]{\textwidth}{%
    \raggedright\normalfont
    {\bfseries
    Xigui Li\textsuperscript{1,2}\quad
    Yuanye Zhou\textsuperscript{2}\quad
    Feiyang Xiao\textsuperscript{1,2}\quad
    Xin Guo\textsuperscript{1,2,\textdagger}\quad
    Chen Jiang\textsuperscript{1,2}\\
    Tan Pan\textsuperscript{1,2}\quad
    Xingmeng Zhang\textsuperscript{1,2}\quad
    Cenyu Liu\textsuperscript{2}\quad
    Zeyun Miao\textsuperscript{1,2}\quad
    Xiansheng Wang\textsuperscript{2}\\
    Qimeng Wang\textsuperscript{2}\quad
    Yichi Zhang\textsuperscript{1,2}\quad
    Wenbo Zhang\textsuperscript{1,2}\quad
    Hongwei Zhang\textsuperscript{1,2}\quad
    Ruoxi Jiang\textsuperscript{1,2}\\
    Fengping Zhu\textsuperscript{1,2}\quad
    Limei Han\textsuperscript{1,2}\quad
    Chensen Lin\textsuperscript{1,2,\textdagger}\quad
    Yuan Cheng\textsuperscript{1,2,\textdagger}}\\[0.3em]
    \nolinkurl{guoxin@sais.org.cn}\quad
    \texttt{\{lixigui, linchensen, cheng\_yuan\}@fudan.edu.cn}\\[0.3em]
    \textsuperscript{1}Fudan University\\
    \textsuperscript{2}Shanghai Academy of Artificial Intelligence for Science
  }%
}

\begin{document}
\begingroup
\setlength{\tabcolsep}{0pt}
\maketitle
\endgroup
\lhead{Preprint}
\begingroup
\renewcommand{\thefootnote}{\fnsymbol{footnote}}
\footnotetext[2]{Corresponding authors.}
\endgroup

\begin{abstract}
Scientific machine learning uses simulation data to train surrogate models for fast physical-field prediction across geometries. Local shape editing can expand limited geometry collections, but whether its variants improve prediction on unseen geometries, and how to allocate them across sources, require controlled evaluation. We introduce \textbf{\mbox{AneumoBench}}, a dataset and benchmark linking 401 source aneurysm geometries to 9,693 locally edited descendant records, with computational fluid dynamics (CFD) fields computed on both. It contains 80,752 steady velocity--pressure cases across eight inlet conditions and 9,715 transient sequences of velocity, pressure, and wall shear stress (WSS). Each sequence contains 100 frames sampled at 0.01-s intervals from a 1-s cardiac cycle. Mesh, point, and voxel interfaces support steady field prediction and WSS forecasting from four observed frames. With family-disjoint splits, we compare source-only training, descendant training, and descendant pretraining followed by source fine-tuning across nine architectures on 79 held-out sources. Under the reported schedules, two-stage training lowers steady-field and reset-window WSS errors relative to source-only training. With the number of sampled fields and training updates fixed within each comparison, GraphSAGE benefits from descendant training and from distributing a fixed number of descendants across more sources. For WSS, reset-window gains do not consistently persist through 96-step rollout, and lower trajectory error need not improve cycle-level shear metrics or hotspot localization. These data and protocols enable researchers to compare descendant selection and training strategies on the same unseen source geometries.
\end{abstract}

\section{Introduction}

Scientific machine learning aims to accelerate physical-field prediction across geometries. For surrogate models trained on simulation data, prediction on unseen geometries depends on model design and the range of geometries in the training set \citep{wu2024transolver}. The cost of preparing, meshing, and simulating geometries limits the amount of training data. Intracranial aneurysms provide a challenging setting: complex vessel and sac shapes affect the flow \citep{7sforza2009hemodynamics,xiang2011hemodynamic}. Extending prediction beyond a limited source collection requires deciding which additional geometries to simulate.

Methods for learning physical fields include physics-informed neural networks (PINNs) \citep{raissi2019physics} and neural operators such as DeepONet and FNO \citep{lu2021deeponet,li2020fourier}. Recent work has advanced geometry-aware prediction through both model design and larger computational fluid dynamics (CFD) datasets \citep{wu2024transolver,garnier2025meshmask}. AirfRANS and DrivAerNet++ support evaluation across engineered geometries \citep{bonnet2022airfrans,elrefaie2024drivaernet}, while AneuG-Flow provides generated aneurysm geometries with steady and pulsatile fields \citep{ding2025aneug}. For local variants of a limited source collection, we ask whether their recomputed fields improve prediction on unseen sources and how a fixed number of variants should be distributed among sources.

Local shape editing creates related variants that share a source geometry's vessel context. We call these variants descendants; a source and its descendants form a source family. A pool of 280 descendant records can contain one variant from each of 280 sources or eight from each of 35. The total is the same, but the number of sources and the number of descendants per source differ. Comparing these choices requires CFD fields linked to each source and descendant, selectable training pools, and test sets that hold out entire source families. Within each comparison, we use the same number of sampled fields and training updates.

\begin{figure}[t]
    \centering
    \includegraphics[width=\linewidth]{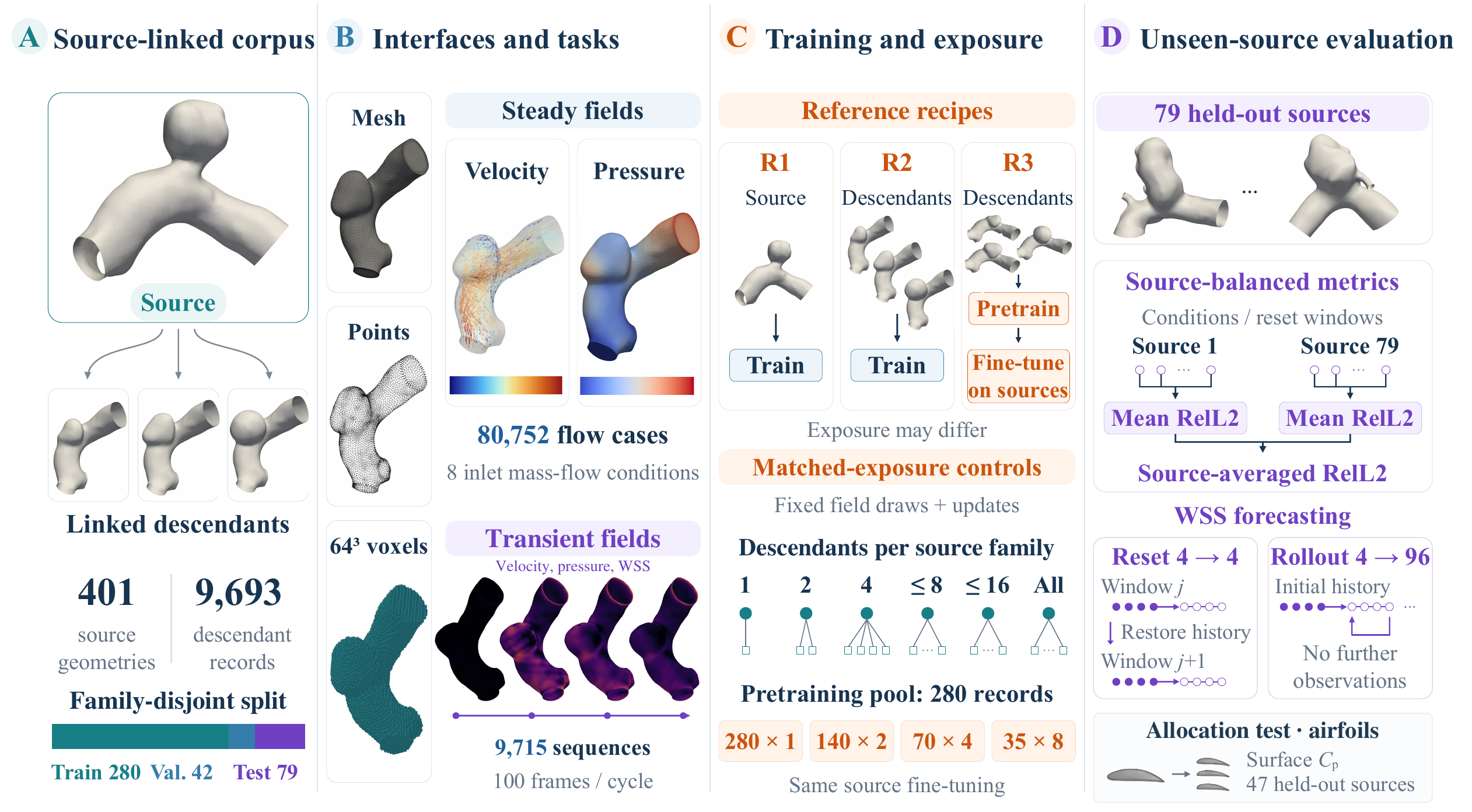}
    \caption{\textbf{AneumoBench corpus and benchmark.}
    \textbf{(A)} Family-disjoint splits contain 401 sources and 9,693 descendants (280/42/79 sources).
    \textbf{(B)} Steady cases provide velocity and pressure at eight inlet conditions; transient sequences provide 100 frames of velocity, pressure, and WSS. Mesh, point, and voxel interfaces are available.
    \textbf{(C)} R1/R2/R3 compare source, descendant, and descendant-to-source training; GraphSAGE controls fix sampling and update counts.
    \textbf{(D)} Source-balanced WSS evaluation uses $4\!\to\!4$ reset and $4\!\to\!96$ rollout; the airfoil extension uses 47 held-out sources.}
    \label{fig:aneumobench_overview}
\end{figure}

We introduce \textbf{AneumoBench}, a dataset and benchmark for studying the training value of locally edited aneurysm geometries (Figure~\ref{fig:aneumobench_overview}). It links each descendant to its source, includes CFD fields on both, and provides mesh, point, and voxel interfaces. The reference tasks are steady velocity--pressure prediction and wall shear stress (WSS) forecasting from four observed frames.

Our contributions are:
\begin{itemize}
    \item We construct \textbf{AneumoBench} with 401 source geometries and 9,693 locally edited descendant records, each linked to its source and inheriting its split. The corpus contains 80,752 steady cases across eight inlet conditions and 9,715 transient sequences. Each sequence records volume velocity and pressure and wall WSS over one 1-s cardiac cycle, with 100 frames sampled at 0.01-s intervals.
    \item We benchmark nine architectures on steady velocity--pressure prediction and WSS forecasting, using source-only, descendant-trained, and descendant-to-source recipes evaluated on 79 held-out source geometries. WSS evaluation includes four-step reset windows, 96-step rollouts, and cycle-level physical metrics.
    \item We use controlled GraphSAGE experiments to compare descendant pool sizes and allocations across sources. Training on descendant fields improves prediction, and distributing a fixed number of descendants across more sources lowers error; airfoil allocation rankings differ by model. We also test whether WSS transfer gains persist across forecast horizons and correspond to improvements in cycle-level shear metrics and hotspot localization.
\end{itemize}

\section{Related Work}

\paragraph{Scientific-ML datasets and benchmarks.}
Scientific-ML benchmarks evaluate variation across geometries and prediction through time. For geometry-based tasks, AirfRANS and DrivAerNet++ evaluate field prediction across engineered shapes \citep{bonnet2022airfrans,elrefaie2024drivaernet}, while AFBench studies airfoil generation and editing \citep{liu2024afbench}.

For temporal prediction, PDEBench provides time-dependent partial differential equation (PDE) simulation tasks, and APEBench emphasizes autoregressive training and rollout \citep{takamoto2022pdebench,koehler2024apebench}. BubbleML and LagrangeBench provide multiphase-flow and particle-based simulation tasks \citep{hassan2023bubbleml,toshev2023lagrangebench}. RealPDEBench pairs numerical simulations with real-world measurements to evaluate simulation-to-measurement transfer \citep{hu2026realpdebench}. Table~\ref{tab:related_benchmarks} compares the data relationships and evaluation settings of selected resources.

\begin{table}[!htbp]
\caption{Data relationships and evaluation settings in selected scientific-ML resources. Appendix Table~\ref{tab:dataset_comparison} details aneurysm-resource coverage and protocol records.}
\label{tab:related_benchmarks}
\centering\normalsize
\setlength{\tabcolsep}{4.2pt}
\renewcommand{\arraystretch}{1.02}
\begin{tabularx}{\linewidth}{@{}
>{\raggedright\arraybackslash}p{0.16\linewidth}
>{\raggedright\arraybackslash}p{0.25\linewidth}
>{\raggedright\arraybackslash}p{0.32\linewidth}
>{\raggedright\arraybackslash}X@{}}
\toprule
\textbf{Resource} & \textbf{Data structure} & \textbf{Evaluation setting} & \textbf{Time dependence} \\
\midrule
AneuG-Flow & Generated aneurysm geometries & Geometry-to-field learning & Steady and pulsatile \\
AirfRANS & Airfoils and operating conditions & Data scarcity and extrapolation & Steady \\
APEBench & PDE trajectories & Training and rollout & Time-dependent \\
RealPDEBench & Simulation--measurement pairs & Simulation-to-measurement transfer & Time-dependent \\
\textbf{AneumoBench} & Source-linked geometry families & Unseen-source prediction and training-pool comparisons & Steady and pulsatile \\
\bottomrule
\end{tabularx}
\end{table}

\paragraph{Aneurysm datasets.}
Aneurysm research includes detection and segmentation methods \citep{bo2021toward,cao2024semi} and annotated imaging or geometry datasets \citep{timmins2021comparing,di2023towards,chen2024,ivantsits2022detection,de2024time,yang2020intra}. Aneurisk and AneuX provide vascular surfaces and morphology records \citep{AneuriskWeb,12Aneux}, while CMHA adds hemodynamic annotations \citep{song2024intracranial}. AneuG-Flow generates aneurysm sacs and parent vessels with steady and pulsatile CFD fields \citep{ding2025aneug}. BenchAnXplore combines patient-derived aneurysm bulges with idealized parent vessels for transient prediction \citep{lannelongue2026physics}. AneumoBench links locally edited descendants to their source geometries and CFD fields for training-pool comparisons under family-disjoint evaluation.

\paragraph{CFD surrogate models.}
FNO predicts fields on regular grids, while Transolver targets general geometries \citep{li2020fourier,wu2024transolver}. MeshMask studies masked mesh-node pretraining, and GA-Field repeatedly injects geometry embeddings during aerodynamic field prediction \citep{garnier2025meshmask,zheng2026gafield}.

\section{AneumoBench Dataset}
\label{sec:dataset}

\subsection{Source geometries and descendant construction}

AneumoBench contains 401 source geometries retained from an archived AneuX snapshot \citep{12Aneux}. Each describes an aneurysm and its adjoining vessel segments. We preserve these geometries and their CFD fields and prepare separate editing surfaces for descendant construction.

To construct descendants, we inspect and repair the editing surface, remove the original aneurysm, and remesh the adjoining wall. A connected patch near the excision site is displaced by 0.5--1.0 mm along local normals and joined to the retained vessel through a transition band. We then smooth and remesh the surface before export. Patch selection and repair use operator-guided Geomagic Wrap operations. The resulting 9,693 descendant records retain source identifiers and CFD fields computed on each geometry. Appendix~\ref{sec:appendix_geometry_construction} gives the construction details.

\begin{figure}[t]
\centering
\includegraphics[width=0.95\linewidth]{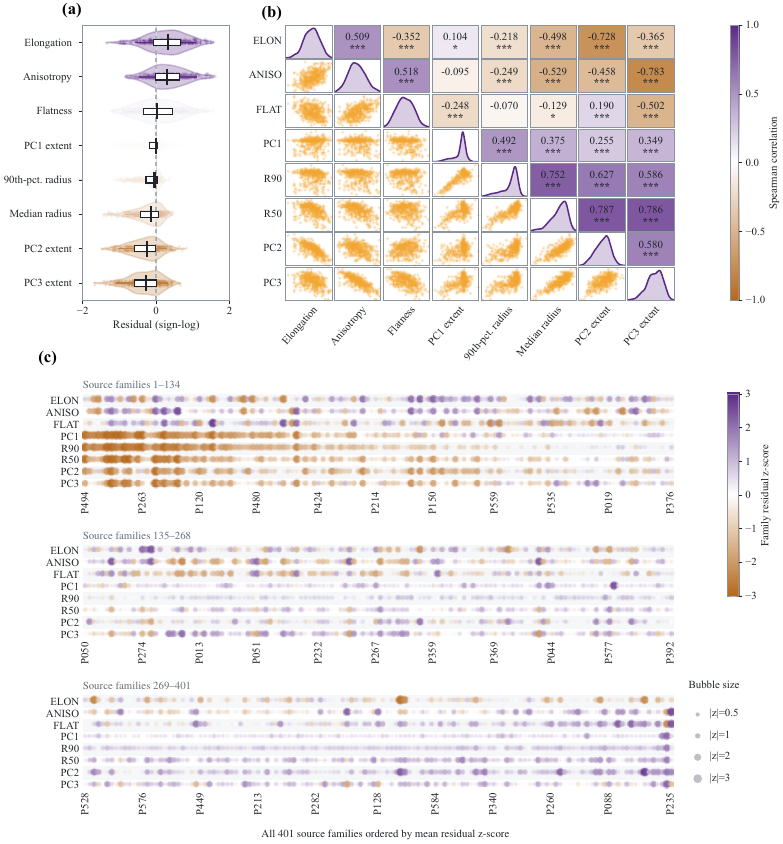}
\caption{Geometry changes relative to each source. Each family contributes one median residual per descriptor. (a) Distributions of these residuals: positive values indicate larger descriptor values in descendants. Residuals use source-IQR scaling and a sign-log transform. (b) Spearman correlations of these family summaries; stars mark FDR-adjusted significance. (c) The residuals standardized across 401 families for each descriptor, showing departures from the mean family response. Color and bubble size encode signed and absolute z-scores. Families are ordered by mean z-score; Appendix~\ref{sec:appendix_geometry_coverage} gives the definitions.}
\label{fig:synthetic_geometry_coverage}
\end{figure}

\begin{samepage}
For 72.5\% of descendants, the normalized distance to their source is below the median source--source distance. Across eight rotation-invariant descriptors, the median fraction of descendants within the sources' 5th--95th percentile ranges is 89.1\%. The largest shifts are toward greater elongation and anisotropy (Figure~\ref{fig:synthetic_geometry_coverage}). Appendix~\ref{sec:appendix_geometry_coverage} defines the surface distances and descriptors.\par
\end{samepage}

\subsection{CFD data generation}

For each geometry, we designate the open end with the largest diameter as the inlet and the others as outlets. We impose zero-pressure outlet boundary conditions and no-slip vessel walls. Steady simulations use eight inlet mass-flow rates from $0.001$ to $0.004\,\mathrm{kg\,s^{-1}}$ and store volume velocity and pressure. The resulting 80,752 cases each correspond to one geometry and one inlet condition.

Transient simulations run five cycles of a common 1-s inlet waveform. We retain 100 fifth-cycle frames at 0.01-s intervals for volume velocity and pressure and wall WSS. Before cohort filtering, the campaign used over 11 million CPU core-hours and produced approximately 23 TB of raw steady output and over 100 TB of raw transient output (Appendix~\ref{sec:appendix_construction}).

\subsection{Data organization and validation}

HDF5 records store fields, coordinates, connectivity, and boundary labels; transient records add timestamps and boundary-to-volume mappings for tracking wall locations (Appendix Table~\ref{tab:appendix_schema}).

Reported steady results use arrays containing zero-valued placeholders at points unreferenced by cells. For new training, cell-connected preprocessing excludes these points from sampling, normalization, and voxel interpolation. Appendix~\ref{sec:appendix_valid_point_domain} defines this workflow and reevaluates six GraphSAGE allocation checkpoints on valid points with weights and normalization fixed.

Of 10,094 geometry records, 9,718 have converted transient HDF5 files; 376 descendants do not. Field and structural checks exclude three training-source sequences, leaving 9,715. Appendix~\ref{sec:appendix_cfd_screening} documents finite-value, timestamp, and boundary-mapping checks and field-level conservation diagnostics.

\subsection{Data splits}

Sources are split into 280/42/79 training, validation, and test geometries, with descendants inheriting the assignment. The three transient exclusions leave 277 WSS training sources; validation and test sets remain unchanged. Appendix Table~\ref{tab:appendix_cohort_ledger} gives cohort counts, and Appendix~\ref{sec:appendix_artifacts} specifies data access and review-package scope.

\section{Benchmark Design}
\label{sec:protocol}

\subsection{Prediction tasks}

The \textbf{steady task} predicts velocity $\mathbf u$ and pressure $p$ from geometry and inlet mass flow. Unstructured models receive coordinates, boundary type, wall distance, and flow rate at sampled volume points; voxel models receive occupancy, grid coordinates, and flow rate. Predictions are evaluated at sampled stored volume vertices or constructed fluid voxels, respectively.

The \textbf{WSS task} forecasts wall shear stress from geometry, four observed WSS frames, and the known target cardiac phase. Each history tracks a fixed wall location within a case. \textbf{Reset evaluation} predicts four frames autoregressively within each window and restores observed history at the next window. \textbf{Rollout evaluation} uses only the initial four observations to predict the remaining 96 frames autoregressively.

\subsection{Training data and comparison conditions}

\textbf{Source-only (R1)} trains on source fields. \textbf{Descendant-trained (R2)} trains on descendants of the training sources. \textbf{Descendant$\rightarrow$Source (R3)} starts from an R2 checkpoint and fine-tunes on source fields. Checkpoint selection uses source-validation data, and evaluation uses the same 79 test sources. For WSS, all three settings train with teacher forcing, appending an observed frame to the history at each step.

The \textbf{R1--R3 comparison} uses each setting's reported training schedule; the number of sampled fields and optimizer updates may differ. The \textbf{GraphSAGE controls} examine descendant pool size and allocation, keeping these counts, normalization, and validation selection fixed within each comparison. The two controls use separate schedules (Section~\ref{sec:results}).

\subsection{Metrics and source-level statistics}

We compute relative $\ell_2$ error (RelL2) for decoded velocity, pressure, and WSS, retaining the simulations' zero-pressure outlet reference. Errors are calculated per steady flow condition or reset window, then averaged within and across sources. For target $\mathbf y_{im}$ and prediction $\hat{\mathbf y}_{im}$ in case $m$ of source $i$,
\begin{equation}
\mathcal E_{\mathrm{rel}}=\frac{1}{N}\sum_{i=1}^{N}\frac{1}{M_i}\sum_{m=1}^{M_i}
\frac{\|\hat{\mathbf y}_{im}-\mathbf y_{im}\|_2}{\|\mathbf y_{im}\|_2+\epsilon_{\mathbf y}},
\label{eq:main_source_rell2}
\end{equation}
where $N=79$, and $M_i$ is eight flow conditions or 19 stride-five reset windows. Thus each test geometry contributes equally. Rollout error compares all 96 predicted frames as one sequence per source.

Pressure diagnostics separate mean bias from error after removing the spatial mean. Time-averaged wall shear stress (TAWSS) is the cycle-average WSS magnitude at each wall location; oscillatory shear index (OSI) measures directional cancellation. We evaluate TAWSS and OSI field errors and overlap of the top-10\% TAWSS locations. The predicted cycle contains four observations and 96 predictions. Appendix~\ref{sec:appendix_metrics} defines these metrics and peak errors; Appendix~\ref{sec:appendix_training_details} details spatial sampling.

Bootstrap intervals resample sources, describing test-source variation at one validation-selected checkpoint per architecture and setting in the R1--R3 comparison. Three-seed GraphSAGE controls and complete Transolver/U-Net WSS training runs assess training variability. The matched WSS experiment repeats only adaptation from fixed upstream checkpoints.

\subsection{Reference models and inputs}

The unstructured models are PointNet, Transolver, and Galerkin Transformer on sampled points, and GraphSAGE on a 16-nearest-neighbor graph \citep{qi2017pointnet,wu2024transolver,cao2021galerkin,hamilton2017graphsage}. The voxel models are U-Net, U-FNO, the Fourier neural operator (FNO), the U-shaped neural operator (U-NO), and the factorized Fourier neural operator (F-FNO) \citep{ronneberger2015unet,wen2022ufno,li2020fourier,rahman2023uno,tran2023ffno}. For WSS, voxel features are queried at wall points and combined with histories and phase. Both tracks use the same WSS wall queries. R1--R3 are compared within each architecture.

The main steady voxel results use unit-box coordinates, which omit absolute physical scale. U-Net and FNO reruns restore this scale with centered physical coordinates in metres (Appendix Table~\ref{tab:appendix_voxel_scale_sensitivity}); this is the recommended coordinate input for new steady voxel experiments.

Persistence and linear extrapolation provide temporal predictions without training. We use history+phase ridge regression and PointwiseMLP to assess prediction from local histories and phase, without explicit coordinates or interactions between wall points. Appendices~\ref{sec:appendix_training_details} and~\ref{sec:appendix_temporal_references} give implementation and training details.

\section{Benchmark Results}
\label{sec:results}

\subsection{Baseline results}

Descendant pretraining followed by source fine-tuning lowers error across all nine architectures under the reported training schedules (Table~\ref{tab:core_benchmark}). Relative to R1, R3 improves every primary endpoint, with median model-wise reductions of 18.8\% for steady velocity, 13.5\% for pressure, and 10.1\% for reset-window WSS. Paired source-bootstrap confidence intervals (CIs) for the validation-selected checkpoints exclude zero for 8/9 velocity, 6/9 pressure, and 9/9 WSS architectures. Appendix~\ref{sec:appendix_source_error} reports the steady error distributions across sources.

\begin{table}[t]
\caption{RelL2 for R1--R3 on 79 held-out sources under the reported training schedules. Each score uses one validation-selected checkpoint. Steady scores use sampled vertices from the original export arrays or the constructed fluid-voxel masks; Appendix~\ref{sec:appendix_valid_point_domain} separately reevaluates six GraphSAGE allocation checkpoints on cell-connected queries. Source-bootstrap intervals describe held-out-source variation. Lower is better; bold and underlined values mark the best and second-best recipes for each model--endpoint pair.}
\label{tab:core_benchmark}
\centering
\setlength{\tabcolsep}{3pt}
\renewcommand{\arraystretch}{1.15}
\begin{tabular}{@{}lrrrrrrrrr@{}}
\toprule
& \multicolumn{3}{c}{\textbf{Steady velocity}} & \multicolumn{3}{c}{\textbf{Stored-gauge pressure}} & \multicolumn{3}{c}{\textbf{Reset-window WSS}} \\
\cmidrule(lr){2-4}\cmidrule(lr){5-7}\cmidrule(lr){8-10}
\textbf{Model} & R1 & R2 & R3 & R1 & R2 & R3 & R1 & R2 & R3 \\
\midrule
\multicolumn{10}{l}{\textit{Unstructured sampled points}} \\
GraphSAGE & 0.8269 & \underline{0.6812} & \textbf{0.6521} & 0.7131 & \textbf{0.5480} & \underline{0.5525} & 0.3571 & \underline{0.2561} & \textbf{0.2235} \\
Transolver & 1.0031 & \underline{0.7329} & \textbf{0.6979} & 0.8672 & \underline{0.5808} & \textbf{0.5783} & 0.1253 & \underline{0.1092} & \textbf{0.1084} \\
Galerkin Transformer & 1.0033 & \underline{0.8477} & \textbf{0.8150} & 0.8682 & \textbf{0.6417} & \underline{0.6667} & 0.1210 & \underline{0.1094} & \textbf{0.1090} \\
PointNet & 0.9859 & \underline{0.9371} & \textbf{0.8899} & 0.8489 & \underline{0.7732} & \textbf{0.7718} & 0.1889 & \textbf{0.1481} & \underline{0.1484} \\
\midrule
\multicolumn{10}{l}{\textit{Regular voxel geometry ($64^3$, unit-box reference inputs)}} \\
U-Net & 0.9867 & \underline{0.7916} & \textbf{0.7050} & 0.8147 & \underline{0.7970} & \textbf{0.7930} & 0.2369 & \underline{0.2216} & \textbf{0.2064} \\
U-FNO & 0.9859 & \underline{0.8272} & \textbf{0.7568} & 0.9167 & \textbf{0.7722} & \underline{0.7925} & 0.2442 & \underline{0.2315} & \textbf{0.2227} \\
FNO & 1.0099 & \underline{0.8592} & \textbf{0.8387} & 1.3334 & \underline{1.0614} & \textbf{0.9126} & 0.2387 & \underline{0.2243} & \textbf{0.2145} \\
U-NO & 0.9957 & \underline{0.8829} & \textbf{0.8303} & 0.9490 & \underline{0.8984} & \textbf{0.8946} & 0.2449 & \underline{0.2320} & \textbf{0.2230} \\
F-FNO & \underline{1.0153} & 1.0755 & \textbf{1.0005} & 1.3032 & \underline{1.1677} & \textbf{1.1466} & 0.2334 & \underline{0.2209} & \textbf{0.2104} \\
\bottomrule
\end{tabular}
\end{table}

Comparing R2 with R3, source fine-tuning further lowers velocity error for every architecture, while pressure error can increase or decrease. For GraphSAGE velocity, most of the reduction from R1 occurs during descendant training, with a smaller gain from fine-tuning.

\begin{samepage}
Pressure results also depend on the metric and coordinate representation. After removing the spatial mean, R3 still lowers pressure RelL2 relative to R1 for GraphSAGE and Transolver (Appendix Table~\ref{tab:appendix_physics_sensitivity}). For U-Net, R3 raises this error with unit-box inputs but lowers it with physical coordinates (Appendix Table~\ref{tab:appendix_voxel_scale_sensitivity}).\par
\end{samepage}

\subsection{Number and allocation of descendants}

\begin{figure}[t]
\centering
\begin{minipage}[t]{0.49\linewidth}
\centering
\includegraphics[width=\linewidth]{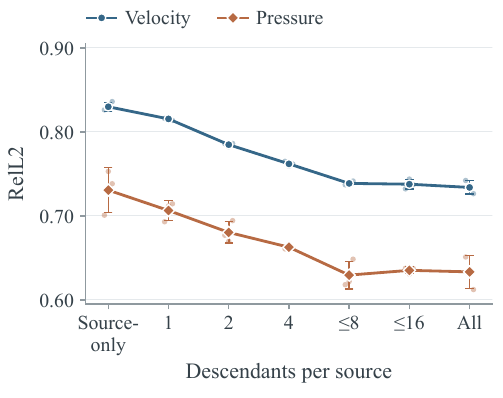}
\par\vspace{3pt}
{\normalsize\textbf{(a)} Pool expansion\par}
{\normalsize 28,000 updates\par}
\end{minipage}\hfill
\begin{minipage}[t]{0.49\linewidth}
\centering
\includegraphics[width=\linewidth]{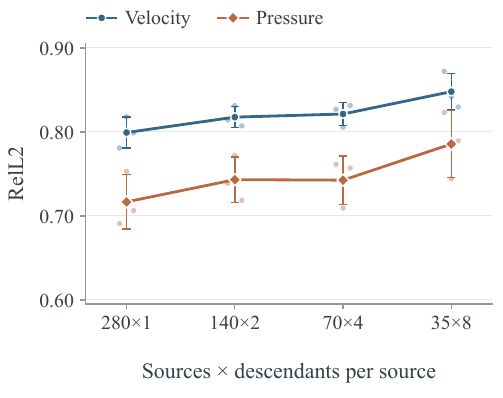}
\par\vspace{3pt}
{\normalsize\textbf{(b)} Fixed-budget allocation\par}
{\normalsize 42,000 updates, including fine-tuning\par}
\end{minipage}
\caption{GraphSAGE pool-size and allocation controls on 79 held-out sources. Small dots show individual seeds; connected markers and error bars give means and sample SD across seeds 42--44. Both panels use the same absolute RelL2 scale (lower is better); pressure uses the stored gauge. (a) Accessible-pool expansion. (b) Allocation of 280 descendants, followed by source fine-tuning. Within each panel, runs use equal numbers of sampled fields and training updates. Zero-velocity prediction has RelL2 approximately one. Complete values are in Appendix Tables~\ref{tab:graphsage_scaling} and~\ref{tab:appendix_lineage_breadth_depth}.}
\label{fig:matched_exposure_scaling}
\end{figure}

To examine data selection, the GraphSAGE controls hold sampling and update counts fixed within each experiment. Increasing the descendant pool improves prediction, with gains that level off as the pool grows (Figure~\ref{fig:matched_exposure_scaling}a). The pool-size control includes all 280 training sources and increases the accessible descendant count per source. Every run, including Source-only, samples 2,240 fields per epoch and performs 28,000 updates. Pool membership is fixed across seeds (Appendix Table~\ref{tab:graphsage_scaling}).

A cap of eight descendants per source reduces velocity and pressure error by 11.0\% and 13.8\% relative to Source-only. Every tested pool improves both endpoints in all three seeds. Expanding beyond approximately eight descendants per source, up to the full pool, yields little additional improvement under this schedule.

At a fixed total of 280 descendants, broader source coverage gives GraphSAGE lower mean errors (Figure~\ref{fig:matched_exposure_scaling}b). We compare 280 sources with one descendant each, 140 with two, 70 with four, and 35 with eight. Each allocation receives 28,000 descendant-training updates followed by 14,000 fine-tuning updates on the same fields from all 280 training sources, for a separate total budget of 42,000 updates.

Both endpoints have their lowest mean at $280\!\times\!1$. Concentrating the pool at $35\!\times\!8$ raises velocity and pressure errors by 6.1\% and 9.6\%, despite the shared source data used for fine-tuning. Reevaluation with fixed checkpoints and normalization also favors $280\!\times\!1$ on cell-connected queries, before and after removing orphan inputs (Appendix~\ref{sec:appendix_valid_point_domain}).

\subsection{Allocation on source-edited airfoils}

\begin{samepage}
To examine whether the allocation preference holds on another task and for another model, we train GraphSAGE and Transolver on source-edited AFBench airfoils~\citep{liu2024afbench}. They predict recomputed surface pressure coefficients on a 256-point contour, using arc-length-weighted relative $\ell_2$ error. Within each model, the four 280-record allocations share a two-stage schedule and are evaluated on 47 held-out sources across three full-pipeline seeds.\par
\end{samepage}

GraphSAGE has similar mean errors across allocations, with its lowest at $280\!\times\!1$; Transolver has its lowest at $35\!\times\!8$ (Table~\ref{tab:appendix_airfoil_transfer}). For Transolver, changing from $280\!\times\!1$ to $35\!\times\!8$ lowers mean error by 17.2\% and improves all three seeds. Compared with the separately trained Source-only reference, GraphSAGE at $280\!\times\!1$ also achieves lower error in every seed. Appendix~\ref{sec:appendix_airfoil_transfer} gives construction and training details.

\begin{table}[t]
\caption{Airfoil pressure-coefficient RelL2 on 47 sources: mean $\pm$ sample SD across three full-pipeline seeds. The four 280-descendant allocations share a training budget; Source-only is a separately trained reference recipe. Bold marks the lowest mean within each model.}
\label{tab:appendix_airfoil_transfer}
\centering
\normalsize
\renewcommand{\arraystretch}{1.08}
\begin{tabularx}{\textwidth}{l*{5}{>{\centering\arraybackslash}X}}
\toprule
Model & Source-only & $280\!\times\!1$ & $140\!\times\!2$ & $70\!\times\!4$ & $35\!\times\!8$ \\
\midrule
GraphSAGE & \makecell{$0.0482$\\{\normalsize $\pm0.0005$}} & \makecell{$\mathbf{0.0450}$\\{\normalsize $\mathbf{\pm0.0001}$}} & \makecell{$0.0463$\\{\normalsize $\pm0.0012$}} & \makecell{$0.0453$\\{\normalsize $\pm0.0017$}} & \makecell{$0.0454$\\{\normalsize $\pm0.0009$}} \\
Transolver & \makecell{$0.0585$\\{\normalsize $\pm0.0115$}} & \makecell{$0.0623$\\{\normalsize $\pm0.0028$}} & \makecell{$0.0541$\\{\normalsize $\pm0.0129$}} & \makecell{$0.0581$\\{\normalsize $\pm0.0089$}} & \makecell{$\mathbf{0.0516}$\\{\normalsize $\mathbf{\pm0.0086}$}} \\
\bottomrule
\end{tabularx}
\end{table}

\subsection{WSS forecasting across time horizons}

For aneurysm WSS forecasting, short-window improvements do not consistently extend to uninterrupted rollout (Table~\ref{tab:rollout_main}). R3 lowers reset error relative to R1 for both Transolver and voxel U-Net across three independent training seeds. In rollout, Transolver improves in every seed, while U-Net has mixed seed-level changes and a $+0.2\%$ mean change. Transolver R3 has lower error than persistence over four steps but higher error over 96.

\begin{table}[t]
\caption{Mean reset-window and recursive-rollout RelL2 across three training seeds.}
\label{tab:rollout_main}
\centering\normalsize
\setlength{\tabcolsep}{3.5pt}
\renewcommand{\arraystretch}{1.08}
\begin{tabular*}{\columnwidth}{@{\extracolsep{\fill}}lrrcrrcl@{}}
\toprule
\multirow{2}{*}{\textbf{Model}} & \multicolumn{3}{c}{\textbf{Reset window}} & \multicolumn{3}{c}{\textbf{96-step rollout}} & \multirow{2}{*}{\textbf{Pattern}} \\
\cmidrule(lr){2-4}\cmidrule(lr){5-7}
& R1 & R3 & $\Delta$ & R1 & R3 & $\Delta$ & \\
\midrule
Transolver & 0.128 & 0.109 & \textbf{$-14.9\%$} & 1.279 & 0.966 & \textbf{$-24.5\%$} & Both improve \\
Voxel U-Net & 0.247 & 0.208 & \textbf{$-15.6\%$} & 0.954 & 0.956 & $+0.2\%$ & Reset gain; rollout mixed \\
PointwiseMLP & 0.146 & -- & -- & 0.735 & -- & -- & Local-history reference \\
Persistence & 0.258 & -- & -- & 0.881 & -- & -- & Training-free \\
\bottomrule
\end{tabular*}
{\raggedright\normalsize Lower is better; $\Delta$ is the R1-to-R3 change. PointwiseMLP uses three R1 training seeds; persistence is deterministic. Sample SD and seed-level results are in Appendix Tables~\ref{tab:appendix_baselines}, \ref{tab:appendix_transfer_rollout}, and~\ref{tab:appendix_transfer_rollout_seeds}.\par}
\end{table}

\begin{samepage}
Standard WSS training uses observed histories, whereas evaluation feeds predictions back into the history. We therefore examine four-step scheduled sampling and longer full-feedback windows with detached predictions (Appendix~\ref{sec:appendix_feedback_training}). For Transolver, longer windows yield lower rollout error after 30 epochs of adaptation from the same R3 checkpoint (Appendix Figure~\ref{fig:appendix_horizon_feedback_grid}).\par
\end{samepage}

R3 retains its advantage under matched feedback adaptation. We adapt the seed-42 Transolver R1/R3 checkpoints for eight epochs with $F=16$ predicted frames per window, repeating adaptation with three paired seeds while keeping the upstream checkpoints fixed. R3 has 20.8\% lower mean rollout RelL2 (Table~\ref{tab:f16_initialization_main}). Before adaptation, R3 outperforms R1 on 27 of 79 sources despite its lower mean error. After adaptation, R3 outperforms R1 on 72, 74, and 72 sources across the three seeds.

\begin{table}[t]
\caption{WSS endpoints after identical eight-epoch $F=16$ feedback adaptation of fixed Transolver R1/R3 checkpoints. Entries are means $\pm$ sample SD over three adaptation seeds on 79 sources. Lower is better. Appendix Table~\ref{tab:appendix_f16_initialization} reports all endpoints.}
\label{tab:f16_initialization_main}
\centering\normalsize
\setlength{\tabcolsep}{7pt}
\begin{tabular}{lrr}
\toprule
Endpoint & R1 initialization & R3 initialization \\
\midrule
96-step rollout RelL2 & $0.46840\pm0.01369$ & $0.37076\pm0.00390$ \\
TAWSS relative error & $0.32239\pm0.02310$ & $0.21591\pm0.00458$ \\
OSI absolute error & $0.02747\pm0.00025$ & $0.02836\pm0.00013$ \\
\bottomrule
\end{tabular}
\end{table}

Lower rollout error does not ensure better cycle-level metrics or high-shear localization. In the standard pipelines, R3 lowers TAWSS error for Transolver but raises it for U-Net on average across seeds. After matched feedback adaptation, Transolver R3 has lower TAWSS error than R1 but slightly higher OSI error. PointwiseMLP also has lower rollout error than persistence, yet its top-10\% TAWSS overlap falls from 0.847 to 0.519 (Appendix~\ref{sec:appendix_extended_transient}).

\section{Discussion and Conclusion}

We introduced AneumoBench, linking source aneurysm geometries and their locally edited descendants to corresponding CFD fields. Its steady-field prediction and WSS forecasting tasks use source-family splits to compare how descendant data and training strategies affect prediction on unseen source geometries.

Under the reported training schedules, descendant pretraining followed by source fine-tuning lowers steady-field and reset-window WSS errors across all nine reference architectures. In the GraphSAGE controls, gains level off as the descendant pool grows; distributing a fixed number of descendants across more sources yields lower mean errors. The airfoil experiments give different allocation preferences for GraphSAGE and Transolver, showing that allocation strategies should be evaluated for each model.

In WSS forecasting, short-window gains do not consistently persist through uninterrupted rollout, and trajectory error alone does not fully describe the recovery of cycle-level shear or high-shear regions. Evaluation therefore needs to specify the observation schedule and forecast horizon and assess TAWSS, OSI, and hotspot overlap alongside trajectory error. AneumoBench provides common data and protocols for studying when training on descendant fields improves prediction on unseen geometries and which flow features benefit.

\FloatBarrier
\clearpage
\section*{Artificial Intelligence Use Statement}
We used generative AI tools to assist with language editing, translation, and relevant literature searches, and used tools such as Codex to run code following the authors’ instructions. All AI-generated or AI-modified content was reviewed by the authors before inclusion in the manuscript. AI tools were not used to generate the synthetic geometries or CFD-derived targets. The authors take responsibility for the final text, claims, code, and artifacts.

\section*{Ethics Statement}
AneumoBench uses public, deidentified source geometries. No new human data were collected, and no identifiable information was accessed. The AneuX source record specifies the Creative Commons Attribution-NonCommercial 4.0 (CC BY-NC 4.0) license together with attribution and article-citation requirements. The release preserves these notices for derived geometry files. CFD targets are generated under standardized research assumptions and are intended for machine-learning evaluation, not clinical decision making. The release includes cohort, availability, and exclusion records so that downstream users can assess the scope of the benchmark.

\section*{Reproducibility Statement}
An anonymized code ZIP is provided as supplementary material. The \href{https://figshare.com/s/f54b0f5b728524e89a5b}{Figshare review collection} hosts the review data and pretrained checkpoints: a full-schema inspection subset, native-resolution WSS arrays for 277 training and 42 validation sources, and a fixed-test inference package. The training arrays support source-only WSS training and feedback adaptation. The inference package supplies all 79 fixed-sample test sources and eight Transolver checkpoints for reset and rollout evaluation. Source-level records reconstruct the reported statistics. The full multimodal corpus will be released after de-anonymization with permanent hosting and the required upstream notices. Appendix~\ref{sec:appendix_artifacts} details the reproduction tasks and data scope.

\subsubsection*{Acknowledgments}
This research was supported by the NovaInspire platform of Shanghai Academy of Artificial Intelligence for Science. The computations in this research were performed using the CFFF platform of Fudan University.

\bibliography{iclr2027_conference}
\bibliographystyle{iclr2027_conference}

\clearpage
\appendix
\section{Dataset Composition and Geometry Construction}
\label{sec:appendix_repro}

Sections~\ref{sec:appendix_repro}--\ref{sec:appendix_construction} describe the geometry records and CFD generation. Representations and training settings are in Section~\ref{sec:appendix_training_details}, and metric definitions are in Section~\ref{sec:appendix_metrics}. Extended reference, training-pool, and WSS results follow in Sections~\ref{sec:appendix_all_baselines}--\ref{sec:appendix_extended_transient}. Section~\ref{sec:appendix_physics_sensitivity} contains representation and field diagnostics; Section~\ref{sec:appendix_artifacts} gives data access and reproduction instructions.

\subsection{Source geometries}

AneumoBench uses data from the AneuX morphology database, an open-access, multi-centric database combining data from three European projects: AneuX project, @neurIST project and Aneurisk \citep{12Aneux}. The archived source surfaces describe aneurysms and adjoining vessel segments. Source geometries and their computed fields remain separate records when descendants are constructed. The acquisition inventory and retained cohort are distinguished in Section~\ref{sec:appendix_release_analysis}.

\subsection{Descendant construction}
\label{sec:appendix_geometry_construction}

Each construction starts from a public, deidentified AneuX triangular surface. The mesh is inspected and repaired for non-manifold edges, self-intersections, spikes, tunnels, and holes. We prepare a separate editing surface by removing the original aneurysm and remeshing the adjoining wall. A connected patch at the excision site is then displaced by $d\in[0.5,1.0]\,\mathrm{mm}$ along local normals. A polygon transition band joins the modified patch to the retained vessel before global remeshing and STL export. Patch selection and repair use operator-guided Geomagic Wrap 2024 operations. The final surfaces and their CFD fields form the benchmark records used for training, with each descendant linked to its source in the manifest.

\subsection{Source relationships and inherited splits}

We sort the 427 source identifiers in the acquisition inventory and shuffle them with Python's \texttt{random.Random(42)}. A 70/10/20 split, with training and validation counts rounded to the nearest integer, initially assigns 299/43/85 sources to training, validation, and testing. Applying the initial eight-flow steady and transient eligibility filters within these partitions removes 19/1/6 sources, leaving 280/42/79 without reshuffling. Descendants inherit their source's assignment and are retained only if that source passes the filters. Training pools use descendants of training sources; validation and testing use held-out source geometries. The scripts \path{data/build_patient_split.py} and \path{data/build_strict_split.py} specify the procedure.

\subsection{Cohort accounting and modality availability}
\label{sec:appendix_release_analysis}

The release manifest records source geometries, eligible descendants, and the modalities available for each record. Every retained geometry has eight steady conditions. The transient archive contains 9,718 HDF5 sequences (401 source and 9,317 descendant); excluding three training-source sequences leaves 9,715 for the benchmark.

\begin{table}[!htb]
\caption{Cohort ledger. ``Eligible'' records belong to the construction manifest and have a retained source; ``available'' records have a readable modality that passes structural validation. Counts refer to the unit named in each row; source geometries are counted in the first row. Parentheses give each split's share of the row total.}
\label{tab:appendix_cohort_ledger}
\centering\normalsize
\setlength{\tabcolsep}{3pt}
\begin{tabularx}{\textwidth}{@{}>{\raggedright\arraybackslash}p{0.28\textwidth}rrrr>{\raggedright\arraybackslash}X@{}}
\toprule
Release unit & Train & Validation & Test & Total & Evaluation treatment \\
\midrule
Source geometries & \makecell{280\\(69.8\%)} & \makecell{42\\(10.5\%)} & \makecell{79\\(19.7\%)} & 401 & Split unit \\
Descendant records, eligible & \makecell{6,687\\(69.0\%)} & \makecell{1,055\\(10.9\%)} & \makecell{1,951\\(20.1\%)} & 9,693 & Inherit source split \\
Distinct wall-coordinate arrays & 6,638 & 1,052 & 1,943 & 9,633 & Coordinate-array equality \\
Descendant transient, available & \makecell{6,438\\(69.1\%)} & \makecell{1,001\\(10.7\%)} & \makecell{1,878\\(20.2\%)} & 9,317 & Readable 101-entry HDF5 \\
Descendant transient, unavailable & 249 & 54 & 73 & 376 & Tracked in manifest; excluded from transient use \\
Source transient, archived/readable & 280 & 42 & 79 & 401 & Three train-only exclusions below \\
Source transient, protocol-included & 277 & 42 & 79 & 398 & Test denominator remains 79 \\
\bottomrule
\end{tabularx}
\end{table}

The cohort is formed in three stages: inventory the collected geometries, apply the cross-modal inclusion rule, then record which targets are readable. The acquisition inventory consists of the archived source surfaces because their record-level correspondence to the complete AneuX collection is unavailable. Table~\ref{tab:appendix_exclusion_ledger} lists the exclusions at each stage. Every raw descendant has a source mapping; field-level CFD checks are described in Section~\ref{sec:appendix_cfd_screening}.

\begin{table*}[!htb]
\caption{Release-cohort accounting for the archived 427-surface acquisition inventory. The archive lacks a record-level mapping to the complete 668-record AneuX collection. Counts in overlapping exclusion categories are unioned before subtraction.}
\label{tab:appendix_exclusion_ledger}
\centering\normalsize
\setlength{\tabcolsep}{5pt}
\begin{tabularx}{\textwidth}{@{}p{0.20\textwidth}>{\raggedright\arraybackslash}Xrr@{}}
\toprule
Population & Operational transition & Removed & Retained \\
\midrule
Source acquisition inventory & Locally archived source surfaces & -- & 427 \\
Sources & $<100$ transient outputs (25) or no complete eight-flow steady set (1) & 26 & 401 \\
\addlinespace[2pt]
Raw descendant inventory & All records; every record has a source mapping & -- & 10,660 \\
Descendants & Incomplete transient index (330) or missing eight-flow steady set (7), overlap 2 & 335 & 10,325 \\
Descendants & Otherwise complete but linked to one of the 26 excluded sources & 632 & 9,693 \\
Descendants & Missing conversion source files & 26 & 9,667 \\
Descendants & Required pressure array unreadable in multimodal conversion & 350 & 9,317 \\
\addlinespace[2pt]
Archived source HDF5 & Structurally readable source records & -- & 401 \\
Source transient protocol & Numerically unusable training-source records (143, 246, 441) & 3 & 398 \\
\bottomrule
\end{tabularx}
\end{table*}

Transient availability is assessed after the cohort is fixed. Missing source files or unreadable pressure arrays can prevent conversion, leaving WSS availability unknown for those records. The manifest records each failure and its reason. Transient availability is 96.27\%, 94.88\%, and 96.26\% in the train, validation, and test splits, respectively.

Three source transient cases are structurally readable but contain divergent fields outside the intended fifth-cycle interval. All three belong to the training partition and are excluded from the transient protocol, leaving the validation and test cohorts unchanged. Their descendants retain the inherited split and remain eligible for pretraining when their own converted fields are available.

\begin{figure*}[!htb]
\centering
\includegraphics[width=\textwidth]{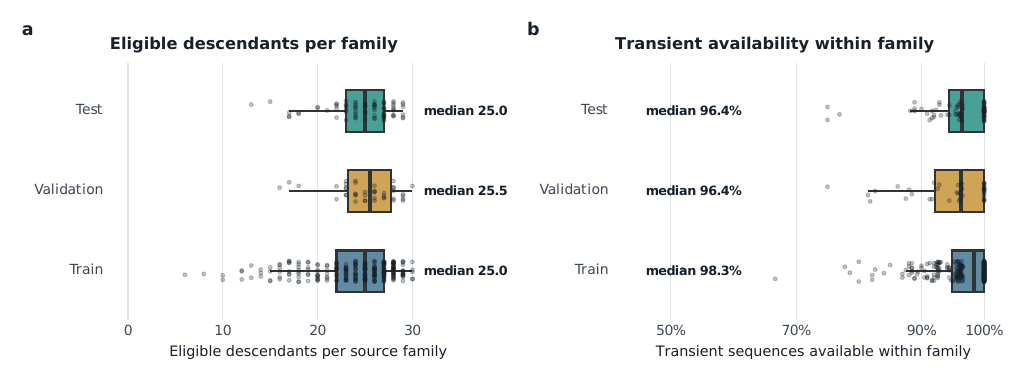}
\caption{Descendant coverage for the 401-source/9,693-descendant manifest. (a) Eligible descendant counts and (b) transient availability within each lineage, grouped by inherited split. Each point represents a source lineage; boxes show the interquartile range and median, and labels give split medians. The manifest retains the source identifiers of the 376 unavailable descendant sequences.}
\label{fig:appendix_lineage_coverage}
\end{figure*}

\begin{figure*}[!htb]
\centering
\includegraphics[width=\textwidth]{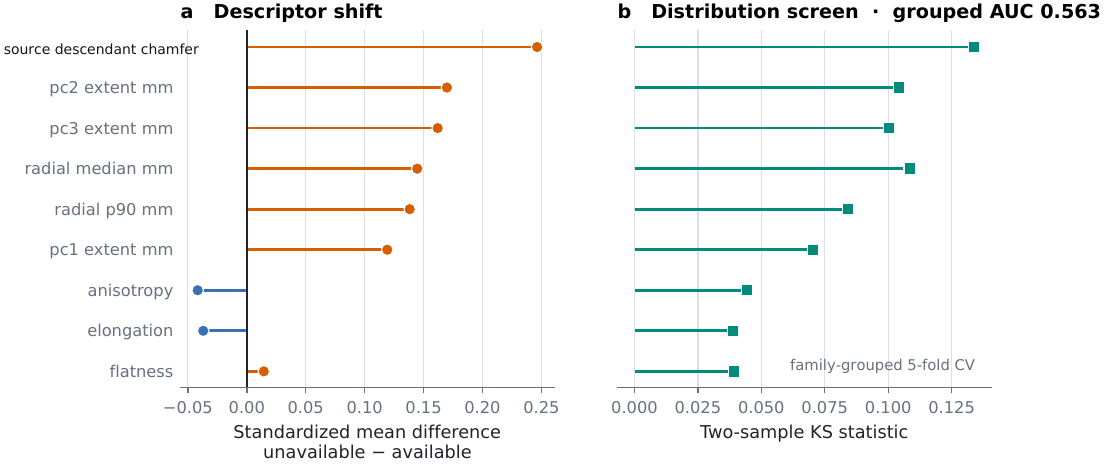}
\caption{Geometry differences between the 376 descendants without a transient HDF5 file and the 9,317 available descendants. Standardized mean differences and Kolmogorov--Smirnov (KS) statistics compare eight rotation-invariant descriptors and source--descendant Chamfer distance. A class-balanced logistic model, evaluated by five-fold GroupKFold over lineages, achieves an out-of-fold area under the receiver operating characteristic curve (AUC) of 0.563. The recorded descriptors therefore predict availability only weakly.}
\label{fig:appendix_availability_bias}
\end{figure*}

\subsection{Geometric coverage and repetition}
\label{sec:appendix_geometry_coverage}

The geometric audit uses one stored wall point set per record. We compute eight descriptors from the complete point set, then draw a deterministic 512-point sample after principal component analysis (PCA) alignment for surface-distance comparisons. For 72.5\% of descendants, the source--descendant distance is below the median random source--source distance, so the synthetic cohort forms predominantly local neighborhoods around its sources. For each descriptor, we compute the fraction of descendants within the 5th--95th percentile range of the 401-source distribution. The median of these eight fractions is 89.1\%. The largest shifts are toward greater elongation and anisotropy.

\paragraph{Descriptor definitions.}
For wall coordinates $\mathbf x_a$ in metres, let $\bar{\mathbf x}$ be their arithmetic mean and $C=(n-1)^{-1}\sum_{a=1}^{n}(\mathbf x_a-\bar{\mathbf x})(\mathbf x_a-\bar{\mathbf x})^\top$ their covariance. Write its eigenvalues as $\lambda_1\geq\lambda_2\geq\lambda_3\geq0$, with unit eigenvectors $\mathbf v_k$. The three PC extents are $10^3[\max_a q_{ak}-\min_a q_{ak}]$ mm, where $q_{ak}=(\mathbf x_a-\bar{\mathbf x})^\top\mathbf v_k$. Median radius and 90th-percentile radius are the corresponding quantiles of $10^3\|\mathbf x_a-\bar{\mathbf x}\|_2$, also in mm. The remaining three descriptors are dimensionless:
\begin{equation}
\text{elongation}=\sqrt{\frac{\lambda_1+\varepsilon}{\lambda_2+\varepsilon}},\qquad
\text{flatness}=\sqrt{\frac{\lambda_2+\varepsilon}{\lambda_3+\varepsilon}},\qquad
\text{anisotropy}=\frac{\lambda_1-\lambda_3}{\lambda_1+\varepsilon}.
\end{equation}
Here $\varepsilon$ is double-precision machine epsilon, added to the numerical eigenvalues computed in square metres. These statistics weight stored wall vertices equally. Length descriptors retain physical scale; source-IQR normalization is applied only when comparing descriptor changes below. The implementation is \path{scripts/analyze_geometry_coverage.py}.

To compare the direction and size of descriptor changes across families, we scale each change by the spread among sources and summarize it within its family. For descendant $j$ in the family $D_i$ of source $i$ and descriptor $k$, let $\mathrm{IQR}_k$ be the descriptor interquartile range across the 401 source geometries and define
\begin{align}
\delta_{ijk}&=\frac{x^{\mathrm{descendant}}_{ijk}-x^{\mathrm{source}}_{ik}}{\mathrm{IQR}_k}, &
r_{ik}&=\operatorname{median}_{j\in D_i}
 \left[\operatorname{sign}(\delta_{ijk})\log(1+|\delta_{ijk}|)\right],\\
z_{ik}&=\frac{r_{ik}-\bar r_k}{s_{r,k}}, &
\bar r_k&=\frac{1}{401}\sum_i r_{ik},\quad
s_{r,k}^2=\frac{1}{401}\sum_i(r_{ik}-\bar r_k)^2.
\end{align}
Figure~\ref{fig:synthetic_geometry_coverage}(a,b) plots the lineage-median residuals $r_{ik}$. Panel (c) plots $z_{ik}$, with lineages ordered by their mean z-score across descriptors; zero corresponds to the mean residual across lineages. Color is clipped at $\pm3$, and bubble size saturates at $|z|=3$. The script \path{scripts/plot_geometry_lineage_residuals.py} provides the figure data and transformations. Table~\ref{tab:appendix_geometry_coverage} summarizes coverage, surface distances, and repeated geometries.

For PCA-canonical 512-point samples $A$ and $B$, the normalized symmetric surface distance is
\begin{equation}
d_{\mathrm{sym}}(A,B)=\frac{1}{2s}\left[
\frac{1}{|A|}\sum_{a\in A}\min_{b\in B}\lVert a-b\rVert_2+
\frac{1}{|B|}\sum_{b\in B}\min_{a\in A}\lVert b-a\rVert_2\right],
\end{equation}
where $s$ is the mean of the two wall bounding-box diagonals. The distance compares the sampled shapes after PCA alignment.

\begin{table*}[!htb]
\caption{Geometric coverage and redundancy audit for the release cohort. Surface distances use 512-point deterministic samples after PCA canonicalization and are normalized by the mean wall bounding-box diagonal. Descriptor distances use median--IQR scaling fitted only on the 401 sources. Values are median [interquartile range]. Duplicate groups use coordinate-array equality; the near-neighbor rows report descriptor distances.}
\label{tab:appendix_geometry_coverage}
\centering\normalsize
\setlength{\tabcolsep}{6pt}
\begin{tabularx}{\textwidth}{@{}>{\raggedright\arraybackslash}p{0.23\textwidth}>{\raggedright\arraybackslash}X>{\raggedright\arraybackslash}p{0.30\textwidth}@{}}
\toprule
Audit quantity & Population / comparison & Result \\
\midrule
Surface variation & Source--descendant ($n=9,693$) & 0.0519 [0.0421--0.0625] \\
Surface variation & Random source--source pairs ($n=20,000$) & 0.0613 [0.0533--0.0743] \\
Descriptor proximity & Nearest sibling & 0.2056 [0.1481--0.2777] \\
Descriptor proximity & Nearest non-sibling & 0.2614 [0.2045--0.3343] \\
Exact coordinate arrays & Groups / affected / excess records & 47 / 107 / 60 \\
Distinct coordinate arrays & One canonical record per exact group & 9,633 \\
Coverage & Descendants inside source 5--95\% range & 89.1\% (median over descriptors) \\
\bottomrule
\end{tabularx}
\end{table*}

Comparing stored coordinate arrays identifies 47 groups of identical arrays among 107 descendant records, leaving 9,633 distinct arrays (60 excess records; 0.62\%). The reported total of 9,693 counts all eligible records, including the excess records. Each group belongs to one source lineage and split and is listed in the release manifest.

The fixed-budget allocation pools each contain 280 distinct stored wall-coordinate arrays for seeds 42--44. In the pool-size controls, $k=1,2,4,8,16$ and full contain 0, 0, 2, 7, 24, and 49 excess records, respectively. These memberships are used for training.\par

\subsection{Comparison with aneurysm resources}

Table~\ref{tab:dataset_comparison} compares resource size, CFD coverage, and the documented training and evaluation settings. BenchAnXplore combines patient-derived aneurysm bulges with idealized parent vessels in 105 geometries, with 80 frames per case \citep{lannelongue2026physics}; our source-linked collection supports selecting multiple descendants per retained source.

\begin{table}[H]
\caption{Comparison with aneurysm resources providing hemodynamic annotations.}
\label{tab:dataset_comparison}
\centering
\normalsize
\setlength{\tabcolsep}{2.5pt}
\renewcommand{\arraystretch}{1.12}
\begin{tabularx}{\textwidth}{@{}
  >{\raggedright\arraybackslash}X
  rrrrcccc@{}}
\toprule
& \multicolumn{2}{c}{Cohort size}
& \multicolumn{2}{c}{CFD coverage}
& \multicolumn{4}{c}{Benchmark protocol} \\
\cmidrule(lr){2-3}\cmidrule(lr){4-5}\cmidrule(l){6-9}
Resource & \makecell{Source\\units} & \makecell{Synthetic\\records}
& Steady & \makecell{Pulsatile\\cases (frames)}
& \makecell{Linked\\lineage} & \makecell{Inherited\\split}
& \makecell{Matched\\exposure} & \makecell{Recursive\\rollout} \\
\midrule
Aneurisk & 103 & -- & -- & -- & n/a & n/a & n/a & -- \\
CMHA & 143 & -- & -- & -- & n/a & n/a & n/a & -- \\
AneuG-Flow & -- & 14,000 & 14,000 & 730 (80) & -- & -- & -- & -- \\
\addlinespace[2pt]
\textbf{AneumoBench} & \textbf{401} & \textbf{9,693}
& \textbf{80,752} & \textbf{9,715 (100)} & \textbf{Yes} & \textbf{Yes}
& \textbf{Yes} & \textbf{Yes} \\
\bottomrule
\end{tabularx}
\vspace{2pt}
\begin{minipage}{\textwidth}
\normalsize
\textit{Notes.} Counts use the units reported by Aneurisk \citep{AneuriskWeb}, CMHA \citep{song2024intracranial}, and AneuG-Flow \citep{ding2025aneug}. Pulsatile entries give cases (frames); the AneumoBench count includes the 9,715 sequences used by the protocol. ``Yes'' indicates a documented property, ``--'' indicates that the cited paper and release documentation do not specify it, and ``n/a'' applies to protocols requiring a synthetic source--descendant cohort.
\end{minipage}
\end{table}

\section{CFD Generation and Field Quality}
\label{sec:appendix_construction}

All velocity, pressure, and WSS targets are numerical CFD outputs. The term synthetic refers to geometry construction, described in Section~\ref{sec:appendix_geometry_construction}.

\paragraph{Generation scale.}
The simulation campaign used 24 processor cores per case and peaked at roughly 10,000 concurrent cores. Its records report approximately 23 terabytes (TB) of raw steady-state output and more than 100 TB of raw transient output, using over 11 million processor core-hours. These totals cover 85,280 steady cases and 10,660 transient cases before the cohort and modality filters in Section~\ref{sec:appendix_repro}. The retained representations are specified in Section~\ref{sec:appendix_training_details}.

\subsection{Domain, mesh, and boundary preparation}

ANSYS 2026 R1 converts the repaired surface to a smooth boundary representation and meshes the vascular lumen as the fluid domain. Case-specific flow measurements are unavailable for this corpus, so we use standardized boundary conditions. Prior studies examine inflow estimation and the effects of modeled or measured boundary conditions in image-based aneurysm CFD \citep{valen2015estimation,marzo2011computational,venugopal2007sensitivity}. During domain preparation, the open cross-section with the largest diameter is designated as the inlet and the remaining open ends as outlets. Each case stores the selected inlet and the complete boundary partition.

\begin{figure*}[!htb]
\centering
\includegraphics[width=0.8\textwidth]{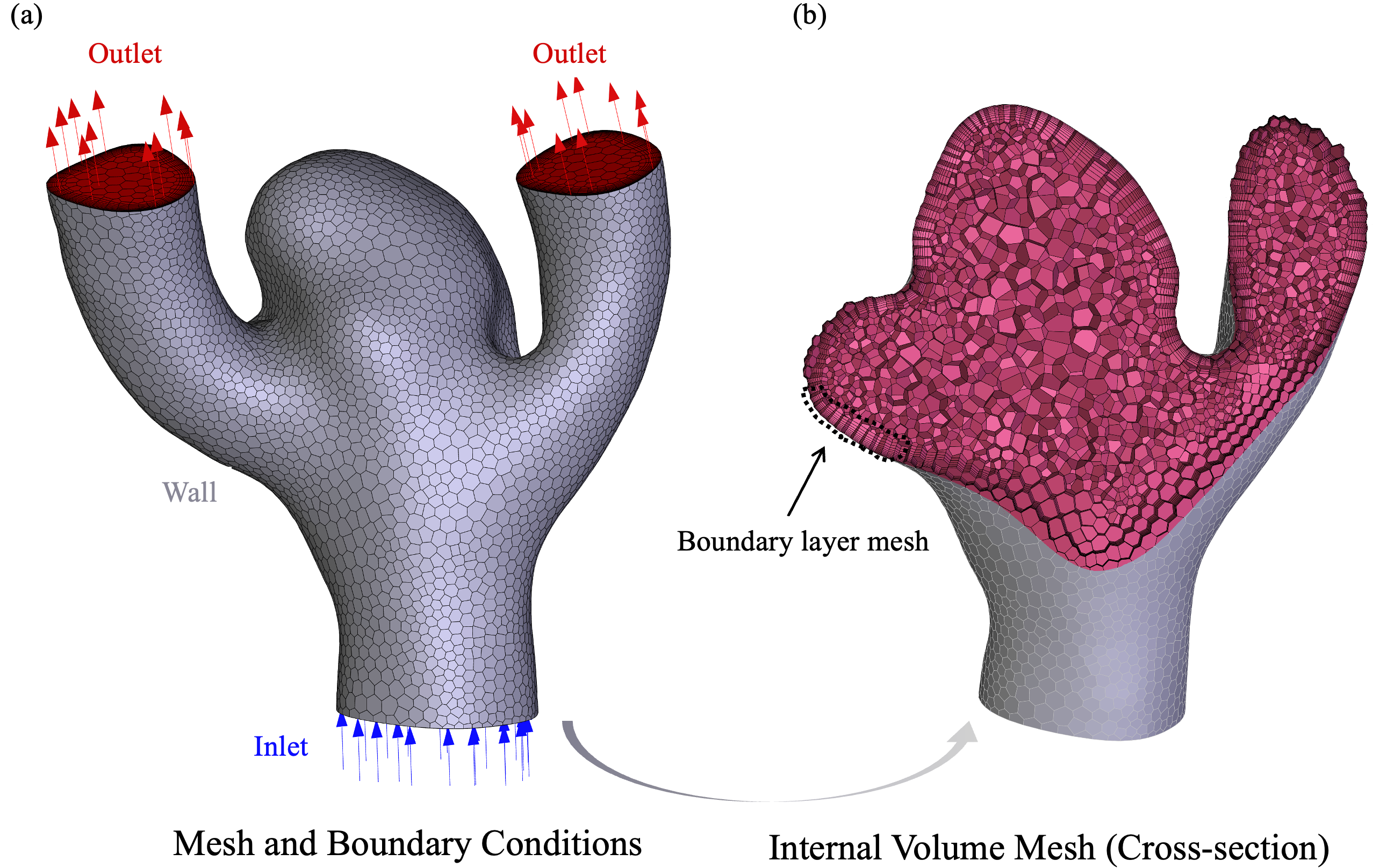}
\caption{CFD domain construction. (a) Inlet, outlet, and wall boundaries on a representative surface mesh. (b) Cross-section of the internal volume mesh and its wall-adjacent boundary-layer cells.}
\label{fig:appendix_mesh_boundary}
\end{figure*}

\begin{figure*}[!htb]
\centering
\begin{minipage}[t]{0.48\textwidth}
\centering
\includegraphics[width=\linewidth]{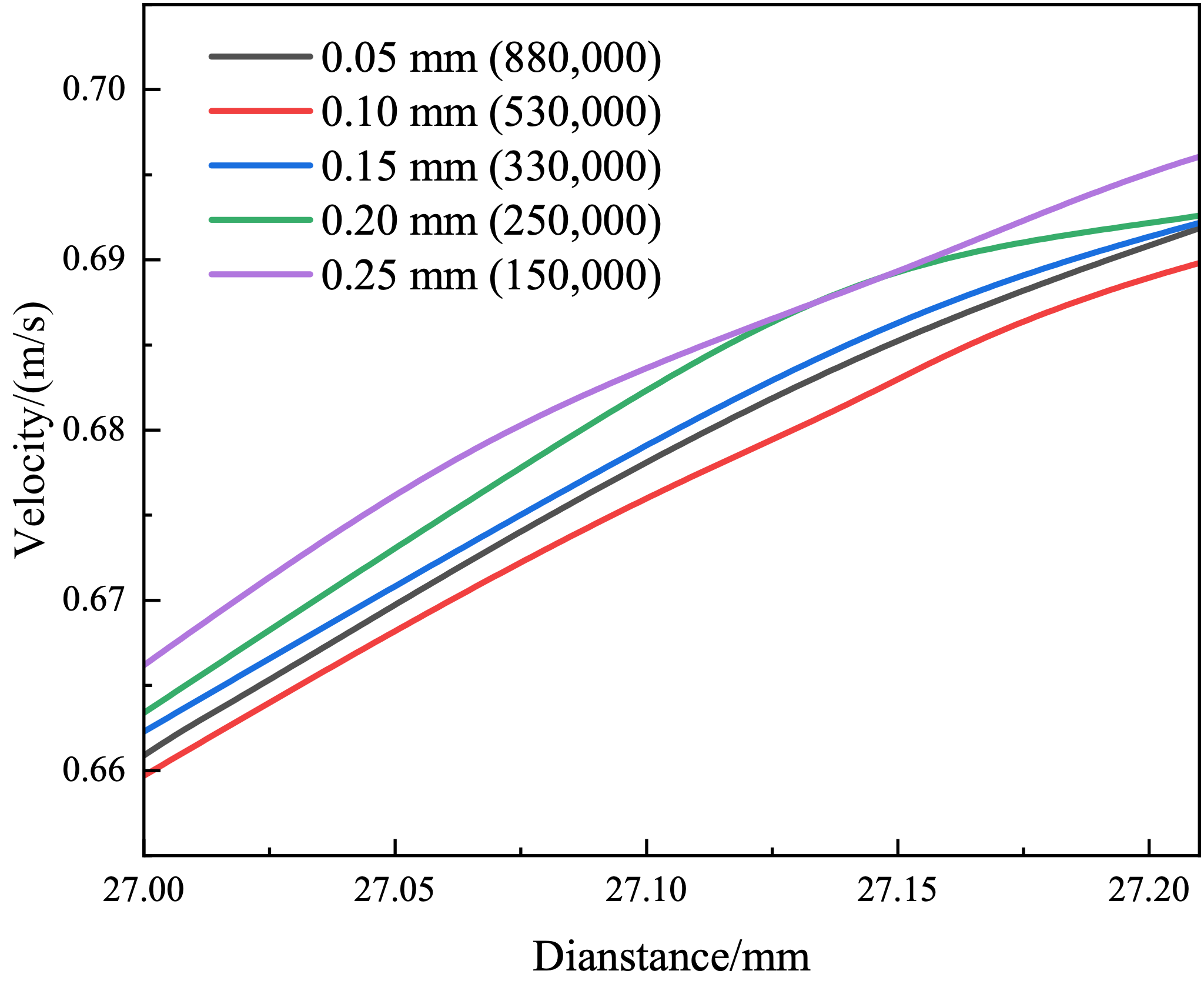}\\[-1mm]
\normalsize (a) Velocity profiles across mesh resolutions.
\end{minipage}\hfill
\begin{minipage}[t]{0.475\textwidth}
\centering
\includegraphics[width=\linewidth]{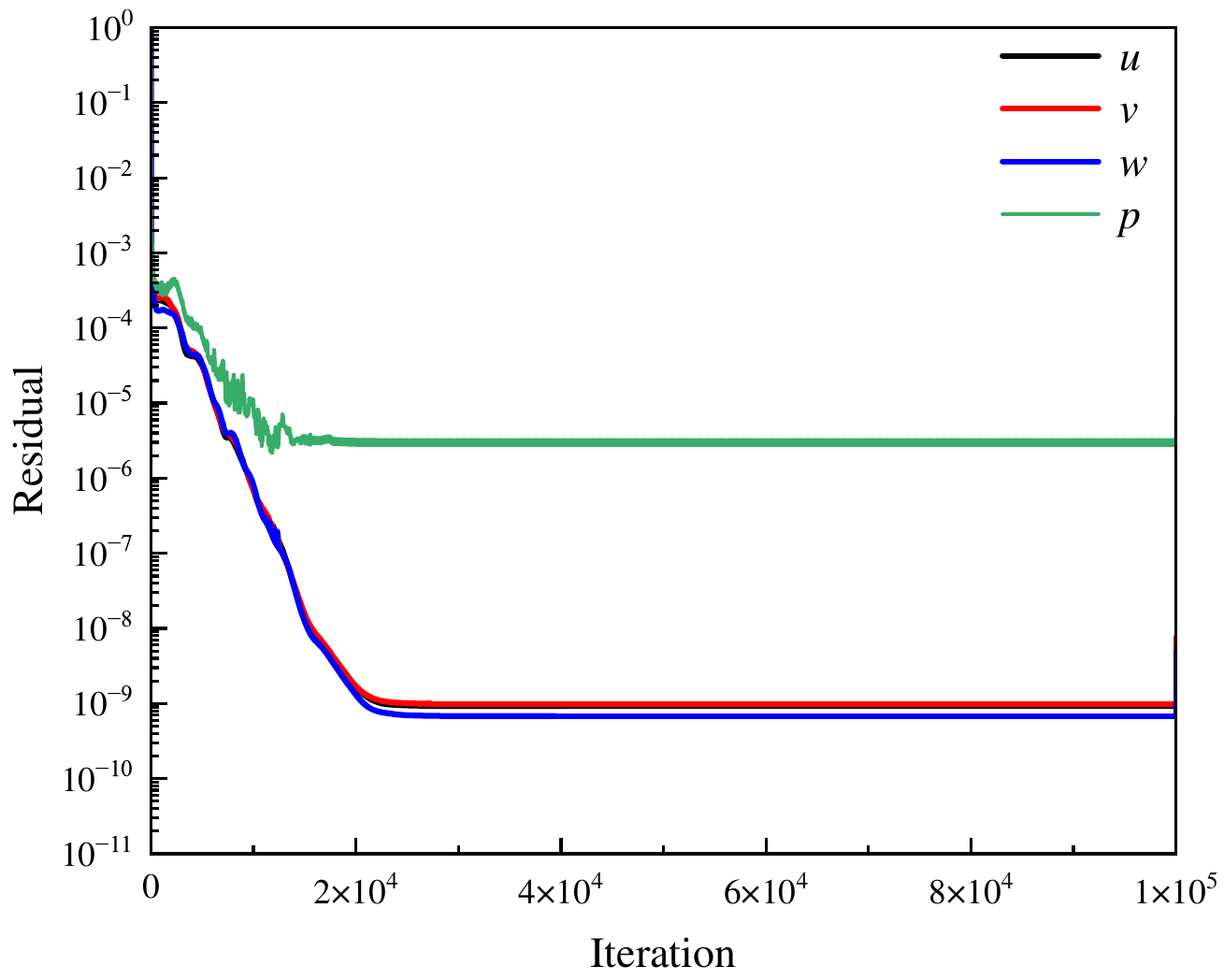}\\[-1mm]
\normalsize (b) Steady-solver residuals.
\end{minipage}
\caption{Mesh-resolution comparison and steady-solver residuals. (a) Velocity profiles for one case at five mesh resolutions, used to inform the 0.15-mm production setting. (b) Velocity-component and pressure residuals over iterations for a representative steady case.}
\label{fig:appendix_mesh_qc}
\end{figure*}

The CFD domains use unstructured polyhedral volume meshes with ten wall-adjacent inflation layers and layer growth rate 1.2. A single-case sensitivity study compared five global minimum sizes: 0.25, 0.20, 0.15, 0.10, and 0.05 mm. These settings produced roughly 0.15, 0.25, 0.33, 0.53, and 0.88 million cells, respectively. The velocity profiles change little between the 0.15- and 0.10-mm settings over the plotted interval (Figure~\ref{fig:appendix_mesh_qc}a). This comparison informed the 0.15-mm production setting.\par

\subsection{Steady solver and inlet conditions}

The simulations solve the incompressible Newtonian Navier--Stokes equations with blood density $\rho=1050\,\mathrm{kg\,m^{-3}}$ and kinematic viscosity $\nu=3.2857\times10^{-6}\,\mathrm{m^2\,s^{-1}}$, corresponding to dynamic viscosity $\mu=0.003449985\;\mathrm{Pa\,s}$. These values are fixed by the archived \texttt{transportProperties} dictionary for every case. Vessel walls are rigid and no-slip; outlets use zero relative pressure.

The steady suite uses eight inlet mass-flow rates: 0.001, 0.0015, 0.002, 0.0025, 0.003, 0.0035, 0.00375, and $0.004\,\mathrm{kg\,s^{-1}}$. Steady simulations use OpenFOAM v2312 \texttt{icoFoam} in pseudo-transient mode with the Pressure-Implicit with Splitting of Operators (PISO) algorithm for pressure--velocity coupling. The generation protocol requires velocity and pressure residuals below $10^{-5}$. Representative completed runs typically reach velocity-component residuals below $10^{-9}$ and pressure residuals below $10^{-5}$, consistent with Figure~\ref{fig:appendix_mesh_qc}.

The project archive retains each case's runtime controls, numerical schemes, linear-solver settings, transport properties, and boundary-condition dictionaries. The steady corpus contains 77,544 descendant and 3,208 source flow-condition files.

\subsection{Transient solver and retained cycle}

For the transient suite, OpenFOAM v2312 \texttt{pimpleFoam} uses the common one-second pulsatile inlet waveform supplied with the release, a fixed $10^{-4}$-s time step, and output every 100 steps. The 119-node inlet table uses \texttt{flowRateInletVelocity}, ranging from $1.50\times10^{-6}$ to $1.45\times10^{-5}\,\mathrm{m^3\,s^{-1}}$ in the first cycle. Velocity is no-slip at walls and \texttt{inletOutlet} at outlets; pressure is zero-gradient at walls and the inlet and fixed to zero at outlets.

The PIMPLE algorithm, which combines PISO with the Semi-Implicit Method for Pressure-Linked Equations (SIMPLE), uses two pressure correctors, one non-orthogonal corrector, and up to 50 outer correctors. Pressure uses geometric agglomerated algebraic multigrid (GAMG; absolute tolerance $10^{-6}$, relative tolerance 0.01), and velocity uses a symmetric Gauss--Seidel smoother (absolute tolerance $10^{-5}$, relative tolerance 0.1).

The simulations use a fixed five-cycle duration, with four warm-up cycles and the fifth retained for analysis.

Each eligible transient HDF5 record contains the volume mesh, time-resolved velocity and pressure, boundary meshes and mappings, wall WSS vectors, and timestamps. One initialization entry precedes the 100 frames retained at 4.01--5.00 s. Field checks determine which converted records enter the benchmark (Section~\ref{sec:appendix_cfd_screening}); modality availability and conversion failures are recorded in Section~\ref{sec:appendix_release_analysis}.

\subsection{Equations and units}
\label{sec:appendix_numerical_specification}

The simulations solve the incompressible Newtonian equations
\begin{align}
\nabla\!\cdot\mathbf{u} &= 0,\\
\frac{\partial\mathbf{u}}{\partial t}
+(\mathbf{u}\!\cdot\nabla)\mathbf{u}
&=-\nabla p_k+\nu\nabla^2\mathbf{u},
\end{align}
where $p_k=p/\rho$ is kinematic pressure and $\nu=3.2857\times10^{-6}\,\mathrm{m^2\,s^{-1}}$. With $\rho=1050\,\mathrm{kg\,m^{-3}}$, the corresponding dynamic viscosity is $\mu=\rho\nu=0.003449985\,\mathrm{Pa\,s}$. Rigid walls use a no-slip condition, and outlets use zero relative pressure. These assumptions are shared across all simulated geometries. The simulations omit wall compliance, non-Newtonian rheology, and case-specific outlet impedance.

OpenFOAM exports $\mathbf{u}$ in $\mathrm{m\,s^{-1}}$ and $p_k$ in $\mathrm{m^2\,s^{-2}}$. The release converts pressure to Pa through $p=\rho p_k$. The stored \texttt{wallShearStress} vector is also in kinematic stress units; multiplying by $\rho$ converts it to Pa. For a Newtonian fluid, the continuous tangential traction has the form
\begin{equation}
\boldsymbol{\tau}_w
=\left(\mathbf{I}-\mathbf{n}\mathbf{n}^{\mathsf T}\right)
\boldsymbol{\sigma}\mathbf{n},\qquad
\boldsymbol{\sigma}=-p\mathbf{I}
+\mu\left(\nabla\mathbf{u}+\nabla\mathbf{u}^{\mathsf T}\right).
\end{equation}
Here $\mathbf n$ is the unit wall normal directed out of the fluid domain and $\mathbf I$ is the identity tensor. The release preserves the OpenFOAM \texttt{wallShearStress} vector convention for $\boldsymbol{\tau}_w$. TAWSS and OSI are derived from this vector time series and remain invariant to a global sign reversal.

\begin{table}[!htb]
\caption{Field and coordinate conventions used by the release and evaluator.}
\label{tab:appendix_units}
\centering\normalsize
\setlength{\tabcolsep}{4pt}
\begin{tabularx}{\textwidth}{>{\raggedright\arraybackslash}p{0.14\textwidth}
>{\raggedright\arraybackslash}p{0.20\textwidth}
>{\raggedright\arraybackslash}X
>{\raggedright\arraybackslash}p{0.19\textwidth}}
\toprule
Quantity & Stored solver form & Released/evaluated form & Domain \\
\midrule
Coordinates & Geometry source units standardized to metres & $x,y,z$ in m;
optional unit-bbox copy for intervention & Volume or wall vertices \\
Velocity & $\mathbf{u}$ & $\mathrm{m\,s^{-1}}$ & Internal volume mesh \\
Pressure & $p_k=p/\rho$ & $p=\rho p_k$ in Pa & Internal volume mesh \\
Wall shear stress & Kinematic wallShearStress vector &
$\boldsymbol{\tau}_w$ in Pa & Wall surface mesh \\
TAWSS & Not stored as a target & Cycle mean of
$\|\boldsymbol{\tau}_w(t)\|_2$ in Pa & Wall surface vertices \\
OSI & Not stored as a target & Dimensionless vector-cycle index in $[0,0.5]$
& Wall surface vertices \\
\bottomrule
\end{tabularx}
\end{table}

\subsection{Solver logs and field checks}
\label{sec:appendix_cfd_screening}

The release connects each CFD target to its geometry, numerical configuration, and eligibility record. Machine-readable summaries report finite-value checks, field shapes, time grids, boundary topology, and mass balance. Retained \texttt{pimpleFoam} logs document each source run; final-time directories match the sources admitted by the transient protocol.

\paragraph{Solver-log audit.}
A separate solver-log audit covers a fixed subset of 12 source geometries. All reached $t=5$ after 50,000 steps, with an \texttt{End} marker and no fatal-error line. The median fraction of steps reporting PIMPLE convergence was 0.99998. In the fifth cycle, the fraction reaching the outer-iteration limit had median zero and maximum 0.0659. The maximum logged Courant number was 48.73. The largest absolute global continuity error over the audited runs was $3.57\times10^{-5}$.

Between cycles four and five, phase-aligned WSS component-extrema trajectories had a median relative $\ell_2$ difference of 0.0302 (maximum 0.1776). Spatial fields are available only for the fifth cycle because \texttt{purgeWrite=100} removes earlier cycles. Selection and per-case records are in \path{data/validation/cfd_qc_selection_12real.json} and \path{data/validation/openfoam_solver_log_audit_12real.json}.

\paragraph{Full-field checks.}
Record validation covers all 9,718 converted transient HDF5 files available within the benchmark manifest: 401 sources and 9,317 descendants. Checks scan velocity, pressure, and WSS across all 101 stored entries and verify array shapes, timestamps, boundary mappings, and boundary polygon topology. Of these records, 9,715 pass. The three excluded source-training records (143, 246, and 441) have non-finite pressure values and invalid time grids; two also contain non-finite WSS values. Velocity is finite in all files. The per-record ledger and \path{data/validation/transient_field_audit_summary.json} report the checks and exclusions.

Field-level diagnostics summarize flow quantities and mass balance over every fifth non-initialization entry of the 9,715 retained records. Table~\ref{tab:appendix_release_qc_distribution} reports their distributions; range flags identify unusual values for inspection and are recorded separately from the field and structural checks.

The release metadata and the Reynolds and Womersley calculations use the transport parameters specified in Section~\ref{sec:appendix_numerical_specification}.

For the screening table, we use the equivalent inlet diameter $D=2\sqrt{A/\pi}$ and framewise mean absolute normal inlet velocity $\bar U$. We compute $\mathrm{Re}=\bar U D/\nu$ and $\alpha=(D/2)\sqrt{2\pi/(T_c\nu)}$ with cardiac period $T_c=1\,\mathrm{s}$.

Peak velocity is the largest internal-volume vertex magnitude over the 20 non-initialization frames retained by the stride-five audit. The 99th percentile of WSS magnitude (WSS p99) pools wall-vertex magnitudes over those frames. Relative residence time (RRT) is defined as $\mathrm{RRT}=1/[(1-2\,\mathrm{OSI})\mathrm{TAWSS}]$. TAWSS, OSI, and RRT are first computed per wall vertex and then area-averaged; the RRT denominator is floored at $10^{-3}$ Pa.

Per-frame mass imbalance is $|\dot m_{\mathrm{in}}+\dot m_{\mathrm{out}}|/|\dot m_{\mathrm{in}}|$; the table reports its maximum, whereas the separate 6\% screen applies to its cycle mean.

\begin{table*}[!htb]
\caption{Transient CFD screening summaries. RRT denotes relative residence time. Values are median [interquartile range] across included records with a finite summary; $n$ gives the case count for each diagnostic. Three divergent source sequences are excluded.}
\label{tab:appendix_release_qc_distribution}
\centering\normalsize
\setlength{\tabcolsep}{4pt}
\begin{tabularx}{\textwidth}{@{}>{\raggedright\arraybackslash}Xrrrr@{}}
\toprule
Diagnostic & Source geometries & $n$ & Descendants & $n$ \\
\midrule
Mean Reynolds number & 608 [479, 748] & 398 & 602 [477, 743] & 9,317 \\
Womersley $\alpha$ & 2.01 [1.63, 2.55] & 398 & 2.03 [1.65, 2.56] & 9,317 \\
Peak velocity ($\mathrm{m\,s^{-1}}$) & 3.53 [2.7, 5.02] & 398 & 3.71 [2.75, 5.26] & 9,317 \\
WSS p99 (Pa) & 146 [83.1, 249] & 398 & 164 [93.4, 290] & 9,317 \\
Mean TAWSS (Pa) & 17 [10.3, 27.8] & 398 & 21.8 [13.2, 37.4] & 9,317 \\
Mean OSI & 0.0224 [0.0144, 0.0356] & 398 & 0.0182 [0.0125, 0.0274] & 9,317 \\
Mean RRT ($\mathrm{Pa^{-1}}$) & 0.166 [0.0879, 0.287] & 398 & 0.0992 [0.0594, 0.165] & 9,317 \\
Maximum mass imbalance & 0.00246 [0.0011, 0.0063] & 398 & 0.00441 [0.00149, 0.0115] & 9,317 \\
\midrule
Records without any range flag & 398
(100.00\%)
& 398 & 9,133
(98.03\%)
& 9,317 \\
\bottomrule
\end{tabularx}
\end{table*}

\begin{figure}[!htb]
\centering
\includegraphics[width=\textwidth]{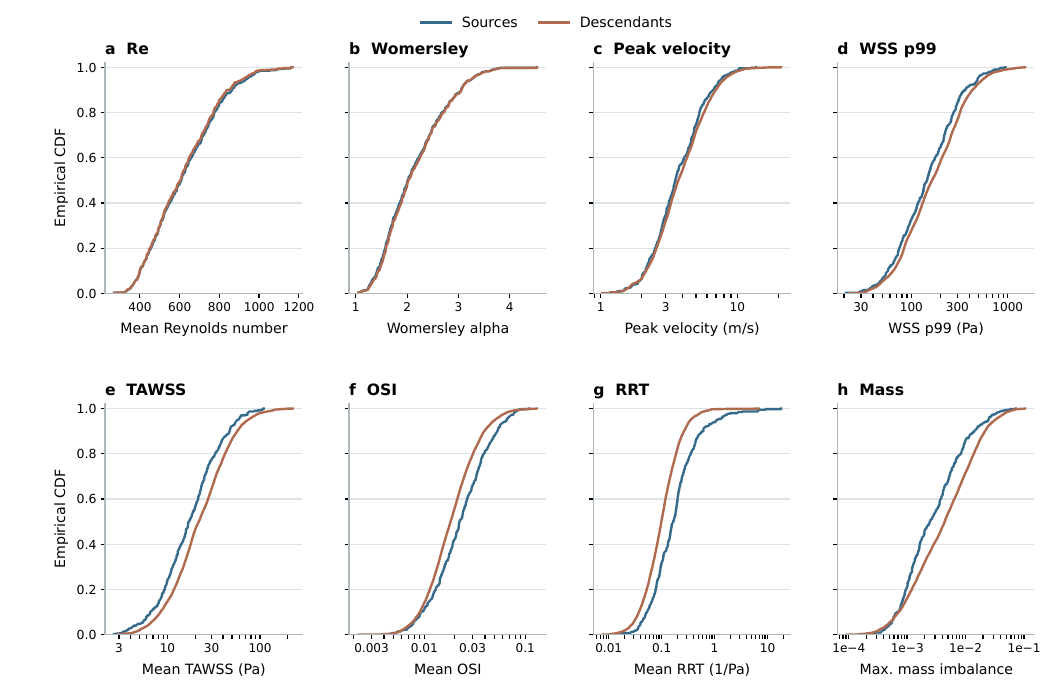}
\caption{Empirical distributions for the audited transient cohort, including dimensionless flow descriptors, field magnitudes, and solver balance. The source curve contains 398 geometries and the descendant curve 9,317 available variants. Positive heavy-tailed quantities use logarithmic horizontal axes.}
\label{fig:appendix_release_qc_distributions}
\end{figure}

The Reynolds and Womersley empirical cumulative distribution functions (CDFs) are similar in Figure~\ref{fig:appendix_release_qc_distributions}, indicating comparable inlet flow regimes under the shared boundary protocol. The wall responses differ: descendants have higher WSS p99 and mean TAWSS but lower median OSI and RRT. Synthetic pretraining therefore uses a related but shifted target distribution.

Screening intervals extend beyond the empirical tails of the source cohort and flag unusual records for inspection. Eligibility is determined by the field and structural checks. Overall, 98.03\% of descendants receive no range flag; mean TAWSS and mean RRT account for the most frequent flags. Mass imbalance is assessed separately from these ranges and the full-field finite-value scan.

\subsection{Ground-truth field diversity}
\label{sec:appendix_field_diversity}

The field galleries show how WSS, pressure, and speed vary spatially on individual source geometries. Sources are selected at approximately even intervals from the ordered test list.

\begin{figure}[!htb]
\centering
\includegraphics[width=0.9\textwidth]{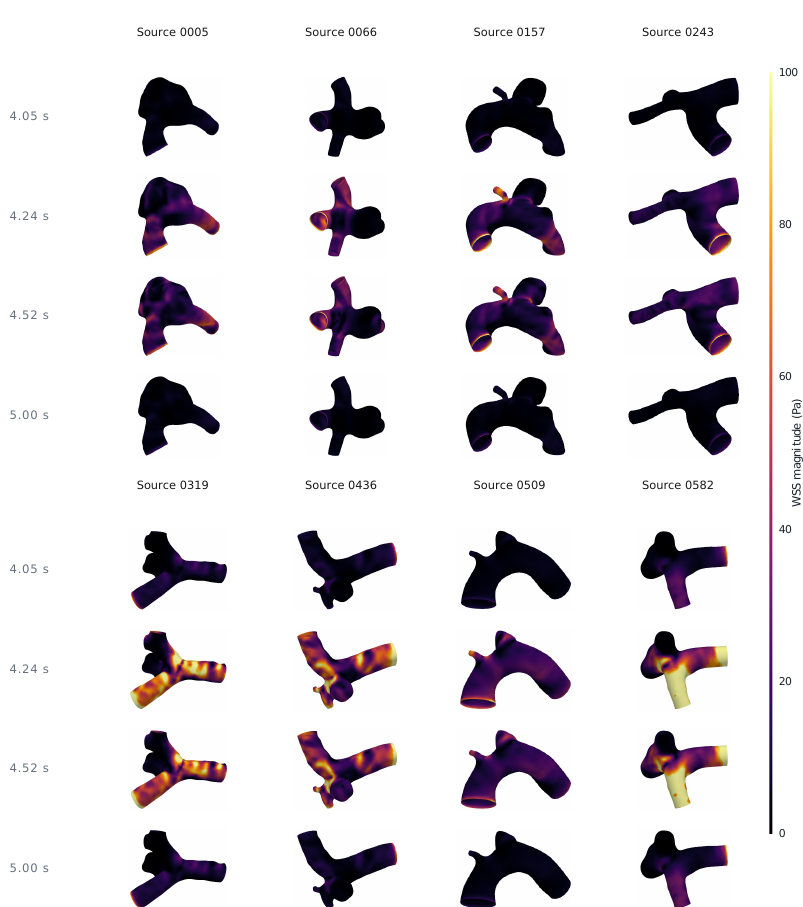}
\caption{Ground-truth WSS for eight source geometries selected at approximately even intervals from the official test list, shown at four cardiac phases. All 32 panels use the released polygonal wall meshes and a shared color scale. Sources are arranged in two groups of four, with the same phase in corresponding rows. The gallery shows how the location and persistence of high-shear regions vary across source geometries.}
\label{fig:appendix_wss_cohort}
\end{figure}

\begin{figure*}[!htb]
\centering
\includegraphics[width=0.9\textwidth]{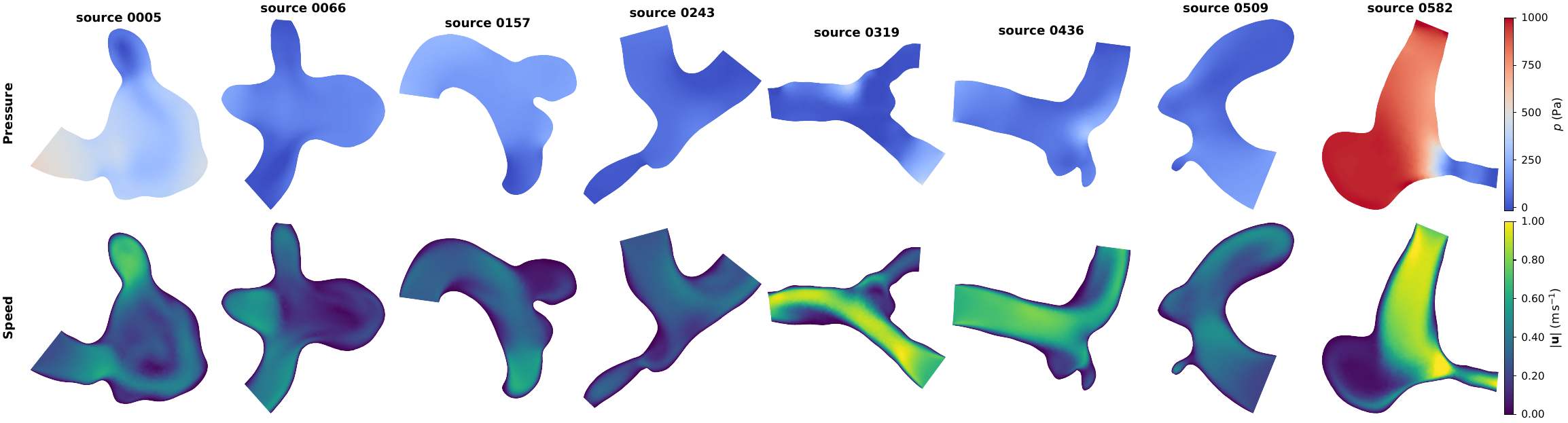}
\caption{Ground-truth steady pressure and speed for eight approximately evenly spaced held-out source geometries at $0.003\,\mathrm{kg\,s^{-1}}$. Each column uses a deterministic geometry-centered PCA plane and the original VTU triangulation. Shared row colorbars show variation in the fields across geometries.}
\label{fig:appendix_steady_cohort}
\end{figure*}

\section{Data Representations and Training Implementations}
\label{sec:appendix_training_details}

Sampling, coordinate transforms, and normalization convert the stored fields into model inputs. This section specifies those choices for each reference implementation.

\subsection{Stored fields, schemas, and preprocessing}

The dataset contains surface meshes (STL), voxel masks (\texttt{.nii.gz}), CFD meshes, mesh-resolved VTU/VTK exports, and ML-ready arrays. The mesh-resolved exports preserve the volume field (\texttt{internal.vtu}) and, when available, named inlet, outlet, and wall fields (\texttt{*.vtp}). These support physical-space inspection and alternative preprocessing. The temporal benchmark reads HDF5 records. Schemas document the arrays and units, while the manifest records split membership and availability.

\begin{table*}[!htb]
\caption{Release representation schema. Counts distinguish geometries from flow-condition files and transient sequences.}
\label{tab:appendix_schema}
\centering
\normalsize
\setlength{\tabcolsep}{3.2pt}
\begin{tabularx}{\textwidth}{@{}>{\raggedright\arraybackslash}p{0.14\textwidth}*{3}{>{\raggedright\arraybackslash}X}@{}}
\toprule
Data form & Principal contents & Benchmark role & Audited accounting \\
\midrule
STL surface & Vascular lumen boundary & Geometry/provenance & 401 sources + 9,693 eligible descendant records \\
NIfTI mask & Binary voxelized lumen & Auxiliary binary geometry mask & Linked by geometry manifest \\
CFD mesh / VTU/VTK & Volume grid, boundary and field exports & Mesh-resolved CFD outputs and provenance & Availability recorded per modality \\
Steady HDF5 & Coordinates, boundary/flow features, $\mathbf{u}$, $p$ & Eight-flow field prediction & 3,208 source + 77,544 descendant files \\
Transient HDF5 & Volume mesh, $\mathbf{u}$, $p$, boundary meshes, WSS, time & WSS 4$\to$4 and 4$\to$96 forecasting; volume fields retained & 401 archived source (398 usable) + 9,317 descendants \\
\bottomrule
\end{tabularx}
\end{table*}

For transient preprocessing, we drop the raw initialization frame, which has zero WSS, before forming prediction windows. Each protocol-included sequence then contains 100 valid cycle frames. The temporal models reported here are trained to predict WSS only.

\begin{table*}[!htb]
\caption{Benchmark task card. The evaluation unit remains the source geometry even when a task contains multiple flow conditions, windows, frames, or points. Extended diagnostics include mean squared error (MSE), root mean squared error (RMSE), and mean absolute error (MAE).}
\label{tab:appendix_task_card}
\centering\normalsize
\setlength{\tabcolsep}{2.5pt}
\begin{tabularx}{\textwidth}{@{}>{\raggedright\arraybackslash}p{0.10\textwidth}>{\raggedright\arraybackslash}p{0.13\textwidth}>{\raggedright\arraybackslash}X>{\raggedright\arraybackslash}p{0.10\textwidth}>{\raggedright\arraybackslash}X>{\raggedright\arraybackslash}p{0.10\textwidth}>{\raggedright\arraybackslash}X@{}}
\toprule
Task & Data form & Model input & Target & Within-case sampling & Primary endpoint & Extended diagnostics \\
\midrule
Steady field & Point / unstructured & $xyz$, boundary type, wall distance, mass flow & $\mathbf{u},p$ & 20,000 sampled volume points; fixed diagnostic subset & Source RelL2 & MSE, RMSE, MAE, $R^2$, direction \\
Steady field & Voxel $64^3$ & Occupancy, query coordinates, mass flow & $\mathbf{u},p$ & Complete constructed fluid mask & Source RelL2 & Same physical-unit metrics \\
Transient reset & Point / surface & 4 WSS frames, wall query, phase & Next 4 WSS frames & 4,096 sampled wall points; fixed diagnostic subset; stride 5 & Source-window RelL2 & RMSE, magnitude and angle \\
Transient reset & Voxel $64^3$ & Occupancy grid, wall queries, 4 WSS frames, phase & Next 4 WSS frames & Same fixed wall subset & Source-window RelL2 & Same physical-unit metrics \\
Transient rollout & Point or voxel & Geometry, phase, and first 4 WSS frames & Next 96 frames & No ground-truth reset & Mean $E_i^{\mathrm{roll}}$ & Mean cumulative error, TAWSS, OSI, top-10\%, peak time/magnitude \\
\bottomrule
\end{tabularx}
\end{table*}

\begin{table}[!htb]
\caption{Stage-wise provenance for the released representations. Source and geometry identifiers are preserved across every stage.}
\label{tab:appendix_provenance_stages}
\centering\normalsize
\setlength{\tabcolsep}{4pt}
\begin{tabularx}{\textwidth}{>{\raggedright\arraybackslash}p{0.14\textwidth}
>{\raggedright\arraybackslash}p{0.22\textwidth}
>{\raggedright\arraybackslash}p{0.28\textwidth}
>{\raggedright\arraybackslash}X}
\toprule
Stage & Input & Output & Recorded checks \\
\midrule
Surface repair & Public-source triangular surface & Watertight repaired source
surface & Topology defects, source identifier \\
Local deformation & Editing surface after sac removal and selected wall patch & Synthetic descendant
surface & Source--descendant mapping and fixed output surface \\
Voxelization & Descendant or source STL & Binary NIfTI lumen mask & Geometry identifier,
foreground/background convention \\
CFD preparation & Repaired STL & Named volume mesh and boundary patches &
ANSYS 2026 R1; inlet/outlet/wall labels and mesh settings \\
Steady solve & Mesh and one mass-flow condition & $\mathbf{u},p$ volume fields
& OpenFOAM v2312 \texttt{icoFoam}; PISO; residual criterion and complete case configuration archived \\
Transient solve & Mesh and periodic waveform & 101-entry raw field sequence &
OpenFOAM v2312 \texttt{pimpleFoam}; $\Delta t=10^{-4}$ s; PIMPLE/convergence logs and complete case configuration archived \\
ML packaging & Solver fields and geometry & HDF5, sampled point, and $64^3$
representations & Units, shapes, split \\
\bottomrule
\end{tabularx}
\end{table}

\subsection{Point sampling, graphs, and loader eligibility}

The \emph{unstructured} track supplies sampled points to point-set and attention models, or a 16-nearest-neighbor graph to GraphSAGE. The \emph{voxel} track encodes geometry on a regular $64^3$ grid. For transient WSS, it queries latent features at the same wall locations used by the unstructured track and passes them to a pointwise history-and-phase decoder. Thus the geometry encoders differ, but the WSS output domain is shared.

Steady unstructured baselines sample 20,000 internal-volume vertices without replacement. Each sample includes physical coordinates, a four-way interior/wall/inlet/outlet code, wall distance, and an inlet-flow feature. Fixed diagnostics derive a stable seed from the case file and evaluate all vertices when a mesh has at most 20,000. Transient fixed evaluation uses seed $42+$case index and samples 4,096 wall vertices without replacement when possible; every test source exceeds this count.

Six training descendants have fewer than 4,096 wall vertices. The cached unstructured R2 loader excludes these 6 of 6,438 available cases (0.09\%; 122,208 reset windows), whereas the case-grouped voxel R2 loader retains them by sampling with replacement (122,322 windows). This difference affects R2 training and, through the inherited R2 checkpoint, R3 initialization. Validation and test cohorts are identical across tracks.

\subsection{Voxel construction, coordinates, and wall queries}

For image-based protocols, 3D Slicer voxelizes the final surface into a binary Neuroimaging Informatics Technology Initiative (NIfTI) label map. The manifest links each mask to the corresponding geometry and CFD record.

The released NIfTI lumen mask differs from the $64^3$ grid used by the baselines. For the baseline grid, internal volume-mesh vertices define a bounding box for each case. Coordinates are centered, divided by the longest bounding-box side, and translated to an endpoint-inclusive $[0,1]^3$ lattice. Each grid location receives velocity and pressure from its nearest volume-mesh vertex. A voxel is marked as fluid when its normalized distance to that vertex is below $1.5/64$; targets outside this mask are zero. The models receive normalized coordinates, occupancy, and inlet flow. The stored bounding-box center and scale are omitted, so the input contains neither absolute physical size nor position.

Transient voxel models reuse the mask and transform of the $0.001\,\mathrm{kg\,s^{-1}}$ steady grid. Their latent features are queried at transformed wall points with trilinear \texttt{grid\_sample}, border padding, and \texttt{align\_corners=False}. All reported temporal voxel checkpoints use this decoder convention. Under the PyTorch sampling geometry, the endpoint-inclusive coordinates differ from endpoint-aligned interpolation by up to half a voxel. An alternative endpoint-aligned query transform is $g=(2u-1)(R-1)/R$ for normalized coordinate $u$ and grid resolution $R$; checkpoint metadata record the sampling convention. Steady voxel endpoints are evaluated directly on the regular grid.

The steady coordinate-scale sensitivity retains the same occupancy grid but replaces each unit-box query coordinate $\mathbf x$ with $(\mathbf x-0.5)s$, where $s$ is the stored longest-side bounding-box scale. This transform restores centered physical coordinates while keeping the bounding-box center at the origin. It is evaluated for U-Net and FNO only; the core matrix and transient voxel track retain unit-box coordinates.

\subsection{Normalization and domain shift}

Each run estimates normalization statistics from its training data. In the core recipes, R1 and the fine-tuning stage of R3 use source-training statistics, whereas R2 uses descendant statistics. The GraphSAGE controls use the same source-training statistics in every branch. For WSS, statistics estimated from source-training windows are reused throughout R3 fine-tuning and evaluation. Predictions are decoded to SI units before auxiliary evaluation on held-out sources. The statistics in Table~\ref{tab:appendix_normalization} also describe the descendant-to-source shift.

\begin{table}[!htb]
\caption{Output normalization statistics in channel order: $u_x,u_y,u_z,p$ for the steady task and the three WSS components for the transient task. Velocity is measured in $\mathrm{m\,s^{-1}}$; pressure and WSS are in Pa.}
\label{tab:appendix_normalization}
\centering\normalsize
\setlength{\tabcolsep}{5pt}
\begin{tabularx}{\textwidth}{@{}Xrr@{}}
\toprule
Channel & Mean & Standard deviation \\
\midrule
\multicolumn{3}{@{}l}{\textbf{Steady source}} \\
\multicolumn{3}{@{}l}{2,240 training flow cases; 514,573,832 points} \\
$u_x$ & $0.00069$ & $0.17669$ \\
$u_y$ & $0.01249$ & $0.16057$ \\
$u_z$ & $0.01157$ & $0.17094$ \\
$p$ & $169.49$ & $266.63$ \\
\addlinespace[4pt]
\multicolumn{3}{@{}l}{\textbf{Steady descendant}} \\
\multicolumn{3}{@{}l}{53,496 training flow cases; 9,962,079,176 points} \\
$u_x$ & $0.00007$ & $0.20798$ \\
$u_y$ & $0.01719$ & $0.18774$ \\
$u_z$ & $0.01553$ & $0.19895$ \\
$p$ & $166.46$ & $303.96$ \\
\addlinespace[4pt]
\multicolumn{3}{@{}l}{\textbf{Transient source}} \\
\multicolumn{3}{@{}l}{Window-weighted $H{=}4\rightarrow F{=}4$ training frames} \\
$\tau_x$ & $-0.2056$ & $38.1109$ \\
$\tau_y$ & $-1.1351$ & $35.3612$ \\
$\tau_z$ & $-1.6628$ & $38.3904$ \\
\addlinespace[4pt]
\multicolumn{3}{@{}l}{\textbf{Transient descendant}} \\
\multicolumn{3}{@{}l}{Window-weighted $H{=}4\rightarrow F{=}4$ training frames} \\
$\tau_x$ & $-0.3560$ & $45.6614$ \\
$\tau_y$ & $-1.6134$ & $42.8244$ \\
$\tau_z$ & $-1.9307$ & $45.7977$ \\
\bottomrule
\end{tabularx}
\end{table}

Component means are small relative to their standard deviations because vessel orientation varies across cases. Synthetic velocity and WSS have consistently larger standard deviations. The pressure means are similar, but the synthetic standard deviation is 14\% larger. The normalization sensitivity experiment holds architecture and training data fixed while changing the normalization statistics.

\subsection{Reference architectures}

The reference suite uses unstructured point sets, explicit k-nearest-neighbor (kNN) graphs, and regular voxel grids. Its models use message passing, attention, convolution, or spectral mixing.

\begin{table}[!htb]
\caption{Trainable parameter counts for completed R3 runs in each task, as recorded in the training logs. M denotes millions.}
\label{tab:appendix_model_inventory}
\centering\normalsize
\setlength{\tabcolsep}{9pt}
\begin{tabular}{lrr}
\toprule
Model & Steady & Transient \\
\midrule
\multicolumn{3}{l}{\textit{Unstructured sampled points}} \\
GraphSAGE & 0.37M & 0.37M \\
Transolver & 3.86M & 3.87M \\
Galerkin Transformer & 1.10M & 1.10M \\
PointNet & 0.23M & 0.23M \\
\multicolumn{3}{l}{\textit{Regular voxel geometry ($64^3$)}} \\
U-Net & 51.89M & 51.92M \\
U-FNO & 217.06M & 217.09M \\
FNO & 113.28M & 113.30M \\
U-NO & 518.65M & 518.68M \\
F-FNO & 1.19M & 1.22M \\
\bottomrule
\end{tabular}
\end{table}

\begin{table*}[!htb]
\caption{Architecture and batch configurations. PointwiseMLP is an auxiliary transient diagnostic; the remaining rows form the nine-architecture reference suite. Width and depth follow the released implementations.}
\label{tab:appendix_model_hyperparameters}
\centering\normalsize
\setlength{\tabcolsep}{4pt}
\begin{tabularx}{\textwidth}{@{}>{\raggedright\arraybackslash}p{0.21\textwidth}rrr>{\centering\arraybackslash}p{0.16\textwidth}>{\raggedright\arraybackslash}X@{}}
\toprule
Model & Width & Depth & Heads & Batch (steady / transient) & Geometry/operator setting \\
\midrule
PointwiseMLP (aux.) & 128 & 4 & -- & -- / 32 & Shared residual multilayer perceptron; no point interaction \\
PointNet & 16 & 8 & -- & 32 / 64 & Point-set multilayer perceptron \\
GraphSAGE & 128 & 8 & -- & 32 / 32 & 16-nearest-neighbor graph \\
Transolver & 256 & 8 & 8 & 16 / 16 & 32 physics-attention slices \\
Galerkin Transformer & 128 & 8 & 8 & 16 / 32 & Galerkin attention \\
U-Net & 64 & 4 & -- & 32 / 16 & $64^3$ voxel grid \\
U-FNO & 64 & 4 & -- & 16 / 16 & 12 Fourier modes \\
FNO & 64 & 4 & -- & 32 / 16 & 12 Fourier modes \\
U-NO & 64 & 4 & -- & 16 / 8 & 12 Fourier modes \\
F-FNO & 64 & 4 & -- & 32 / 16 & 12 factorized modes \\
\bottomrule
\end{tabularx}

{\raggedright\normalsize\textit{Excluded implementation.} The archived adapter for Graph U-Net~\citep{gao2019graphunet} passed feature channels as spatial coordinates to its pooling and interpolation layers. Its steady and transient results are therefore excluded from the reference comparisons and cross-model summaries; the archived records retain their original identifiers.\par}
\end{table*}

\subsection{Training recipes and checkpoint selection}

R1 and R2 use the training sources and their descendants, respectively. R2 checkpoints are selected on source-validation data before R3 fine-tunes them on the source-training fields. The steady R1 corpus contains 2,240 fields per epoch; R2 contains 53,496. R1--R3 therefore differ in the number of sampled fields and training updates, initialization, and data composition.

All core runs use AdamW with weight decay $10^{-5}$, a OneCycle schedule with warm-up fraction 0.3, zero dropout, and default seed 42. Table~\ref{tab:appendix_training_protocol} gives the epoch counts and learning rates for R1--R3.

\begin{table*}[!htb]
\caption{Optimization schedules for the core R1--R3 comparison. Table~\ref{tab:appendix_model_hyperparameters} lists batch sizes, widths, and key operator settings; the released configurations and logs give the full settings.}
\label{tab:appendix_training_protocol}
\centering\normalsize
\setlength{\tabcolsep}{3pt}
\begin{tabularx}{\textwidth}{l
>{\raggedright\arraybackslash}X
>{\raggedright\arraybackslash}X
ccc
>{\raggedright\arraybackslash}X}
\toprule
Regime & Training data & Initial weights & Steady epochs & Transient epochs
& Learning rate & Selection / evaluation \\
\midrule
R1 & Source geometries & Random & 200 & 200 & $10^{-3}$ & Minimum
validation endpoint; held-out sources \\
R2 & Synthetic descendants of training sources & Random & 200 & 50 & $10^{-3}$
& Source-validation selection; common source test \\
R3 & Source training geometries & Best R2 checkpoint & 100 & 100 &
$10^{-4}$ & Minimum validation endpoint; same test cohort \\
\bottomrule
\end{tabularx}
\end{table*}

Steady training minimizes a single per-example RelL2 over velocity and pressure outputs concatenated after per-channel normalization. Core unstructured checkpoints are selected by validation velocity RelL2, while core voxel checkpoints use masked validation velocity RelL2. The GraphSAGE pool-size and fixed-budget allocation controls select on the mean of validation velocity and pressure RelL2. GraphSAGE uses batch size 32 in the core matrix and batch size 16 in its matched controls. Test endpoints are decoded to physical units and are not used for checkpoint selection.

Core transient R1/R2/R3 training appends observed WSS frames to the history at each step. Evaluation feeds back predictions, with observed histories restored between reset windows but not during the 96-step rollout. Section~\ref{sec:appendix_extended_transient} examines training with prediction feedback over longer horizons.

Checkpoint selection in the core unstructured steady and transient runs resampled point subsets on each validation pass; steady voxel validation used the complete regular grid. All extended diagnostics use fixed case-indexed subsets. Steady unstructured metrics are defined on internal volume-mesh vertices, steady voxel metrics on the constructed fluid mask, and both transient tracks on the same wall-query domain. Primary metrics use uniform vertex or mask weighting. Appendix Section~\ref{sec:appendix_physics_sensitivity} reports quadrature-weighted diagnostics.

Pool-size and allocation controls use the specific schedules and pool memberships in Section~\ref{sec:appendix_synthetic_pools}. Additional temporal baselines and feedback-training settings are defined with their comparisons in Section~\ref{sec:appendix_extended_transient}.

\section{Evaluation Metrics and Statistical Methods}
\label{sec:appendix_metrics}

Metrics are computed from decoded physical fields. The definitions below distinguish field errors, recursive prediction, pressure diagnostics, and cycle-level WSS quantities before specifying source-level aggregation.

\subsection{Field and directional errors}

For a decoded steady field or reset window $m$ from held-out source geometry $i$, let $j=1,\ldots,J_{im}$ index the evaluated scalar entries, counting vector components separately. The metrics are
\begin{align}
\mathrm{MSE}_{im}
  &=J_{im}^{-1}\sum_j(\hat y_{imj}-y_{imj})^2, &
\mathrm{RMSE}_{im}
  &=\sqrt{\mathrm{MSE}_{im}},\\
\mathrm{MAE}_{im}
  &=J_{im}^{-1}\sum_j|\hat y_{imj}-y_{imj}|, &
e_{im}
  &=\frac{\|\hat{\mathbf y}_{im}-\mathbf y_{im}\|_2}
          {\|\mathbf y_{im}\|_2+\epsilon_{\mathbf y}},\\
R^2_{im}
  &=1-\frac{\sum_j(\hat y_{imj}-y_{imj})^2}
  {\sum_j(y_{imj}-\bar y_{im})^2+\eta_{R^2}}, &
\bar y_{im}
  &=J_{im}^{-1}\sum_j y_{imj}.
\end{align}

Norm denominators use $\epsilon_z=10^{-8}$ in the physical units of field $z$. The $R^2$ variance denominator adds $\eta_{R^2}=10^{-8}$ in squared field units. Primary norms and means assign equal weight to the deterministically sampled entries; Appendix~\ref{sec:appendix_physics_sensitivity} reports the corresponding quadrature-weighted sensitivity analysis.

For a vector entry $\ell$, directional metrics retain only
\begin{equation}
\mathcal D_{im}=\{\ell:\|\hat{\mathbf y}_{im\ell}\|_2>10^{-8}
       \ \land\ \|\mathbf y_{im\ell}\|_2>10^{-8}\}.
\end{equation}
Cosine similarity averages the normalized dot product over $\mathcal D_{im}$. Angular error averages the arccosine of the clipped cosine similarities, converted to degrees. All thresholds and errors are applied after decoding to physical units; numerical denominator guards use the metric-specific constants above.

\subsection{Reset-window, rollout, and cumulative errors}

For a 4$\rightarrow$4 reset window, the model receives four ground-truth WSS frames and predicts four future frames autoregressively. A 4$\rightarrow$96 rollout uses the first four valid-cycle frames as observations, together with geometry and phase, and feeds predictions back for the remaining 96 steps. Let $\hat{\mathbf W}_{i,s}$ and $\mathbf W_{i,s}$ denote the predicted and target wall vectors, flattened over sampled vertices and components, for source $i$ at step $s$. The primary source-level rollout error is
\begin{equation}
E_i^{\mathrm{roll}}=
\frac{\left(\sum_{s=1}^{96}\|\hat{\mathbf W}_{i,s}-\mathbf W_{i,s}\|_2^2\right)^{1/2}}
     {\left(\sum_{s=1}^{96}\|\mathbf W_{i,s}\|_2^2\right)^{1/2}+\epsilon_{\mathbf W}}.
\end{equation}
Its mean over the 79 sources is reported as rollout RelL2. Error accumulation is summarized by
\begin{align}
C_i(s)&=\left(
\frac{\sum_{r=1}^{s}\lVert\hat{\mathbf W}_{i,r}-\mathbf W_{i,r}\rVert_2^2}
{\sum_{r=1}^{s}\lVert\mathbf W_{i,r}\rVert_2^2+\eta_C}
\right)^{1/2}, &
\overline C_i&=\frac{1}{96}\sum_{s=1}^{96}C_i(s).
\end{align}
The cumulative evaluator uses $\eta_C=10^{-8}\,\mathrm{Pa^2}$ in its squared-energy denominator. Without numerical guards, $C_i(96)$ equals $E_i^{\mathrm{roll}}$. The mean cumulative error $\overline C_i$ averages cumulative errors over forecast steps.

\subsection{Quadrature, centered pressure, and pressure drop}

Quadrature and pressure-gauge sensitivity are evaluated separately from the primary, uniformly weighted evaluation. Let $k=1,\ldots,K_{im}$ index spatial locations and let $q_{imk}\geq0$ be normalized quadrature weights, $\sum_kq_{imk}=1$. For steady unstructured fields, VTK computes cell volumes from the original \texttt{internal.vtu}. Each volume is divided equally among the cell's stored vertices, with contributions summed at shared vertices. All 79 held-out HDF5 geometries exactly match their source VTU point order, cell offsets, connectivity, and VTK cell types. Within each source, all eight steady conditions share that HDF5 mesh, and no nonpositive cell volume occurs.

Wall weights are obtained by fan-triangulating the native boundary polygons and lumping triangle area to their vertices.

Across the 79 source VTUs, 1,199,045 of 19,985,647 stored vertices (6.0\%) are orphan points in the exported point arrays: they are not referenced by the volume-cell connectivity. Audited orphan entries store zero velocity and pressure, including positions inside active fluid cells; these are export placeholders rather than valid CFD point supervision. They remain in the uniform vertex-sampled endpoint and receive zero weight in the native-volume diagnostic. Volume weighting consequently removes the contribution of orphan points as well as changing the integration weights. Uniform voxel volumes cancel after normalization.

In the weighted error, $\mathbf y_{imk}$ collects the components at location $k$, including all four predicted frames for a reset window. The squared-energy guard is $\eta_q=10^{-12}$ in squared field units for both quadrature-weighted and centered-pressure errors:

\begin{equation}
e^{(q)}_{im}=
\left(
\frac{\sum_kq_{imk}\lVert\hat{\mathbf y}_{imk}-\mathbf y_{imk}\rVert_2^2}
     {\sum_kq_{imk}\lVert\mathbf y_{imk}\rVert_2^2+\eta_q}
\right)^{1/2}.
\end{equation}
Centered-pressure RelL2 subtracts each field's spatial mean to separate spatial mismatch from mean bias:
\begin{align}
e^{\mathrm{center}}_{im}&=
\left(
\frac{\sum_kq_{imk}\{(\hat p_{imk}-\hat\mu_{im})
                    -(p_{imk}-\mu_{im})\}^2}
     {\sum_kq_{imk}(p_{imk}-\mu_{im})^2+\eta_q}
\right)^{1/2},\\
\hat\mu_{im}&=\sum_kq_{imk}\hat p_{imk}, &
\mu_{im}&=\sum_kq_{imk}p_{imk}.
\end{align}
With $b_{im}=\hat\mu_{im}-\mu_{im}$ and $d_{imk}=(\hat p_{imk}-\hat\mu_{im})-(p_{imk}-\mu_{im})$, the pressure error decomposes as
\begin{equation}
\sum_k q_{imk}(\hat p_{imk}-p_{imk})^2
=b_{im}^2+\sum_k q_{imk}d_{imk}^2.
\end{equation}
This decomposes stored-gauge squared pressure error into mean-bias and centered-field contributions under the same spatial weights. The analysis reports $b_{im}$ separately, together with a sampled-edge pressure-gradient error and inlet--outlet pressure-drop error. The gradient diagnostic uses a fixed 16-nearest-neighbor graph as a local proxy. For pressure drop, $\mathcal I_{im}$ and $\mathcal O_{im}$ contain the evaluated vertices marked by the released inlet and outlet boundary codes. Cases with multiple outlets pool all outlet vertices. Using uniform means over the sampled vertices, the pressure-drop metric is
\begin{equation}
\Delta p_{im}=|\mathcal I_{im}|^{-1}\!\sum_{k\in\mathcal I_{im}}p_{imk}
-|\mathcal O_{im}|^{-1}\!\sum_{k\in\mathcal O_{im}}p_{imk},\qquad
E^{\mathrm{drop}}_{im}=
\frac{|\widehat{\Delta p}_{im}-\Delta p_{im}|}{|\Delta p_{im}|+\epsilon_p}.
\end{equation}
The absolute pressure-drop error is $|\widehat{\Delta p}_{im}-\Delta p_{im}|$ in Pa. Voxel inputs do not retain inlet and outlet query sets, so this diagnostic is reported only for the unstructured track.

\subsection{Cycle-level WSS, hotspots, and peaks}

TAWSS and OSI are evaluated over the retained fifth cycle. The predicted cycle combines four observed frames with 96 predictions; the target cycle uses all 100 CFD frames. Let $\mathbf w_{t,v}$ denote the WSS vector at wall vertex $v$ and cycle frame $t$. For $N_t=100$ equally spaced frames,

\begin{align}
\mathrm{TAWSS}(v) &= \frac{1}{N_t}\sum_{t=1}^{N_t}\|\mathbf{w}_{t,v}\|_2,\\
\mathrm{OSI}(v) &= \frac12\left(1-
\frac{\|\sum_t \mathbf{w}_{t,v}\|_2}{\sum_t\|\mathbf{w}_{t,v}\|_2+\epsilon_{\mathbf W}}\right).
\end{align}
Let $V_i$ be the set of sampled wall vertices. The target and predicted TAWSS fields are $\tau_i,\hat\tau_i$, and the corresponding OSI fields are $o_i,\hat o_i$. Their source-level errors are
\begin{align}
E_i^{\mathrm{TAWSS}}
  &=\frac{\|\hat\tau_i-\tau_i\|_2}{\|\tau_i\|_2+\epsilon_{\tau}}, &
E_i^{\mathrm{OSI}}
  &=|V_i|^{-1}\sum_{v\in V_i}|\hat o_i(v)-o_i(v)|.
\end{align}
If $G_i$ and $P_i$ are the equal-cardinality ground-truth and predicted top-10\% TAWSS vertex sets, then hotspot overlap is $|G_i\cap P_i|/|G_i|$; hotspot error is TAWSS relative $\ell_2$ error restricted to $G_i$. For peak timing and magnitude, use the spatial-mean magnitude curve $a_i(t)=|V_i|^{-1}\sum_v\|\mathbf w_{t,v}\|_2$ and define $t_i^*=\operatorname*{arg\,max}_t a_i(t)$ and $\hat t_i^*=\operatorname*{arg\,max}_t\hat a_i(t)$. Peak timing is the circular frame distance $\min(|\hat t_i^*-t_i^*|,N_t-|\hat t_i^*-t_i^*|)$ and peak-magnitude error is $|\hat a_i(\hat t_i^*)-a_i(t_i^*)|/(|a_i(t_i^*)|+\epsilon_a)$.

\subsection{Source-level aggregation and uncertainty}

Each reported metric $z$ uses the nested reduction

\begin{equation}
\bar z=\frac{1}{N}\sum_{i=1}^{N}
       \frac{1}{M_i}\sum_{m=1}^{M_i}z_{im},\qquad
\mathcal E_{\mathrm{rel}}=\frac{1}{N}\sum_{i=1}^{N}
       \frac{1}{M_i}\sum_{m=1}^{M_i}e_{im}.
\end{equation}
Here $M_i=8$ for steady flow conditions and $M_i=19$ for the stride-five $4\!\rightarrow\!4$ reset windows. Full-cycle metrics are computed once per source and averaged over the 79 sources. With this nested reduction, mean RMSE generally differs from the square root of mean MSE.

Core endpoint confidence intervals use 1,000 source-geometry cluster-bootstrap resamples and the 2.5th/97.5th percentiles; analyses with a different resampling count state it locally. Points, flow conditions, windows, and frames remain nested within their source geometries.

Training-seed summaries report sample SD across the runs identified in each table. Repeated adaptation from fixed initial weights describes adaptation-stage variability; it is reported separately from independent full-pipeline repeats.

\section{Extended Reference Results}
\label{sec:appendix_all_baselines}

The reference scores are reported in main-text Table~\ref{tab:core_benchmark}; Appendix Table~\ref{tab:appendix_model_inventory} gives parameter counts for both tasks. The following comparisons examine source-level differences and variability in source fine-tuning.

\subsection{Paired-source comparisons}

\begin{figure}[!htb]
\centering
\includegraphics[width=\textwidth]{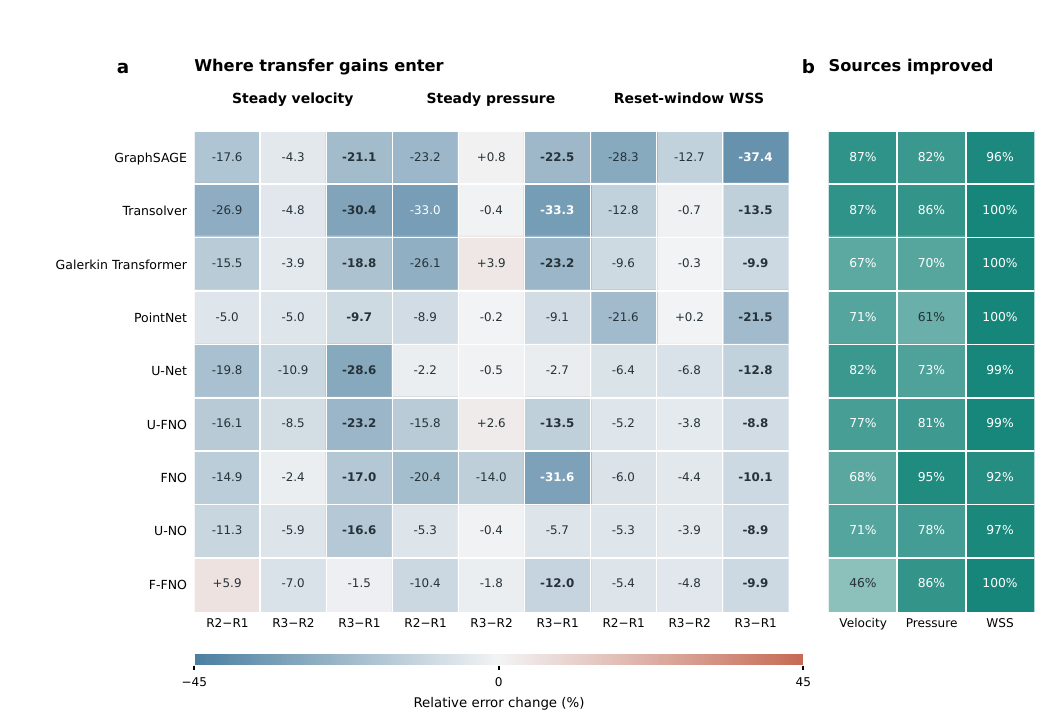}
\caption{Paired comparisons of the three training settings across nine reference architectures. (a) Cells report relative error changes for R2 versus R1, R3 versus R2, and R3 versus R1; negative values denote lower error. Bold R3--R1 values have paired-source 95\% bootstrap intervals excluding zero. The upper four rows are point/unstructured architectures and the lower five are voxel architectures. (b) Cells give the percentage of the same 79 source geometries improved by R3 relative to R1; darker fill denotes broader consistency. Figure~\ref{fig:matched_exposure_scaling} presents the GraphSAGE pool-size and allocation controls.}
\label{fig:transfer_breadth}
\end{figure}

A lower source-mean error can combine improvements on some geometries with deteriorations on others. Figure~\ref{fig:transfer_breadth} compares R1, R2, and R3 on the same sources and reports the fraction that improve. R1 and R2 are trained independently; R3 fine-tunes the R2 checkpoint on source fields.

Source fine-tuning lowers velocity error relative to R2 for every architecture and lowers reset-window WSS error for 8 of 9. Pressure changes are smaller and mixed. Most of the full R1-to-R3 difference is already present after descendant training.

\subsection{Steady fine-tuning repeats}

We measure source fine-tuning variability across random seeds while holding one descendant-pretrained checkpoint fixed for each architecture.

\begin{table*}[!htb]
\caption{Steady fine-tuning variability for three representative architectures. Each architecture uses one fixed seed-42 descendant-pretraining checkpoint and source fine-tuning seeds 42--44. Entries are held-out-source means $\pm$ sample SD across these seeds, conditional on the shared pretrained weights. RMSE is computed from decoded fields and reported in $\mathrm{m\,s^{-1}}$ for velocity and Pa for pressure. Lower is better.}
\label{tab:appendix_steady_finetune_seeds}
\centering\normalsize
\setlength{\tabcolsep}{5pt}
\begin{tabular}{lllrr}
\toprule
Model & Representation & Field & RelL2 & Physical-unit RMSE \\
\midrule
PointNet & Point set & Velocity & 0.891 $\pm$ 0.001 & 0.12606 $\pm$ 0.00009 \\
 & & Pressure & 0.773 $\pm$ 0.003 & 130.7 $\pm$ 0.2 \\
\addlinespace[1pt]
GraphSAGE & 16-NN graph & Velocity & 0.654 $\pm$ 0.001 & 0.09486 $\pm$ 0.00001 \\
 & & Pressure & 0.560 $\pm$ 0.006 & 96.4 $\pm$ 0.3 \\
\addlinespace[1pt]
U-Net & $64^3$ voxel & Velocity & 0.704 $\pm$ 0.002 & 0.09795 $\pm$ 0.00010 \\
 & & Pressure & 0.795 $\pm$ 0.012 & 125.4 $\pm$ 0.2 \\
\bottomrule
\end{tabular}
\end{table*}

\subsection{Steady source distributions and field examples}

\label{sec:appendix_source_error}

\begin{figure}[!htb]
\centering
\includegraphics[width=0.82\textwidth]{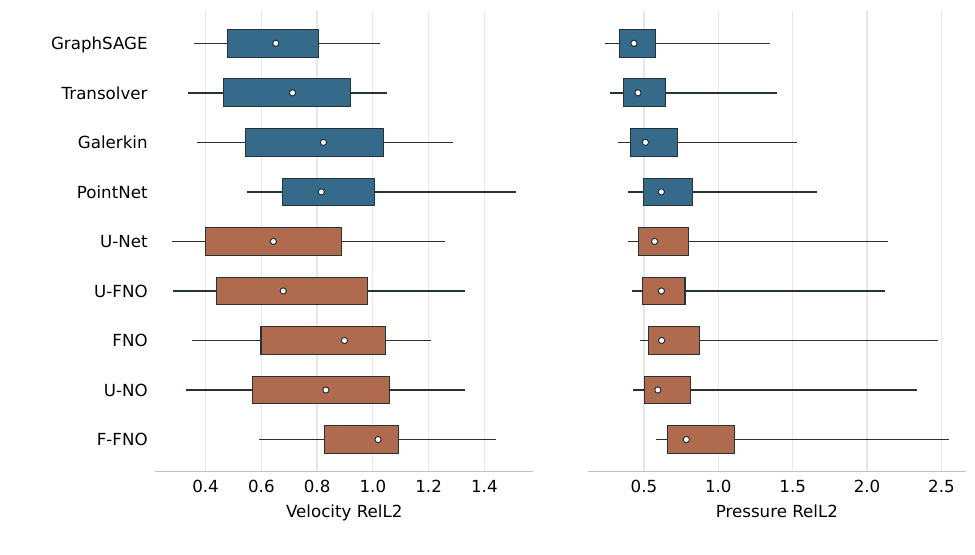}
\caption{Source-level R3 steady relative-$\ell_2$ distributions for the nine reference architectures. Blue denotes unstructured sampled-point architectures and orange voxel architectures. Each source contributes once after averaging its eight mass-flow conditions; boxes show quartiles and whiskers the 5th--95th percentiles.}
\label{fig:appendix_steady_source_distributions}
\end{figure}

Figure~\ref{fig:appendix_steady_source_distributions} shows R3 error distributions across sources, with eight flow conditions averaged within each source.

F-FNO's R3 source-mean velocity RelL2 is 1.0005, close to the zero-velocity reference of one. GraphSAGE, Transolver, and voxel U-Net have lower medians, with substantial variation across sources. Pressure errors have longer right tails for nearly all architectures. Centered-pressure diagnostics in Appendix~\ref{sec:appendix_physics_sensitivity} separate spatial mismatch from the overall errors shown here.

\begin{figure*}[!htb]
\centering
\includegraphics[width=0.9\textwidth]{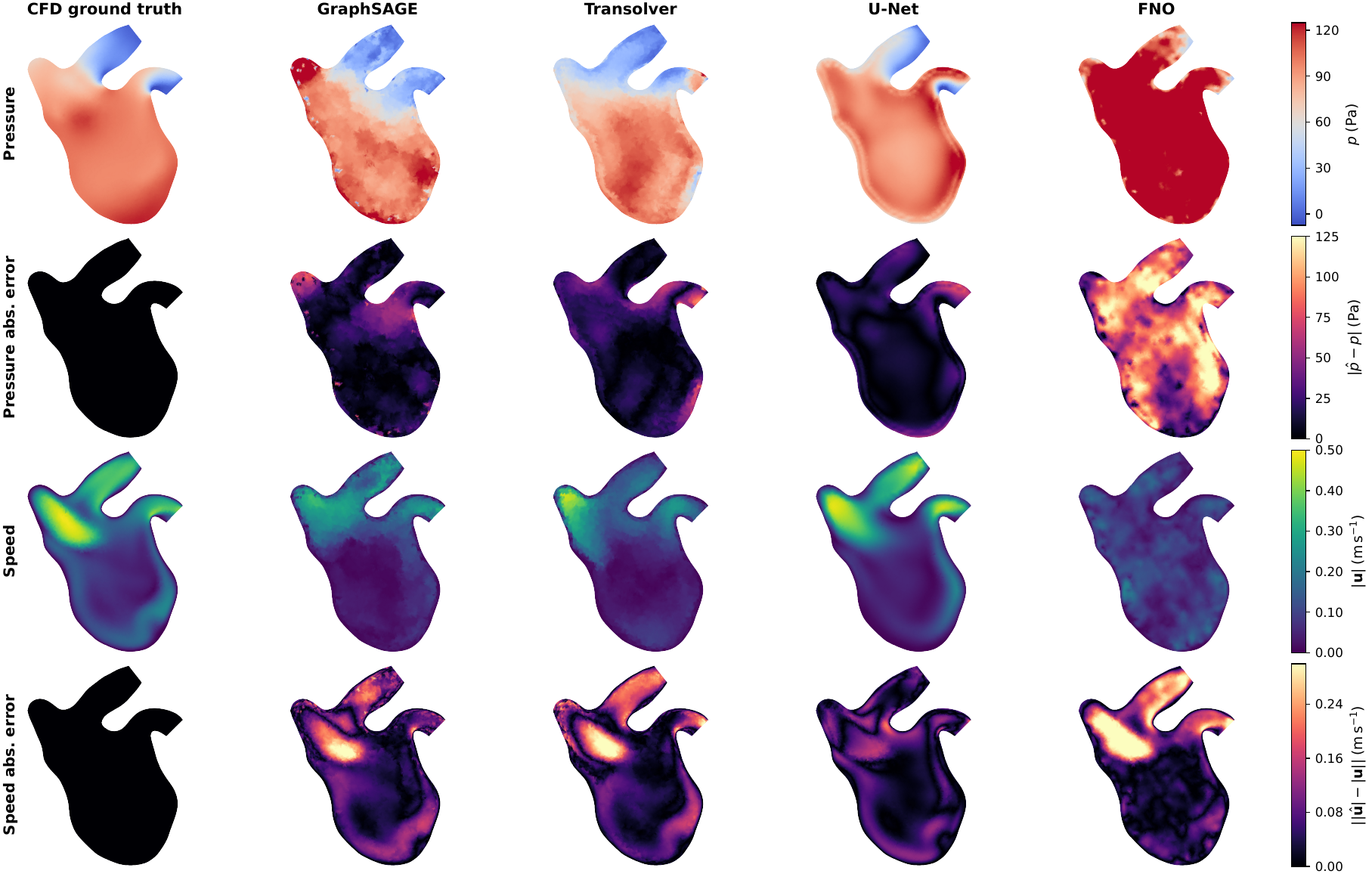}
\caption{Steady-field comparison on a common plane through the 410,202-vertex, 137,745-cell internal VTU for held-out \texttt{real\_0072} at $0.003\,\mathrm{kg\,s^{-1}}$. Rows show pressure in Pa, absolute pressure error, speed, and absolute speed error. The mostly-polyhedral mesh is tetrahedralized before slicing; all panels use the same plane triangulation and share a colorbar within each row. Point-model outputs are evaluated directly at 20,000 deterministic vertices and nearest-neighbor interpolated elsewhere for display. Voxel outputs are sampled trilinearly from the $64^3$ grid. Cohort-level metrics appear in Tables~\ref{tab:appendix_steady_velocity_metrics} and~\ref{tab:appendix_steady_pressure_metrics}.}
\label{fig:appendix_steady_fields}
\end{figure*}

Figure~\ref{fig:appendix_steady_fields} shows where velocity and pressure predictions depart from the target fields. The source, camera, slice, and color scale are shared across models.

\section{Descendant Pool Size and Allocation}
\label{sec:appendix_synthetic_pools}

The two aneurysm controls use GraphSAGE and the same source split. Pool-size runs change the available descendants while keeping sampling and update counts fixed. Allocation runs redistribute a fixed descendant count, then share a source fine-tuning stage. The airfoil study applies the allocation principle to a separate dataset and independently trained models.

\subsection{Pool-size settings and results}
\label{sec:appendix_fixed_compute}

The GraphSAGE pool-size control includes all 280 training sources and expands the accessible descendant pool from 280 to 560, 1,120, 2,238, 4,422, and 6,687 records. Each run draws 2,240 fields per epoch and performs 28,000 optimizer updates. Source-training normalization, validation selection, and deterministic test sampling are shared across the compared pools.

For each fixed pool, seeds 42--44 vary initialization and stochastic field sampling; sample SD summarizes the resulting training variability. Table~\ref{tab:graphsage_scaling} gives the pool means.

\begin{table*}[!htb]
\caption{GraphSAGE pool-size comparison on 79 held-out source geometries. All branches use 280 source lineages, 2,240 fields per epoch, 28,000 updates, and identical deterministic test samples. Values are source-level means $\pm$ training-seed SD across three seeds. $\Delta_{\mathrm{rel}}$ is the paired relative change from source-only; ``Wins'' counts seeds improved on both fields. Pool membership is fixed across seeds at every size.}
\label{tab:graphsage_scaling}
\centering
\normalsize
\setlength{\tabcolsep}{3.2pt}
\begin{tabular}{lrrrrrr}
\toprule
Training pool & Records & Velocity RelL2 & $\Delta_{\mathrm{rel}}$ & Pressure RelL2 & $\Delta_{\mathrm{rel}}$ & Wins \\
\midrule
Source geometries & 280 & 0.830 $\pm$ 0.005 & -- & 0.731 $\pm$ 0.027 & -- & -- \\
Descendant $k{=}1$ & 280 & 0.815 $\pm$ 0.002 & -1.7\% & 0.706 $\pm$ 0.012 & -3.3\% & 3/3 \\
Descendant $k{=}2$ & 560 & 0.785 $\pm$ 0.002 & -5.4\% & 0.680 $\pm$ 0.013 & -6.8\% & 3/3 \\
Descendant $k{=}4$ & 1,120 & 0.762 $\pm$ 0.003 & -8.2\% & 0.663 $\pm$ 0.003 & -9.2\% & 3/3 \\
Descendant $k{\leq}8$ & 2,238 & 0.739 $\pm$ 0.002 & -11.0\% & \textbf{0.630 $\pm$ 0.017} & -13.8\% & 3/3 \\
Descendant $k{\leq}16$ & 4,422 & 0.738 $\pm$ 0.006 & -11.1\% & 0.635 $\pm$ 0.004 & -13.0\% & 3/3 \\
Descendant full pool & 6,687 & \textbf{0.734 $\pm$ 0.008} & -11.6\% & 0.633 $\pm$ 0.020 & -13.2\% & 3/3 \\
\bottomrule
\end{tabular}
\end{table*}

For GraphSAGE, most of the observed reduction occurs by approximately eight descendants per lineage, after which changes are small. All six tested descendant pools improve both endpoints in every paired seed. Each descendant is paired with a CFD field recomputed on its geometry, so the control evaluates the training value of these geometry--field pairs.

Table~\ref{tab:graphsage_lineage_control} lists the source-only and descendant results for each paired seed at $k=4$. All three seeds use the same capped descendant membership.

\begin{table*}[!htb]
\caption{Seed-level GraphSAGE $k{=}4$ control on held-out source geometries. Both branches cover the same 280 source families and use equal numbers of sampled fields and updates, with the same normalization, validation, and deterministic test samples. The synthetic $k=4$ membership is fixed across seeds. Metrics average eight flows per source; the final row reports mean (sample SD). Negative paired changes favor synthetic training.}
\label{tab:graphsage_lineage_control}
\centering
\normalsize
\setlength{\tabcolsep}{2.8pt}
\begin{tabular}{lrrrrrr}
\toprule
& \multicolumn{3}{c}{Velocity RelL2}
& \multicolumn{3}{c}{Pressure RelL2} \\
\cmidrule(lr){2-4}\cmidrule(lr){5-7}
Seed & \makecell{Source\\only} & \makecell{Descendant\\$k{=}4$} & \makecell{Paired\\$\Delta_{\mathrm{rel}}$}
& \makecell{Source\\only} & \makecell{Descendant\\$k{=}4$} & \makecell{Paired\\$\Delta_{\mathrm{rel}}$} \\
\midrule
42 & 0.8263 & \textbf{0.7649} & \textbf{-7.4\%} & 0.7008 & \textbf{0.6607} & \textbf{-5.7\%} \\
43 & 0.8274 & \textbf{0.7593} & \textbf{-8.2\%} & 0.7529 & \textbf{0.6660} & \textbf{-11.6\%} \\
44 & 0.8361 & \textbf{0.7616} & \textbf{-8.9\%} & 0.7383 & \textbf{0.6615} & \textbf{-10.4\%} \\
\addlinespace[2pt]
Mean (SD) & 0.830 (0.005) & \textbf{0.762 (0.003)} & \textbf{-8.2 (0.7)\%} & 0.731 (0.027) & \textbf{0.663 (0.003)} & \textbf{-9.2 (3.1)\%} \\
\bottomrule
\end{tabular}
\end{table*}

\subsection{Fixed-count allocation and common source fine-tuning}

The fixed-budget lineage-allocation control instead holds the stage-1 pool at exactly 280 descendant records and compares $280\!\times\!1$, $140\!\times\!2$, $70\!\times\!4$, and $35\!\times\!8$ allocations, written as source lineages times descendants per lineage. Within each seed, the smaller lineage sets are nested inside the larger sets and use a shared within-lineage descendant ordering. Each branch trains on 2,240 descendant fields per epoch for 200 epochs (28,000 updates). All then fine-tune on the same 2,240 source fields per epoch for 100 epochs (14,000 updates). Normalization, validation, deterministic test sampling, and flow-condition sampling are fixed. GraphSAGE uses 4,096 points and is repeated with seeds 42--44.

\begin{table*}[!htb]
\caption{GraphSAGE fixed-budget lineage breadth--depth allocation using 280 descendant records on the same 79 held-out source geometries. All branches use 2,240 stage-1 fields per epoch, 28,000 stage-1 updates, and the same 14,000-update source fine-tuning stage. Values are mean $\pm$ sample SD across seeds 42--44. $\Delta$ is the mean paired absolute change from $280\!\times\!1$; ``wins'' counts seeds with lower error than that reference. Lower is better.}
\label{tab:appendix_lineage_breadth_depth}
\centering\normalsize
\setlength{\tabcolsep}{3.2pt}
\begin{tabular}{lrrrrrr}
\toprule
& \multicolumn{3}{c}{Velocity RelL2} & \multicolumn{3}{c}{Pressure RelL2} \\
\cmidrule(lr){2-4}\cmidrule(lr){5-7}
Allocation & Mean $\pm$ SD & $\Delta$ & Wins & Mean $\pm$ SD & $\Delta$ & Wins \\
\midrule
$280\!\times\!1$ & \textbf{0.799 $\pm$ 0.019} & -- & -- & \textbf{0.717 $\pm$ 0.032} & -- & -- \\
$140\!\times\!2$ & 0.818 $\pm$ 0.012 & +0.018 & 0/3 & 0.743 $\pm$ 0.027 & +0.026 & 0/3 \\
$70\!\times\!4$ & 0.821 $\pm$ 0.014 & +0.022 & 1/3 & 0.743 $\pm$ 0.029 & +0.026 & 1/3 \\
$35\!\times\!8$ & 0.848 $\pm$ 0.022 & +0.049 & 0/3 & 0.786 $\pm$ 0.040 & +0.069 & 1/3 \\
\bottomrule
\end{tabular}
\end{table*}

GraphSAGE has its lowest mean velocity and stored-gauge pressure errors at the broadest of the four allocations. Table~\ref{tab:allocation_pressure_structure} separates stored-gauge pressure error from spatial pressure structure.

\begin{table*}[!htb]
\caption{GraphSAGE pressure-structure diagnostics for the fixed-280-descendant allocation control. Entries average source-balanced endpoints across seeds 42--44. Each checkpoint uses the same uniformly weighted fixed point samples as its primary allocation evaluation. Centered-pressure metrics subtract each field's spatial mean; a spatially constant pressure prediction has centered-pressure RelL2 approximately one. Lower is better.}
\label{tab:allocation_pressure_structure}
\centering\normalsize
\setlength{\tabcolsep}{5pt}
\begin{tabular}{lrrrr}
\toprule
Allocation & \makecell{Stored-gauge\\RelL2} & \makecell{Centered-pressure\\RelL2} & \makecell{Centered-pressure\\RMSE (Pa)} & \makecell{Pressure-drop\\relative error} \\
\midrule
$280\!\times\!1$ & 0.717 & 1.082 & 96.47 & 0.514 \\
$140\!\times\!2$ & 0.743 & 1.117 & 99.07 & 0.505 \\
$70\!\times\!4$ & 0.743 & 1.127 & 98.60 & 0.546 \\
$35\!\times\!8$ & 0.786 & 1.174 & 101.94 & 0.650 \\
\bottomrule
\end{tabular}
\end{table*}

GraphSAGE's centered-pressure RelL2 exceeds one across these allocations, while its velocity predictions distinguish the allocations more clearly.

\subsection{Airfoil construction, training, and allocation}
\label{sec:appendix_airfoil_transfer}

The cross-domain study applies the lineage and exposure controls to airfoil geometries from AFBench~\citep{liu2024afbench}. Its construction names link each class--shape transformation (CST) variant to a source airfoil, allowing descendants to inherit the source split directly. XFoil 6.97 computes the surface pressure coefficient at Reynolds number $10^6$, Mach number 0.2, and $4^{\circ}$ angle of attack. The result is represented on a canonical 256-point contour and evaluated by arc-length-weighted pressure-coefficient RelL2, averaged over source airfoils.

The split contains 280 training sources and 42 source-validation airfoils. Each descendant branch contains 280 fields allocated as $280\!\times\!1$, $140\!\times\!2$, $70\!\times\!4$, or $35\!\times\!8$. These four branches use 7,000 updates on the declared descendant allocation followed by 3,500 updates on the same 280 sources, with batch size 16 and source-only target normalization; the Source-only column is the separately trained reference recipe. Seeds 42--44 independently determine the nested lineage pool, data order, parameter initialization, and optimization. Evaluation uses all 47 unique solver-eligible sources remaining after cohort construction; none shares an exact geometry with the training or validation cohort.

GraphSAGE uses the airfoil implementation in \path{experiments/airfoil_protocol_transfer/train_airfoil.py}, with hidden width 128, depth parameter \texttt{n\_layers=8}, and a symmetrized 16-nearest-neighbor graph on the contour coordinates. Both training stages use AdamW with weight decay $10^{-5}$ and a OneCycle schedule with peak learning rate $10^{-3}$ and warm-up fraction 0.3. Training minimizes arc-length-weighted MSE on normalized pressure coefficients.

\begin{figure*}[!htb]
\centering
\includegraphics[width=0.96\textwidth]{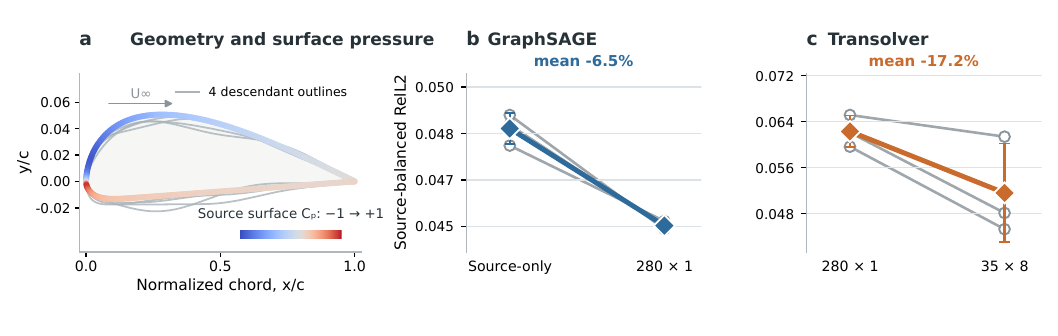}
\caption{Cross-domain protocol transfer on source-edited airfoils. (a) The first training lineage in the frozen manifest: source airfoil \texttt{ag03} and its first four accepted CST descendants. The source surface is colored by its XFoil pressure coefficient at the declared flow condition; gray lines show descendant outlines. Separate $x/c$ and $y/c$ axis scales make the local edits visible. (b) GraphSAGE with broad descendant pretraining versus source-only training. (c) Transolver with the same 280-descendant budget allocated across 280 or 35 source lineages. Gray paths connect the same full-pipeline seed across conditions; colored diamonds and whiskers show mean $\pm$ training-seed SD for seeds 42--44. Both result panels use the same 47 held-out source airfoils; lower is better.}
\label{fig:appendix_airfoil_transfer}
\end{figure*}

A sweep on source-validation data selected a peak learning rate of $3\!\times\!10^{-4}$ for Transolver. The 15 Transolver runs are evaluated on the 47 held-out airfoils. Figure~\ref{fig:appendix_airfoil_transfer} shows two comparisons, and Table~\ref{tab:appendix_airfoil_transfer} in the main text gives the full allocation sweep. Relative reductions are calculated from the unrounded means across seeds.

Among the four descendant allocations, GraphSAGE has its lowest mean at $280\!\times\!1$, whereas Transolver has its lowest at $35\!\times\!8$.

\section{WSS Forecasting and Temporal Diagnostics}
\label{sec:appendix_extended_transient}

The WSS analyses compare local-history predictors with spatial models, then examine how training feedback and forecast length affect rollout. All use the same held-out source geometries. Cycle metrics and input interventions assess what the resulting forecasts preserve and which observations they use.

\subsection{Temporal reference baselines}
\label{sec:appendix_temporal_references}

Copy-last persistence repeats the final observed WSS frame. Linear extrapolation fits a line through the current four-frame history independently at each point and component, without fitting parameters across training cases.

History+phase ridge regression instead fits shared coefficients in closed form on deterministic subsets of the source-training cohort. Its 16 features comprise 12 normalized WSS history components, phase, sine and cosine of phase, and an intercept. The normal-equation penalty is $10^{-6}N_{\mathrm{fit}}$, where $N_{\mathrm{fit}}=13{,}615{,}104$ is the number of fitting rows; the intercept is unpenalized. These coefficients remain fixed during evaluation.

All three baselines use each wall point's own history throughout the sequence, without explicit coordinates or interactions between points.

\paragraph{PointwiseMLP.}

PointwiseMLP provides a nonlinear residual counterpart to the ridge baseline. It shares weights across points, while each output depends only on that point's registered history and phase. It maps the 13 history-and-phase inputs to width 128, applies four pre-normalized residual multilayer perceptron (MLP) blocks with expansion two and Gaussian error linear unit (GELU) activation, and decodes one WSS vector per point (267,139 trainable parameters).

R1 uses AdamW, weight decay $10^{-5}$, a 200-epoch OneCycle schedule with peak learning rate $10^{-3}$, batch size 32, and independent seeds 42, 43, and 44. The checkpoint with the lowest four-step autoregressive RelL2 on validation sources is selected. Rollout predicts one frame at a time for 96 steps.

\begin{table*}[!htb]
\caption{Temporal reference baselines and an exploratory PointwiseMLP diagnostic. Persistence and linear extrapolation require no training-data fit; history+phase ridge regression is fit in closed form on source-training data. Bracketed entries give source means and 95\% source-bootstrap CIs over the 79-source test set. The final row reports mean $\pm$ training-seed SD over three independently trained and validation-selected PointwiseMLP pipelines.}
\label{tab:appendix_baselines}
\centering
\normalsize
\begin{tabularx}{\textwidth}{@{}>{\raggedright\arraybackslash}Xccc@{}}
\toprule
Method & Window RelL2 & Rollout RelL2 & \makecell{Mean cumulative\\RelL2} \\
\midrule
Copy-last persistence & \makecell{0.2580\\{[0.2482, 0.2679]}} & \makecell{0.8806\\{[0.8785, 0.8828]}} & \makecell{0.7389\\{[0.7370, 0.7407]}} \\
Linear extrapolation & \makecell{0.3426\\{[0.3229, 0.3618]}} & \makecell{0.9929\\{[0.9802, 1.0135]}} & \makecell{0.8040\\{[0.7962, 0.8152]}} \\
History+phase ridge, no coordinates & \makecell{0.2706\\{[0.2613, 0.2810]}} & \makecell{0.9911\\{[0.9888, 0.9937]}} & \makecell{0.8624\\{[0.8582, 0.8668]}} \\
PointwiseMLP R1, seed 42 & \makecell{0.1463\\{[0.1341, 0.1596]}} & \makecell{0.7797\\{[0.7614, 0.7986]}} & \makecell{0.6446\\{[0.6282, 0.6612]}} \\
PointwiseMLP R1, 3-seed mean $\pm$ SD & $0.1459\pm0.0022$ & $0.7346\pm0.1047$ & $0.6013\pm0.0869$ \\
\bottomrule
\end{tabularx}

\end{table*}

PointwiseMLP seed 42 improves rollout error over persistence on 72/79 sources. Top-10\% overlap decreases on every source. The paired records are in \path{data/validation/pointwise_persistence_paired.json}. All three PointwiseMLP seeds have lower mean rollout error than persistence, although the metric values vary across seeds.

\subsection{Standard reset-window and rollout results}

Under reset-window evaluation, Transolver and Galerkin Transformer are nearly tied within the unstructured track under both RelL2 and decoded RMSE (Table~\ref{tab:appendix_transient_multimetric}). U-Net has the lowest error among the voxel models.

We evaluate the standard R1, R2, and R3 checkpoints over the 96-step rollout. Their largest error increase occurs near the rapid transition at predicted step 19.

Table~\ref{tab:appendix_transfer_rollout} reports seed-42 R1--R3 rollouts for Transolver and voxel U-Net, the two models with rollout results for all three settings. R2 predictions are decoded with descendant-training normalization; R1 and R3 use source-training statistics.

Panel (b) reports paired rollout differences, intervals, relative changes, and source improvement fractions. The intervals describe variation among test sources for these fixed checkpoints. Statistics for all metrics are retained in \path{data/validation/transient_transfer_rollout.json}. R2 rollout is available only for seed 42; independent R1/R3 training repeats follow below.

\begin{table*}[!htb]
\caption{R1--R3 checkpoints under reset-window and 96-step WSS evaluation on the same 79 held-out sources, using seed 42. Lower is better except top-10\% overlap. Panel (b) reports paired-source rollout changes; negative values favor the later regime. Table~\ref{tab:appendix_transfer_rollout_seeds} reports independent pipeline repeats across training seeds.}
\label{tab:appendix_transfer_rollout}
\centering\normalsize
\textbf{(a) Endpoint estimates}\par\vspace{2pt}
\setlength{\tabcolsep}{3pt}
\begin{tabular}{llrrrrr}
\toprule
Model & Regime & Reset RelL2 & Rollout RelL2 & TAWSS rel. & OSI abs. & Top-10\% overlap \\
\midrule
Transolver & R1 & 0.1253 & 1.5907 & 1.2793 & 0.0383 & 0.7898 \\
 & R2 & 0.1092 & 0.9810 & 0.8873 & 0.0602 & 0.8211 \\
 & R3 & 0.1084 & 0.9836 & 0.8785 & 0.0569 & 0.8240 \\
\addlinespace[2pt]
Voxel U-Net & R1 & 0.2369 & 0.9118 & 0.7673 & 0.0461 & 0.6128 \\
 & R2 & 0.2216 & 0.9450 & 0.8296 & 0.0380 & 0.4118 \\
 & R3 & 0.2064 & 0.9824 & 0.8740 & 0.0381 & 0.6117 \\
\bottomrule
\end{tabular}

\vspace{5pt}\textbf{(b) Paired rollout changes}\par\vspace{2pt}
\begin{tabular}{llrrcr}
\toprule
Model & Comparison & \makecell{Rollout\\difference} & \makecell{Relative\\change} & \makecell{95\% paired-\\source CI} & \makecell{Sources\\improved} \\
\midrule
Transolver & R2--R1 & -0.6097 & -38.3\% & [-1.2064, -0.1579] & 31.6\% \\
 & R3--R1 & -0.6071 & \textbf{-38.2\%} & [-1.1686, -0.1435] & 34.2\% \\
 & R3--R2 & +0.0026 & +0.3\% & [-0.0284, +0.0329] & 50.6\% \\
\addlinespace[2pt]
Voxel U-Net & R2--R1 & +0.0333 & +3.6\% & [+0.0159, +0.0506] & 27.8\% \\
 & R3--R1 & +0.0707 & \textbf{+7.8\%} & [+0.0404, +0.1071] & 21.5\% \\
 & R3--R2 & +0.0374 & +4.0\% & [+0.0132, +0.0636] & 32.9\% \\
\bottomrule
\end{tabular}

\end{table*}

For Transolver seed 42, R3 lowers the mean rollout error, although only 34.2\% of sources improve. Each of the three independent R3 runs also has lower source-mean error than its paired R1 run (Table~\ref{tab:appendix_transfer_rollout_seeds}).

\subsection{Independent R1/R3 training repeats}

To measure training variability, we independently train R1 and the complete descendant-to-source pipeline with seeds 42--44 (Table~\ref{tab:appendix_transfer_rollout_seeds}). Each R3 run starts from its own R2 checkpoint, and sample SD summarizes variability across the resulting pipelines.

\begin{table*}[!htb]
\caption{Independent full-pipeline replications for transient transfer. Each seed trains R1 from random initialization, trains its own R2 checkpoint, and initializes R3 from that checkpoint. Panel (a) reports vertex-sampled RelL2; $\Delta$ is paired within seed. The final rows give mean $\pm$ sample SD across three training seeds. Panel (b) reports cycle endpoints as mean $\pm$ sample SD. All evaluations use the same 79 sources. R2 rollout was not repeated and remains in Table~\ref{tab:appendix_transfer_rollout} as a seed-42 diagnostic.}
\label{tab:appendix_transfer_rollout_seeds}
\centering\normalsize
\textbf{(a) Reset-window and 96-step rollout RelL2}\par\vspace{2pt}
\setlength{\tabcolsep}{3pt}
\begin{tabular}{llrrrrrr}
\toprule
Model & Seed & R1 reset & R3 reset & $\Delta$ & R1 rollout & R3 rollout & $\Delta$ \\
\midrule
Transolver & 42 & 0.1253 & 0.1084 & -0.0169 & 1.5907 & 0.9836 & -0.6071 \\
 & 43 & 0.1272 & 0.1091 & -0.0181 & 1.1231 & 0.9467 & -0.1764 \\
 & 44 & 0.1305 & 0.1083 & -0.0222 & 1.1219 & 0.9673 & -0.1546 \\
 & \makecell{Mean\\$\pm$ SD} & \makecell{$0.1276$\\$\pm 0.0026$} & \makecell{$0.1086$\\$\pm 0.0004$}
 & \makecell{$-0.0190$\\$\pm 0.0028$} & \makecell{$1.2786$\\$\pm 0.2703$} & \makecell{$0.9659$\\$\pm 0.0185$}
 & \makecell{$-0.3127$\\$\pm 0.2552$} \\
\addlinespace[2pt]
Voxel U-Net & 42 & 0.2369 & 0.2064 & -0.0304 & 0.9118 & 0.9824 & +0.0707 \\
 & 43 & 0.2469 & 0.2098 & -0.0371 & 0.9835 & 0.9509 & -0.0326 \\
 & 44 & 0.2565 & 0.2085 & -0.0480 & 0.9676 & 0.9348 & -0.0328 \\
 & \makecell{Mean\\$\pm$ SD} & \makecell{$0.2468$\\$\pm 0.0098$} & \makecell{$0.2083$\\$\pm 0.0017$}
 & \makecell{$-0.0385$\\$\pm 0.0089$} & \makecell{$0.9543$\\$\pm 0.0377$} & \makecell{$0.9560$\\$\pm 0.0242$}
 & \makecell{$+0.0018$\\$\pm 0.0597$} \\
\bottomrule
\end{tabular}

\vspace{5pt}\textbf{(b) Cycle endpoints across training seeds}\par\vspace{2pt}
\setlength{\tabcolsep}{3pt}
\begin{tabular}{llrrrrrr}
\toprule
& & \multicolumn{2}{c}{TAWSS rel.} & \multicolumn{2}{c}{OSI abs.}
& \multicolumn{2}{c}{Top-10\% overlap} \\
\cmidrule(lr){3-4}\cmidrule(lr){5-6}\cmidrule(lr){7-8}
Model & Seeds & R1 & R3 & R1 & R3 & R1 & R3 \\
\midrule
Transolver & 42--44 & \makecell{$1.0263$\\$\pm 0.2354$} & \makecell{$0.7900$\\$\pm 0.0893$}
& \makecell{$0.0446$\\$\pm 0.0055$} & \makecell{$0.0393$\\$\pm 0.0154$}
& \makecell{$0.7923$\\$\pm 0.0087$} & \makecell{$0.7875$\\$\pm 0.0335$} \\
Voxel U-Net & 42--44 & \makecell{$0.7756$\\$\pm 0.0330$} & \makecell{$0.8474$\\$\pm 0.0264$}
& \makecell{$0.0482$\\$\pm 0.0029$} & \makecell{$0.0384$\\$\pm 0.0064$}
& \makecell{$0.5198$\\$\pm 0.0853$} & \makecell{$0.5739$\\$\pm 0.0331$} \\
\bottomrule
\end{tabular}

\end{table*}

\subsection{Feedback training and forecast horizon}
\label{sec:appendix_feedback_training}

Training feedback determines which frames enter the history. The forecast horizon determines how many future frames contribute to each update. We compare both choices before evaluating fixed R1 and R3 initializations under the same long-window adaptation (Section~\ref{sec:appendix_f16_initialization}).

Scheduled sampling starts from the same architecture-specific R2 checkpoint as core R3 fine-tuning and retains its 100-epoch source-geometry schedule, $H=F=4$, and default seed 42. At epoch $e$, the feedback probability is
\begin{equation}
\pi_e=0.5\,e/100,\qquad e=1,\ldots,100.
\end{equation}
At each autoregressive target step in a minibatch, one Bernoulli draw is shared by the batch; success appends the detached prediction and failure appends the ground-truth frame. Each reported SS row is one validation-selected checkpoint; its rollout endpoints are reported with the other temporal interventions in Table~\ref{tab:appendix_rollout_extended}.

The long-window variants retain four history frames, predict $F=16$ targets, and reuse detached predictions as full feedback. The Transolver variant is the $F=16$ full-feedback cell of the factorial below. The U-Net replication starts from its corresponding fixed R3 checkpoint and uses the same batch size eight, learning rate $10^{-4}$, and validation selection with a 20-epoch schedule. For each model, the reported mean and sample SD summarize three adaptation seeds from one fixed R3 initialization.

\subsubsection{Horizon--feedback factorial}

Every run in the horizon--feedback factorial starts from the same Transolver R3 checkpoint. The factorial crosses training horizon $F\in\{4,8,16\}$ with teacher forcing or full stop-gradient feedback. All 18 runs use four history frames, batch size eight, AdamW with maximum learning rate $10^{-4}$, and a complete 30-epoch OneCycle schedule; the three repeats vary only the adaptation seed.

\begin{figure}[!htb]
\centering
\includegraphics[width=0.76\linewidth]{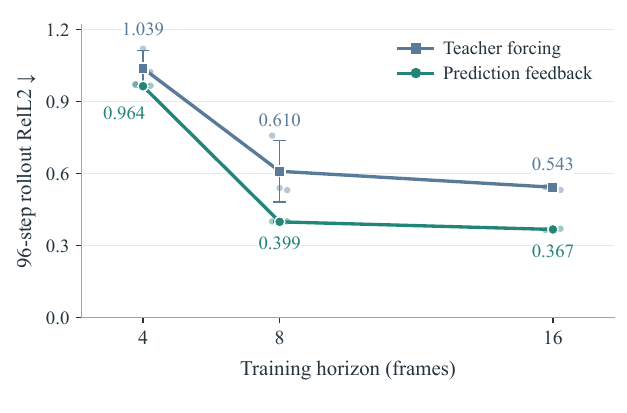}
\caption{Transolver rollout accuracy after 30 epochs of adaptation from one shared R3 checkpoint. Each training window predicts $F$ frames, with either observed or predicted frames appended to the history. Small dots are individual adaptation seeds; markers show means with sample SD. Table~\ref{tab:appendix_horizon_feedback_grid} gives the complete metrics.}
\label{fig:appendix_horizon_feedback_grid}
\end{figure}

Longer horizons predict more training frames within the common 30-epoch schedule. Figure~\ref{fig:appendix_horizon_feedback_grid} compares rollout errors, and Table~\ref{tab:appendix_horizon_feedback_grid} gives the reset and cycle-level endpoints.

\begin{table*}[!htb]
\caption{Three-seed Transolver horizon--feedback factorial from one shared R3 initialization. All settings use 30 epochs. Entries are mean $\pm$ adaptation-seed sample SD. Lower is better except top-10\% overlap.}
\label{tab:appendix_horizon_feedback_grid}
\centering\normalsize
\setlength{\tabcolsep}{1.3pt}
\begin{tabular}{@{}rlrrrrr@{}}
\toprule
$F$ & Feedback & Reset $4\!\rightarrow\!4$ & 96-step rollout & TAWSS rel. & OSI abs. & Top-10\% overlap \\
\midrule
4 & Teacher forcing & \makecell{$0.1087$\\$\pm 0.0001$} & \makecell{$1.0387$\\$\pm 0.0747$} & \makecell{$0.9245$\\$\pm 0.0693$} & \makecell{$0.0561$\\$\pm 0.0002$} & \makecell{$0.816$\\$\pm 0.003$} \\
4 & Full feedback & \makecell{\textbf{0.1073}\\$\mathbf{\pm 0.0001}$} & \makecell{$0.9640$\\$\pm 0.0064$} & \makecell{$0.8809$\\$\pm 0.0126$} & \makecell{$0.0422$\\$\pm 0.0005$} & \makecell{\textbf{0.830}\\$\mathbf{\pm 0.003}$} \\
8 & Teacher forcing & \makecell{$0.1088$\\$\pm 0.0001$} & \makecell{$0.6096$\\$\pm 0.1286$} & \makecell{$0.4263$\\$\pm 0.1264$} & \makecell{\textbf{0.0273}\\$\mathbf{\pm 0.0003}$} & \makecell{$0.794$\\$\pm 0.007$} \\
8 & Full feedback & \makecell{$0.1078$\\$\pm 0.0003$} & \makecell{$0.3990$\\$\pm 0.0035$} & \makecell{$0.2416$\\$\pm 0.0093$} & \makecell{$0.0282$\\$\pm 0.0000$} & \makecell{$0.827$\\$\pm 0.001$} \\
16 & Teacher forcing & \makecell{$0.1090$\\$\pm 0.0001$} & \makecell{$0.5431$\\$\pm 0.0109$} & \makecell{$0.3573$\\$\pm 0.0148$} & \makecell{$0.0281$\\$\pm 0.0003$} & \makecell{$0.802$\\$\pm 0.004$} \\
16 & Full feedback & \makecell{$0.1084$\\$\pm 0.0001$} & \makecell{\textbf{0.3673}\\$\mathbf{\pm 0.0025}$} & \makecell{\textbf{0.2088}\\$\mathbf{\pm 0.0017}$} & \makecell{$0.0283$\\$\pm 0.0001$} & \makecell{$0.830$\\$\pm 0.001$} \\
\bottomrule
\end{tabular}
\end{table*}

At the common 30-epoch budget, increasing $F$ lowers rollout error under both teacher forcing and full feedback, and full feedback is lower at every tested horizon. Reset-window RelL2 remains between 0.1073 and 0.1090, so the separation appears primarily in uninterrupted rollout.

\subsection{Initialization under matched long-window adaptation}
\label{sec:appendix_f16_initialization}

The initialization comparison gives fixed seed-42 Transolver R1 and R3 checkpoints identical $F=16$ feedback training. The six runs use paired adaptation seeds 42--44, eight epochs, batch size eight, and a complete OneCycle schedule with peak learning rate $10^{-4}$. Each run predicts 602,752 training frames. Checkpoints are selected by source-validation error at epochs five and eight; test evaluation is omitted during training. Table~\ref{tab:appendix_f16_initialization} reports results on the same 79 sources.

\begin{table}[!htb]
\caption{Transolver initialization comparison after identical eight-epoch $F=16$ full-feedback adaptation. Entries are source-level means $\pm$ sample SD over three adaptation seeds, conditional on one fixed upstream checkpoint per initialization. The final column is R3 minus R1. Lower is better except hotspot overlap.}
\label{tab:appendix_f16_initialization}
\centering\normalsize
\setlength{\tabcolsep}{5pt}
\begin{tabular}{lrrr}
\toprule
Endpoint & R1 initialization & R3 initialization & Difference \\
\midrule
Reset RelL2 & $0.12811\pm0.00060$ & $0.10826\pm0.00004$ & $-0.01986$ \\
96-step rollout RelL2 & $0.46840\pm0.01369$ & $0.37076\pm0.00390$ & $-0.09764$ \\
Mean cumulative RelL2 & $0.38256\pm0.00826$ & $0.30933\pm0.00297$ & $-0.07323$ \\
TAWSS relative error & $0.32239\pm0.02310$ & $0.21591\pm0.00458$ & $-0.10649$ \\
OSI absolute error & $0.02747\pm0.00025$ & $0.02836\pm0.00013$ & $+0.00089$ \\
Top-10\% overlap & $0.83226\pm0.00015$ & $0.83737\pm0.00027$ & $+0.00512$ \\
\bottomrule
\end{tabular}
\end{table}

R3 initialization lowers rollout error in every paired seed, with improvements on 72, 74, and 72 of the 79 sources. Its mean reduction is 20.8\%; a source-bootstrap interval for the seed-averaged R3--R1 difference is $[-0.1250,-0.0735]$. TAWSS error also decreases, while OSI error rises slightly. The complete protocol and per-source results are recorded under \path{experiments/transient_r1_r3_f16_20260907/}.

\subsection{Cycle metrics, source distributions, and field examples}

For uninterrupted forecasts, Table~\ref{tab:appendix_rollout_extended} compares cycle and spatial endpoints for the temporal references and feedback variants.

\begin{table*}[!htb]
\caption{Exploratory full-cycle transient diagnostics. Lower is better except top-10\% overlap; peak timing uses circular frame distance. PointwiseMLP entries average three independently trained R1 pipelines. F=16 entries average three adaptation seeds from one fixed R3 initialization: 30 epochs for Transolver and 20 for U-Net. The eight-epoch initialization comparison is in Table~\ref{tab:appendix_f16_initialization}. The remaining neural-model rows use single validation-selected checkpoints.}
\label{tab:appendix_rollout_extended}
\centering\normalsize
\setlength{\tabcolsep}{4pt}
\textbf{(a) Trajectory and cycle errors}\par\vspace{3pt}
\begin{tabular}{@{}lrrrr@{}}
\toprule
Method & Rollout RelL2 & \makecell{Mean cumulative\\RelL2} & TAWSS rel. & OSI abs. \\
\midrule
Persistence & 0.8806 & 0.7389 & 0.7254 & 0.0298 \\
Linear extrapolation & 0.9929 & 0.8040 & 0.8349 & 0.1482 \\
History+phase ridge & 0.9911 & 0.8624 & 0.9045 & 0.0696 \\
PointwiseMLP R1 & 0.7346 & 0.6013 & 0.5694 & 0.0287 \\
U-Net R3 & 0.9824 & 0.8171 & 0.8740 & 0.0381 \\
U-Net SS(0.5) & 0.9711 & 0.8084 & 0.8670 & 0.0364 \\
U-Net F=16 & 0.9326 & 0.7809 & 0.8486 & 0.0395 \\
Transolver R3 & 0.9836 & 0.8017 & 0.8785 & 0.0569 \\
Transolver SS(0.5) & 0.9475 & 0.7771 & 0.8836 & 0.0550 \\
Transolver F=16 & 0.3673 & 0.3073 & 0.2088 & 0.0283 \\
Galerkin R3 & 1.1258 & 0.8519 & 0.9340 & 0.0371 \\
Galerkin SS(0.5) & 1.0729 & 0.8155 & 0.8242 & 0.0356 \\
\bottomrule
\end{tabular}
\par\vspace{6pt}
\textbf{(b) Hotspot and peak errors}\par\vspace{3pt}
\begin{tabular}{@{}lrrrr@{}}
\toprule
Method & Top-10\% rel. & Top-10\% overlap & \makecell{Peak time\\(frames)} & Peak mag. rel. \\
\midrule
Persistence & 0.7158 & 0.8466 & 26.97 & 0.9415 \\
Linear extrapolation & 0.8421 & 0.6388 & 27.62 & 0.9117 \\
History+phase ridge & 0.9278 & 0.7562 & 21.57 & 0.9213 \\
PointwiseMLP R1 & 0.5640 & 0.5194 & 0.89 & 0.3558 \\
U-Net R3 & 0.8541 & 0.6117 & 17.82 & 0.9236 \\
U-Net SS(0.5) & 0.8504 & 0.6294 & 17.96 & 0.9275 \\
U-Net F=16 & 0.8476 & 0.7371 & 26.25 & 0.9410 \\
Transolver R3 & 0.8749 & 0.8240 & 24.58 & 0.9222 \\
Transolver SS(0.5) & 0.8814 & 0.8297 & 25.92 & 0.9242 \\
Transolver F=16 & 0.1769 & 0.8299 & 0.20 & 0.1256 \\
Galerkin R3 & 0.8276 & 0.7155 & 18.00 & 0.8933 \\
Galerkin SS(0.5) & 0.7147 & 0.7519 & 18.00 & 0.7372 \\
\bottomrule
\end{tabular}
\end{table*}

\begin{figure}[!htb]
\centering
\includegraphics[width=\textwidth]{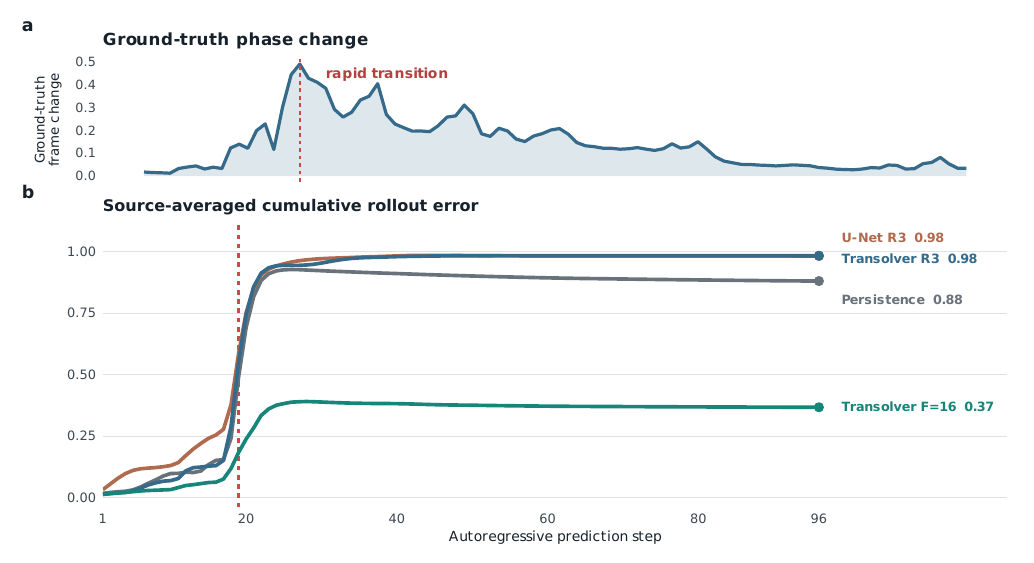}
\caption{Exploratory 96-step rollout diagnostics aligned with the cardiac phase. (a) Mean ground-truth frame change peaks near predicted step 19. (b) Source-mean cumulative error for four reference methods over the same rollout, with step-96 values labeled directly. The standard neural pipelines separate from 16-step full-feedback Transolver near the rapid transition. Additional exploratory methods are reported in Table~\ref{tab:appendix_rollout_extended}.}
\label{fig:rollout_phase_dynamics}
\end{figure}

U-Net has low adaptation-seed variability but still has large errors through the cardiac transition, with a cumulative-error rise at the same phase as the short-horizon checkpoint.

\begin{figure}[!htb]
\centering
\includegraphics[width=\textwidth]{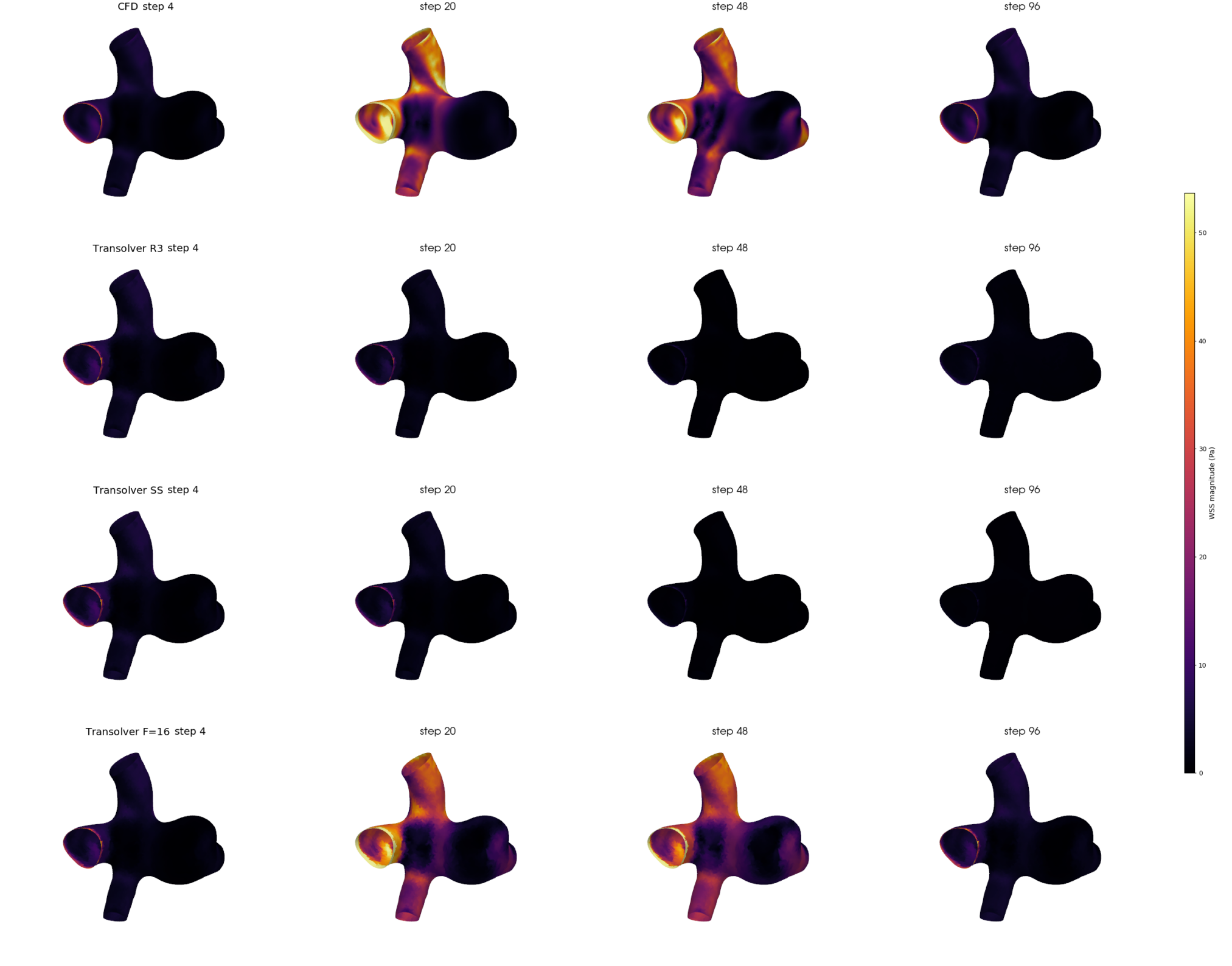}
\caption{WSS evolution for an illustrative held-out case at rollout steps 4, 20, 48, and 96. Rows compare CFD, standard Transolver R3, scheduled sampling, and $F=16$ full-feedback adaptation. All panels use a shared scale derived from ground truth. Values at evaluation vertices are shown directly; values elsewhere are nearest-neighbor interpolated for display. Table~\ref{tab:appendix_rollout_extended} reports cohort-level trajectory and hemodynamic metrics.}
\label{fig:appendix_wss_rollout_phases}
\end{figure}

\begin{figure}[!htb]
\centering
\includegraphics[width=0.7\textwidth]{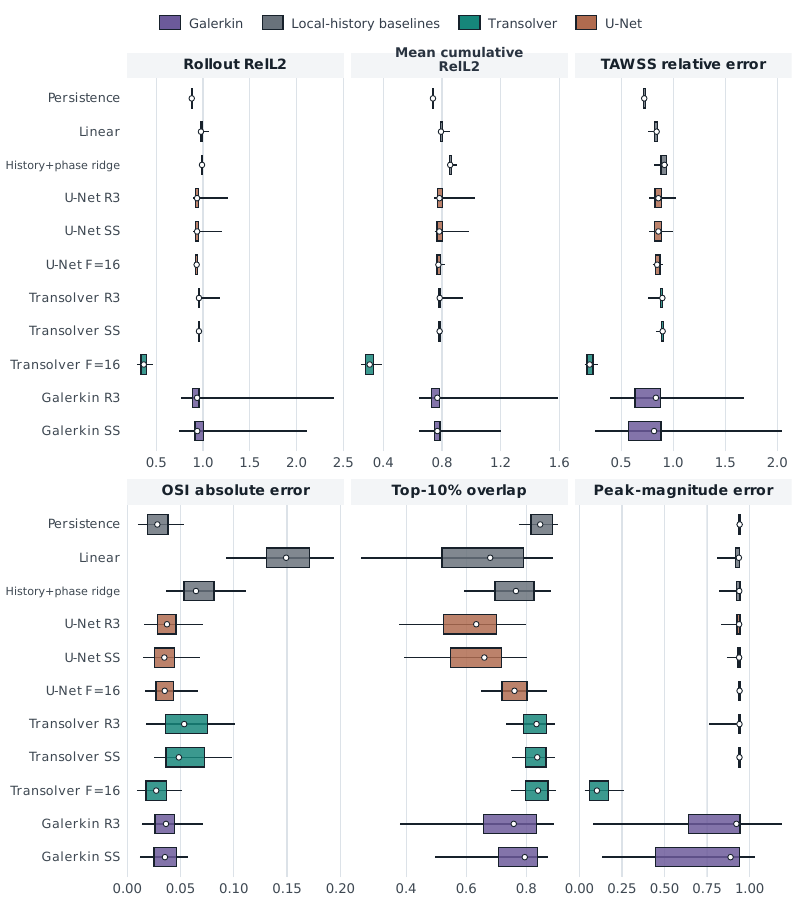}
\caption{Exploratory source-level 96-step rollout distributions for temporal reference baselines and the R3, scheduled-sampling, and F=16 interventions. Each panel uses the metric's natural scale; ``Trajectory AUC'' denotes mean cumulative RelL2. Boxes show quartiles and whiskers the 5th--95th percentiles. Top-10\% overlap is higher-is-better; all other panels are errors.}
\label{fig:appendix_rollout_source_distributions}
\end{figure}

Most sources have high rollout errors and low predicted amplitudes under standard R3 or scheduled sampling. Galerkin's pronounced right tail raises its mean above its median. Transolver F=16 lowers rollout, TAWSS, and peak-magnitude errors. Persistence has the highest median top-10\% overlap, despite a larger mean trajectory error than Transolver F=16.

These distributions complement the mean trajectories in Figure~\ref{fig:rollout_phase_dynamics}b. Most standard neural models show an abrupt error increase near the rapid transition; Transolver F=16 accumulates error more gradually.

\subsection{Coordinate and history interventions}

The intervention study evaluates fixed reset windows after perturbing coordinates, histories, or both at inference. Coordinate permutation reassigns $xyz$ across wall points while leaving each WSS history and target paired. Coordinate zeroing sets the explicit position inputs to zero. History permutation instead breaks the correspondence between each wall point and its local WSS trace. A joint coordinate--history permutation, undone before scoring, checks permutation equivariance. These interventions evaluate the trained Transolver and Galerkin Transformer checkpoints.

For the raw-coordinate Transolver checkpoint, coordinate permutation changes source-mean error by $-2.2\times10^{-6}$ and zeroing by $-1.4\times10^{-5}$; both paired intervals cross zero. Galerkin Transformer changes by $+2.7\times10^{-6}$ and $-3.4\times10^{-7}$. History permutation raises the two errors by 1.2552 and 1.2540, and every source degrades. Under the reset protocol, these checkpoints draw most of their predictive signal from the registered WSS histories, with little incremental use of explicit $xyz$.

\begin{figure*}[!htb]
\centering
\includegraphics[width=0.8\textwidth]{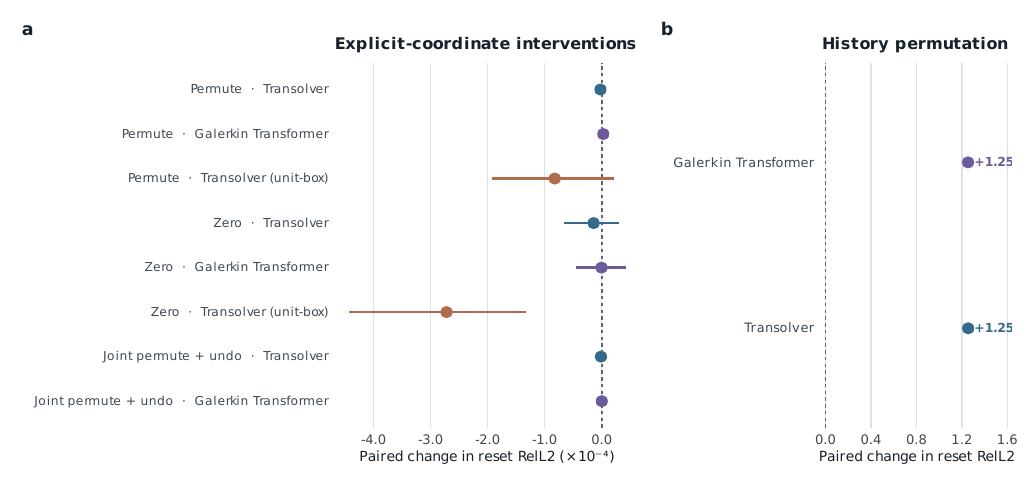}
\caption{Explicit-coordinate and registered-history interventions on point-attention reset checkpoints. (a) Changes from coordinate permutation, coordinate zeroing, and the joint permutation--undo check for the two raw-coordinate checkpoints, shown in units of $10^{-4}$. The panel also includes a unit-box Transolver sensitivity. (b) History-only permutation on the natural RelL2 scale. Breaking the WSS-history correspondence raises error by about 1.255, whereas explicit-coordinate and joint-equivariance changes remain near zero. Points are mean source-paired changes and bars are 95\% source-bootstrap intervals over the same 79 held-out sources.}
\label{fig:coordinate_history_intervention}
\end{figure*}

\section{Representation and Physical-Field Diagnostics}
\label{sec:appendix_physics_sensitivity}

This section reports how weighting, pressure definitions, input coordinates, and point sampling affect the reference results. Definitions are in Section~\ref{sec:appendix_metrics}; the original sampling and input conventions are in Section~\ref{sec:appendix_training_details}.

\subsection{Native-volume and wall-area weighting}

The sensitivity analysis reevaluates representative R1/R3 checkpoints with native-mesh weights and pressure diagnostics (Table~\ref{tab:appendix_physics_sensitivity}). Sampling domains and source-level reduction follow Appendices~\ref{sec:appendix_training_details} and~\ref{sec:appendix_metrics}; intervals resample the 79 test sources. The comparison measures sensitivity to uniform weighting of stored or sampled vertices.

\begin{table*}[!htb]
\caption{Representative quadrature and pressure-gauge sensitivity on the 79 held-out sources. Panel (a) compares the primary, uniformly weighted vertex/mask evaluation with native-VTU lumped cell-volume weighting; uniform voxel volumes cancel exactly. Centered-pressure RelL2 ($\mathcal E_p^{\mathrm{center}}$) subtracts each field's weighted spatial mean. The final two columns are sampled inlet--outlet pressure-drop absolute error in Pa and relative error; both are unavailable for the voxel representation. Panel (b) uses native wall-polygon area weights. Negative change favors R3.}
\label{tab:appendix_physics_sensitivity}
\centering\normalsize
\textbf{(a) Steady velocity and pressure}\par\vspace{2pt}
\setlength{\tabcolsep}{2.5pt}
\begin{tabular}{lllrrrrrrr}
\toprule
Model & Track & Regime & $\mathcal E_u$ & $\mathcal E_u^{(V)}$
& $\mathcal E_p$ & $\mathcal E_p^{(V)}$ & $\mathcal E_p^{\mathrm{center}}$
& Drop abs. (Pa) & Drop rel. \\
\midrule
GraphSAGE & Points & R1 & 0.8269 & 0.8011 & 0.7131 & 0.6769 & 1.1994 & 186.6 & 0.5317 \\
GraphSAGE & Points & R3 & 0.6521 & 0.6395 & 0.5525 & 0.5287 & 0.9211 & 137.7 & 0.4176 \\
Transolver & Points & R1 & 1.0031 & 1.0009 & 0.8672 & 0.8021 & 1.0228 & 401.6 & 0.9962 \\
Transolver & Points & R3 & 0.6979 & 0.6714 & 0.5783 & 0.5062 & 0.8614 & 124.5 & 0.3864 \\
U-Net & Voxel & R1 & 0.9867 & 0.9867 & 0.8147 & 0.8147 & 1.0174 & -- & -- \\
U-Net & Voxel & R3 & 0.7050 & 0.7050 & 0.7930 & 0.7930 & 1.1333 & -- & -- \\
FNO & Voxel & R1 & 1.0099 & 1.0099 & 1.3334 & 1.3334 & 1.2366 & -- & -- \\
FNO & Voxel & R3 & 0.8387 & 0.8387 & 0.9126 & 0.9126 & 1.1802 & -- & -- \\
\bottomrule
\end{tabular}

\vspace{5pt}\textbf{(b) Reset-window WSS}\par\vspace{2pt}
\begin{tabular}{llrrrrr}
\toprule
Model & Track & R1 vertex & R1 area & R3 vertex & R3 area & Area-weighted R3--R1 \\
\midrule
Transolver & Points & 0.1253 & 0.1303 & 0.1084 & 0.1136 & -12.9\% \\
Galerkin Transformer & Points & 0.1210 & 0.1251 & 0.1090 & 0.1139 & -9.0\% \\
U-Net & Voxel & 0.2369 & 0.2404 & 0.2064 & 0.2102 & -12.6\% \\
FNO & Voxel & 0.2387 & 0.2412 & 0.2145 & 0.2186 & -9.4\% \\
\bottomrule
\end{tabular}
\end{table*}

With volume weighting, R3 has lower velocity and stored-gauge pressure error for all four representative steady models. Wall-area weighting gives lower R3 WSS error for all four transient models.

\subsection{Pressure structure and pressure-drop results}

Centered-pressure RelL2 varies across models: GraphSAGE and Transolver improve, FNO has a lower unit-box R3 point estimate, and voxel U-Net does not improve. The FNO rerun with centered physical coordinates has a paired interval spanning zero. Both unstructured models also have lower pressure-drop error under R3.

The models with the lowest pressure RelL2 have negative mean per-case $R^2$. RelL2 normalizes error by the target-field norm, RMSE retains physical units, and $R^2$ uses centered target energy (Section~\ref{sec:appendix_metrics}). For an individual case, negative $R^2$ means that residual energy exceeds the squared error of predicting its target mean. Centered-pressure error and sampled-edge gradients assess errors in spatial variation; the common-plane comparison shows where predictions depart from the target fields.

\subsection{Voxel coordinate-scale sensitivity}

The unit-box voxel reference omits the stored bounding-box scale. The physical-coordinate sensitivity reruns the complete steady R1/R2/R3 pipelines for U-Net and FNO with centered physical coordinates, retaining the original occupancy, grid resolution, exported point domain, splits, normalization protocol, optimizer schedules, and test sources. Checkpoint selection uses the source-validation cohort. These runs change coordinate scale; the cell-connected point-domain evaluation follows in Section~\ref{sec:appendix_valid_point_domain}.

\begin{table*}[!htb]
\caption{Sensitivity to centered physical coordinates on 79 held-out sources. Panel (a) reports source-averaged point estimates. Panel (b) gives paired R3--R1 changes with 10,000 source-bootstrap resamples; negative values favor R3. These reruns restore the stored longest-side physical scale. The main voxel table reports the original unit-box checkpoints.}
\label{tab:appendix_voxel_scale_sensitivity}
\centering\normalsize
\textbf{(a) R1/R2/R3 endpoints}\par\vspace{2pt}
\setlength{\tabcolsep}{5pt}
\begin{tabular}{llrrr}
\toprule
Model & Regime & Velocity RelL2 & \makecell{Stored-gauge\\pressure RelL2} & \makecell{Centered-pressure\\RelL2} \\
\midrule
U-Net & R1 & 0.9803 & 1.5510 & 1.3582 \\
      & R2 & 0.8209 & 0.7442 & 1.0601 \\
      & R3 & 0.7264 & 0.6702 & 0.9786 \\
\addlinespace[2pt]
FNO   & R1 & 1.0093 & 1.3407 & 1.2235 \\
      & R2 & 0.8585 & 1.0012 & 1.2836 \\
      & R3 & 0.8403 & 0.9276 & 1.2037 \\
\bottomrule
\end{tabular}

\vspace{5pt}\textbf{(b) Paired R3--R1 effects}\par\vspace{2pt}
\setlength{\tabcolsep}{3.5pt}
\begin{tabular}{llrrrr}
\toprule
Model & Endpoint & \makecell{R3--R1\\difference} & \makecell{Relative\\change} & \makecell{95\% paired-\\source CI} & \makecell{Sources\\improved} \\
\midrule
U-Net & Velocity & $-0.2539$ & $-25.9\%$ & [$-0.3118,-0.1950$] & 82.3\% \\
      & Stored-gauge pressure & $-0.8809$ & $-56.8\%$ & [$-1.0439,-0.7367$] & 98.7\% \\
      & Centered-pressure RelL2 & $-0.3796$ & $-28.0\%$ & [$-0.4967,-0.2793$] & 89.9\% \\
\addlinespace[2pt]
FNO   & Velocity & $-0.1690$ & $-16.7\%$ & [$-0.2303,-0.1077$] & 68.4\% \\
      & Stored-gauge pressure & $-0.4131$ & $-30.8\%$ & [$-0.5924,-0.2803$] & 94.9\% \\
      & Centered-pressure RelL2 & $-0.0198$ & $-1.6\%$ & [$-0.0835,+0.0438$] & 60.8\% \\
\bottomrule
\end{tabular}
\end{table*}

With centered physical coordinates, U-Net's R3 endpoint is lower than R1 for velocity, stored-gauge pressure, and centered-pressure RelL2, with paired intervals excluding zero; its unit-box centered-pressure RelL2 instead rises. FNO retains velocity and stored-gauge pressure reductions, but its centered-pressure RelL2 change is inconclusive. Coordinate scale changes the direction of U-Net's centered-pressure comparison in this steady task. Centered physical coordinates in metres are recommended for new experiments.

\subsection{Cell-connected point-domain evaluation}
\label{sec:appendix_valid_point_domain}

Some exported volume points are not referenced by any CFD cell and store zero velocity and pressure. Across all 280 training and 42 validation sources at $0.002\,\mathrm{kg\,s^{-1}}$, their median shares are 6.28\% and 6.26\%. Native-cell probes on four sources locate more than 99\% of these coordinates inside active cells, with median interpolated speeds of $0.0377$--$0.1164\,\mathrm{m\,s^{-1}}$.

We define valid supervised points as the unique vertices referenced by mesh cells, preserving connected wall vertices and their physical zero velocities. Selection depends on mesh connectivity, not on field values. Cell-connected preprocessing applies this domain before point sampling, estimation of training normalization statistics, and voxel interpolation.

For new steady point-model training, \path{scripts/train_steady_physical.py} selects the \texttt{cell\_connected} domain. Voxel construction through \path{data/build_voxel.py} uses the same default. The workflow in \path{docs/STEADY_POINT_DOMAIN.md} specifies normalization and the matching training and evaluation commands; voxel training additionally selects \texttt{physical\_centered} coordinates. The archived reproduction commands use \texttt{all}, matching the original arrays used to train the reported steady reference and allocation checkpoints. The physical-coordinate reruns above also use this original point domain.

To separate scoring effects from changes in model inputs, we hold the six GraphSAGE allocation checkpoints and their normalization fixed and evaluate all 79 sources at eight flow conditions. From each original deterministic 4,096-point sample, we first restrict scoring to cell-connected entries while retaining all inputs. We then remove orphan inputs and rebuild the graph, evaluating at the same active query IDs and targets. The second comparison changes graph neighborhoods and GraphSAGE batch statistics as well as input membership.

\begin{table}[!htb]
\caption{Effects of scoring and input point domains on the two GraphSAGE allocations. Values are source-mean RelL2 $\pm$ sample SD across three training seeds. Model weights are fixed during evaluation. The two cell-connected evaluations use the same subset of the original query points.}
\label{tab:appendix_valid_point_domain}
\centering\normalsize
\setlength{\tabcolsep}{3pt}
\begin{tabularx}{\textwidth}{Xrrrr}
\toprule
& \multicolumn{2}{c}{Velocity} & \multicolumn{2}{c}{Stored-gauge pressure} \\
\cmidrule(lr){2-3}\cmidrule(l){4-5}
Evaluation & $280\times1$ & $35\times8$ & $280\times1$ & $35\times8$ \\
\midrule
Original inputs and scoring & $0.7993\pm0.0186$ & $0.8480\pm0.0218$ & $0.7167\pm0.0324$ & $0.7856\pm0.0397$ \\
Active scoring, original inputs & $0.7766\pm0.0178$ & $0.8265\pm0.0205$ & $0.6620\pm0.0291$ & $0.7327\pm0.0431$ \\
Active inputs and scoring & $0.7788\pm0.0186$ & $0.8274\pm0.0216$ & $0.6705\pm0.0334$ & $0.7363\pm0.0443$ \\
\bottomrule
\end{tabularx}
\end{table}

The point domain changes the absolute errors, while both mean endpoints remain lower for $280\times1$ in all three evaluations. Velocity is lower in all three paired seeds and pressure in two, matching their original directions. These comparisons establish the ordering for the six fixed checkpoints on valid queries. Training on the cell-connected domain remains to be evaluated under the same allocation budgets.

\subsection{Wall-point sampling sensitivity}

The main transient protocol evaluates 4,096 deterministic wall vertices per source. Table~\ref{tab:appendix_sampling_sensitivity} repeats the reset-window evaluation with 8,192 vertices and with every available wall vertex. The two attention models change by at most $2.4\times10^{-4}$ RelL2, while full-surface evaluation spans 5,437--29,686 vertices per source. The reset-window ranking is unchanged across these evaluation point budgets.

\begin{table}[!htb]
\caption{Transient reset-window sensitivity to wall-point evaluation density. All columns evaluate the same R3 checkpoint on the same 79 held-out sources; lower RelL2 is better.}
\label{tab:appendix_sampling_sensitivity}
\centering\normalsize
\setlength{\tabcolsep}{8pt}
\begin{tabular}{lrrr}
\toprule
Model & Fixed 4,096 & Fixed 8,192 & Full surface \\
\midrule
Transolver & 0.108406 & 0.108419 & 0.108648 \\
Galerkin Transformer & 0.109043 & 0.108883 & 0.108975 \\
\bottomrule
\end{tabular}
\end{table}

\subsection{Steady fine-tuning recipe sensitivity}

The steady R3 comparisons vary the fine-tuning learning rate or target normalization one at a time (Table~\ref{tab:appendix_steady_robustness}). Across the seven models, each alternative changes velocity RelL2 by at most 0.0158 and pressure RelL2 by at most 0.0354. The direction depends on the model and endpoint: a setting that lowers pressure error can raise velocity error.

\begin{table*}[!htb]
\caption{Targeted steady R3 sensitivity runs. Cells are velocity / pressure RelL2. The base setting uses learning rate $10^{-4}$ and source-training normalization; the other columns change one choice at a time.}
\label{tab:appendix_steady_robustness}
\centering\normalsize
\setlength{\tabcolsep}{5pt}
\begin{tabular}{llccc}
\toprule
Track & Model & \makecell{Base\\R3} & \makecell{Learning rate\\$5\times10^{-4}$} & \makecell{Synthetic\\normalization} \\
\midrule
Unstructured & \makecell[l]{Galerkin\\Transformer} & 0.8150 / 0.6667 & 0.8227 / 0.6880 & 0.8143 / 0.6548 \\
Unstructured & GraphSAGE & 0.6521 / 0.5525 & 0.6550 / 0.5465 & 0.6542 / 0.5421 \\
Unstructured & PointNet & 0.8899 / 0.7718 & 0.8945 / 0.7598 & 0.9021 / 0.7687 \\
Unstructured & Transolver & 0.6979 / 0.5783 & 0.6997 / 0.5875 & 0.6997 / 0.5920 \\
Voxel & FNO & 0.8387 / 0.9126 & 0.8405 / 0.9278 & 0.8414 / 0.9106 \\
Voxel & U-FNO & 0.7568 / 0.7925 & 0.7603 / 0.8168 & 0.7622 / 0.8279 \\
Voxel & U-Net & 0.7050 / 0.7930 & 0.7208 / 0.8013 & 0.7086 / 0.7874 \\
\bottomrule
\end{tabular}

\end{table*}

\subsection{Auxiliary metrics}

The following tables report amplitude errors, directional errors, and pressure $R^2$ using the definitions in Appendix~\ref{sec:appendix_metrics}. Metrics are computed from decoded fields: pressure RMSE is in Pa and velocity RMSE in $\mathrm{m\,s^{-1}}$. Each model is evaluated on its specified domain before any interpolation for visualization.

\begin{table*}[!htb]
\caption{R3 steady velocity diagnostics on 79 held-out source geometries. Metrics are computed from decoded fields, averaged over eight flow conditions within each source, and then averaged across sources. RelL2 brackets are 95\% source-bootstrap CIs. MSE has units $\mathrm{m^2\,s^{-2}}$; RMSE, MAE, and magnitude MAE have units $\mathrm{m\,s^{-1}}$.}
\label{tab:appendix_steady_velocity_metrics}
\centering\normalsize
\setlength{\tabcolsep}{4pt}
\textbf{(a) Field errors}\par\vspace{3pt}
\begin{tabular}{@{}lrrrr@{}}
\toprule
Model & \makecell{RelL2\\95\% CI} & Mean MSE & Mean RMSE & Mean MAE \\
\midrule
\multicolumn{5}{l}{\emph{Unstructured}} \\
GraphSAGE & 0.6521 [0.6070, 0.6982] & 1.342e-02 & 0.0948 & 0.0622 \\
Transolver & 0.6979 [0.6467, 0.7510] & 1.677e-02 & 0.1033 & 0.0673 \\
Galerkin Transformer & 0.8150 [0.7536, 0.8769] & 2.332e-02 & 0.1206 & 0.0786 \\
PointNet & 0.8899 [0.8257, 0.9593] & 2.291e-02 & 0.1260 & 0.0880 \\
\addlinespace[3pt]
\multicolumn{5}{l}{\emph{Voxel}} \\
U-Net & 0.7050 [0.6306, 0.7753] & 1.482e-02 & 0.0981 & 0.0578 \\
U-FNO & 0.7568 [0.6709, 0.8478] & 1.653e-02 & 0.1037 & 0.0602 \\
FNO & 0.8387 [0.7798, 0.8952] & 1.915e-02 & 0.1167 & 0.0728 \\
U-NO & 0.8303 [0.7591, 0.9089] & 1.899e-02 & 0.1142 & 0.0696 \\
F-FNO & 1.0005 [0.9462, 1.0601] & 2.787e-02 & 0.1397 & 0.0872 \\
\bottomrule
\end{tabular}
\par\vspace{6pt}
\textbf{(b) Magnitude, direction, and explained variance}\par\vspace{3pt}
\begin{tabular}{@{}lrrrr@{}}
\toprule
Model & Mean $|\mathbf u|$ MAE & Angle ($^\circ$) & Cosine & Mean case $R^2$ \\
\midrule
\multicolumn{5}{l}{\emph{Unstructured}} \\
GraphSAGE & 0.0788 & 48.26 & 0.5629 & 0.5189 \\
Transolver & 0.0901 & 52.78 & 0.4902 & 0.4354 \\
Galerkin Transformer & 0.1008 & 60.91 & 0.3837 & 0.2270 \\
PointNet & 0.1177 & 62.09 & 0.3830 & 0.0764 \\
\addlinespace[3pt]
\multicolumn{5}{l}{\emph{Voxel}} \\
U-Net & 0.0796 & 52.21 & 0.4674 & 0.3602 \\
U-FNO & 0.0805 & 52.88 & 0.4654 & 0.2236 \\
FNO & 0.1009 & 66.15 & 0.3070 & 0.1936 \\
U-NO & 0.0945 & 61.56 & 0.3624 & 0.1593 \\
F-FNO & 0.1197 & 75.73 & 0.1925 & -0.1169 \\
\bottomrule
\end{tabular}
\end{table*}

\begin{table*}[!htb]
\caption{R3 steady stored-gauge pressure diagnostics on the same 79 held-out sources. MSE is in $\mathrm{Pa^2}$; RMSE and MAE are in Pa. Columns average per-condition metrics within sources, then across sources. Per-case $R^2$ values are averaged without clipping; negative per-case values indicate greater squared error than prediction of that case's target mean.}
\label{tab:appendix_steady_pressure_metrics}
\centering\normalsize
\setlength{\tabcolsep}{2.5pt}
\begin{tabular}{@{}lrrrrr@{}}
\toprule
Model & \makecell{RelL2\\95\% CI} & \makecell{Mean MSE\\($\mathrm{Pa^2}$)} & \makecell{Mean RMSE\\(Pa)} & \makecell{Mean MAE\\(Pa)} & \makecell{Mean case\\$R^2$} \\
\midrule
\multicolumn{6}{l}{\emph{Unstructured}} \\
GraphSAGE & 0.5525 [0.4679, 0.6570] & 2.938e+04 & 96.151 & 75.164 & -0.9774 \\
Transolver & 0.5783 [0.5083, 0.6681] & 3.697e+04 & 107.712 & 85.817 & -1.0354 \\
Galerkin Transformer & 0.6667 [0.5741, 0.7917] & 4.912e+04 & 122.183 & 97.056 & -1.8980 \\
PointNet & 0.7718 [0.6782, 0.8885] & 4.202e+04 & 130.467 & 104.165 & -2.6911 \\
\addlinespace[3pt]
\multicolumn{6}{l}{\emph{Voxel}} \\
U-Net & 0.7930 [0.6700, 0.9208] & 4.344e+04 & 125.318 & 99.013 & -2.6535 \\
U-FNO & 0.7925 [0.6849, 0.9198] & 4.833e+04 & 130.274 & 102.988 & -2.6587 \\
FNO & 0.9126 [0.7668, 1.0761] & 4.305e+04 & 130.897 & 105.238 & -3.8250 \\
U-NO & 0.8946 [0.7478, 1.0627] & 4.303e+04 & 129.908 & 104.490 & -3.8761 \\
F-FNO & 1.1466 [0.9527, 1.3685] & 6.332e+04 & 155.237 & 126.218 & -7.0027 \\
\bottomrule
\end{tabular}
\end{table*}

\begin{table*}[!htb]
\caption{R3 reset-window $4\rightarrow4$ WSS diagnostics on 79 held-out source geometries, with observed histories restored between windows. Values are source means of per-window metrics. MSE is in $\mathrm{Pa^2}$; RMSE, component MAE, and magnitude MAE are in Pa. Direction metrics include only locations at which both vector magnitudes exceed $10^{-8}$ Pa.}
\label{tab:appendix_transient_multimetric}
\centering\normalsize
\setlength{\tabcolsep}{4pt}
\textbf{(a) Field errors}\par\vspace{3pt}
\begin{tabular}{@{}lrrrr@{}}
\toprule
Model & \makecell{RelL2\\95\% CI} & Mean MSE & Mean RMSE & Mean MAE \\
\midrule
\multicolumn{5}{l}{\emph{Unstructured}} \\
Transolver & 0.1084 [0.0969, 0.1218] & 9.021e+01 & 3.694 & 2.075 \\
Galerkin Transformer & 0.1090 [0.0977, 0.1221] & 9.595e+01 & 3.747 & 2.091 \\
PointNet & 0.1484 [0.1373, 0.1610] & 1.332e+02 & 4.582 & 2.648 \\
GraphSAGE & 0.2235 [0.2014, 0.2498] & 1.633e+02 & 5.577 & 3.240 \\
\addlinespace[3pt]
\multicolumn{5}{l}{\emph{Voxel}} \\
U-Net & 0.2064 [0.1958, 0.2191] & 2.019e+02 & 5.919 & 3.392 \\
F-FNO & 0.2104 [0.2002, 0.2225] & 2.153e+02 & 6.037 & 3.445 \\
FNO & 0.2145 [0.2044, 0.2263] & 2.275e+02 & 6.194 & 3.504 \\
U-FNO & 0.2227 [0.2123, 0.2346] & 2.335e+02 & 6.437 & 3.678 \\
U-NO & 0.2230 [0.2122, 0.2354] & 2.419e+02 & 6.491 & 3.699 \\
\bottomrule
\end{tabular}
\par\vspace{6pt}
\textbf{(b) Magnitude and direction}\par\vspace{3pt}
\begin{tabular}{@{}lrrr@{}}
\toprule
Model & Mean magnitude MAE & Angle ($^\circ$) & Cosine \\
\midrule
\multicolumn{4}{l}{\emph{Unstructured}} \\
Transolver & 2.849 & 7.78 & 0.9558 \\
Galerkin Transformer & 2.878 & 7.83 & 0.9556 \\
PointNet & 3.983 & 9.15 & 0.9464 \\
GraphSAGE & 4.914 & 14.46 & 0.9144 \\
\addlinespace[3pt]
\multicolumn{4}{l}{\emph{Voxel}} \\
U-Net & 5.816 & 9.01 & 0.9457 \\
F-FNO & 5.918 & 9.42 & 0.9436 \\
FNO & 6.042 & 9.09 & 0.9459 \\
U-FNO & 6.405 & 8.84 & 0.9479 \\
U-NO & 6.437 & 9.25 & 0.9450 \\
\bottomrule
\end{tabular}
\end{table*}

\FloatBarrier
\section{Release Artifacts, Data Access, and Reproduction}
\label{sec:appendix_artifacts}

\paragraph{Review packages and supported tasks.}

The inspection subset contains four training sources and eight descendants, with all eight steady conditions and full transient fields. Selection spans quantiles of transient HDF5 file size, first among eligible training sources and then among descendants within each selected family.

Native-resolution wall coordinates and 100-frame WSS arrays for 277 training and 42 validation sources support source-only WSS training and feedback adaptation with the supplied recipes and normalization statistics. The fixed-test package provides 4,096 wall points and 100 WSS frames for each of the 79 test sources, the R1/R3 Transolver weights, and six matched $F=16$ adaptation checkpoints. Its script recomputes reset, rollout, and cycle metrics.

The supplementary code ZIP includes training and evaluation scripts, splits, availability records, normalization statistics, and source-level results for reproducing reported aggregates. A compact HDF5 sample exercises the loaders. The \href{https://figshare.com/s/f54b0f5b728524e89a5b}{Figshare review collection} hosts the data and checkpoint ZIPs.

GraphSAGE allocation retraining requires each pool's selected descendants, common source fine-tuning fields, and validation and test fields under all eight steady inlet conditions. These data are provided in the full release. The full VTU/VTK and HDF5 corpus will receive permanent hosting after de-anonymization, with upstream notices. Complete OpenFOAM case configurations are archived separately.

\paragraph{Reproduction scope and attribution.}

Release versions identify the source split and preprocessing used for each result. The data documentation specifies physical units and time grids. Code uses Apache-2.0, and AneuX-derived data carry CC BY-NC 4.0 terms. The upstream notices specify AFBench's CC BY-NC-SA 4.0 data license~\citep{liu2024afbench}, the airfoil geometry archive, and the XFoil version and source revision used to recompute pressure coefficients.

\begin{table}[!htb]
\caption{Reproduction tasks supported by the anonymous review packages and the full release. For numerical provenance, ``Spec.'' denotes the solver specification; ``Archive'' denotes retained OpenFOAM case configurations.}
\label{tab:artifact_reproduction_levels}
\centering\normalsize
\setlength{\tabcolsep}{3pt}
\begin{tabularx}{\textwidth}{@{}>{\raggedright\arraybackslash}p{0.19\textwidth}Xcc@{}}
\toprule
Reproduction task & Materials and use & Review & Full \\
\midrule
Cohort and schema & Source-family splits, availability records, and the compact HDF5 sample; verifies identifiers, inherited splits, array shapes, units, and time grids. & \checkmark & \checkmark \\
Metric and report & Source-level result JSON files and evaluators; regenerates reported aggregates, paired intervals, tables, and figures. & \checkmark & \checkmark \\
Loader and evaluator example & Baseline configurations, loaders, and the compact sample; exercises data loading and endpoint computation on sample fields. & \checkmark & \checkmark \\
Checkpoint inference & Fixed 79-source WSS samples and actual Transolver weights; recomputes reset, rollout, and cycle metrics. & \checkmark & \checkmark \\
Source-only WSS training and adaptation & Native-resolution WSS arrays for 277 training and 42 validation sources, with training recipes and normalization statistics. & \checkmark & \checkmark \\
Steady allocation retraining & Pool manifests, selected descendant fields, common source fine-tuning fields, and complete steady validation and test fields. & -- & \checkmark \\
Full retraining & Complete VTU/VTK and HDF5 corpus with training recipes; recreates selected checkpoints and source-level records. & -- & \checkmark \\
Numerical provenance & Solver assumptions in this appendix and retained OpenFOAM case configurations; documents upstream field generation. & Spec. & Archive \\
\bottomrule
\end{tabularx}
\end{table}

\end{document}